\documentclass [12pt,preprint]{aastex}
\usepackage{epsfig}
\begin{document}
\voffset-0.5cm
\newcommand{\gsim}{\hbox{\rlap{$^>$}$_\sim$}}
\newcommand{\lsim}{\hbox{\rlap{$^<$}$_\sim$}}

\title{The diverse broad-band lightcurves of Swift GRBs\\ 
reproduced with the cannonball model}

\author{Shlomo Dado\altaffilmark{1}, Arnon Dar\altaffilmark{2},
A. De R\'ujula\altaffilmark{3}}

\altaffiltext{1}{dado@phep3.technion.ac.il,
Physics Department, Technion, Haifa 32000,
Israel}
\altaffiltext{2}{ arnon@physics.technion.ac.il,
Physics Department, Technion, Haifa 32000, Israel\\
~~~~dar@cern.ch, Theory Unit, CERN,1211 Geneva 23, Switzerland}

\altaffiltext{3}{alvaro.derujula@cern.ch; Theory Unit, CERN,
1211 Geneva 23, Switzerland \\
Physics Department, Boston University, USA}

\begin{abstract} 

Two radiation mechanisms, inverse Compton scattering (ICS) and synchrotron 
radiation (SR), suffice within the cannonball (CB) model of long gamma ray 
bursts (LGRBs) and X-ray flashes (XRFs) to provide a very simple and 
accurate description of their observed prompt emission and afterglows. 
Simple as they are, the two mechanisms and the burst environment generate 
the rich structure of the light curves at all frequencies and times. This 
is demonstrated for 33 selected Swift LGRBs and XRFs, which are  
well sampled from early until late time and faithfully represent the entire 
diversity of the broad-band light curves of Swift LGRBs and XRFs. Their 
prompt gamma-ray and X-ray emission is dominated by ICS of `glory' light. 
During their fast decline phase, ICS is taken over by SR, which 
dominates their broad-band afterglow. The pulse shape and spectral 
evolution of the gamma-ray peaks and the early-time X-ray flares, and even 
the delayed optical `humps' in XRFs, are correctly predicted. The 
`canonical' and non-canonical X-ray light curves and the chromatic behaviour 
of the broad-band afterglows are well reproduced. In particular, in 
canonical X-ray light curves, the initial fast decline and rapid 
softening of the prompt emission, the transition to the plateau phase, the 
subsequent gradual steepening of the plateau to an asymptotic power-law 
decay, and the transition from chromatic to achromatic behaviour of the 
light curves agrees well with those predicted by the CB model. The Swift 
early-time data on XRF 060218 are inconsistent with a black-body emission 
from a shock break-out through a stellar envelope. Instead, they are well 
described by ICS of glory light by a jet breaking out from SN2006aj. 

\end{abstract}


\maketitle

\section{Introduction}

Since the launch of the Swift satellite, precise data from its Burst Alert 
Telescope (BAT) and X-Ray Telescope (XRT) have been obtained on the 
spectral and temporal behaviour of the X-ray emission of long-duration 
$\gamma$-ray bursts (LGRBs) and X-ray flashes (XRFs) from their beginning 
until late times.  The early data are often complemented by the 
ultraviolet-optical telescope (UVOT) on board Swift, and by ground-based 
{\it UVO} and $NIR$ robotic and conventional telescopes. The ensemble of these 
data have already been used to test the most-studied theories of long 
duration GRBs and their afterglows (AGs), the {\it Fireball} (FB) models 
(see, e.g.~Zhang \& M\'esz\'aros~2004, Zhang~2007, and references 
therein) 
and the {\it Cannonball} (CB) model [see, e.g.~Dar \& De R\'ujula~2004 
(hereafter DD2004), Dado, Dar \& De R\'ujula (hereafter DDD)~2002a, 2003a, 
and references therein].

The Swift X-ray light curves of LGRBs roughly divide into two classes, 
`canonical' and non-canonical (Nousek et al.~2006, O'Brien et al.~2006, 
Zhang~2007).  When measured early enough, the observed X-ray emission has 
prompt peaks which coincide with the $\gamma$-ray peaks of the GRB, and a 
rapidly declining light curve with a fast spectral softening after the 
last detectable peak of the GRB. This rapid decline and spectral softening 
of the prompt emission end within a few hundreds of seconds. In 
canonical LGRBs the X-ray light curve turns sharply into a much flatter 
`plateau' with a much harder power-law spectrum, typically lasting 
thousands to tens of thousands of seconds, and within a time of order one 
day it steepens into a power-law decay, which lasts until the X-ray AG 
becomes too dim to be detected (Fig.~\ref{f1}). 

The plateau phase is missing in non canonical GRBs,
and the asymptotic power-law decline begins the decay of the
prompt emission and lasts until the X-ray become too dim to be detected
(Fig.~\ref{f2}) without any observable break.

In an significant fraction of otherwise canonical GRBs, 
the rapid decay and fast spectral softening of the prompt emission
changes to a slower power-law decay, $\sim t^{-2.1}$, and a harder 
spectrum, before it reaches the plateau (Fig.~\ref{f3}).
We shall refer to such light curves as `semi-canonical'.

The Swift X-ray data show a flaring activity in a large fraction of GRBs, 
both at early and late times. The X-ray peaks during the prompt $\gamma$- 
ray emission follow the pattern of the $\gamma$-ray pulses, they must have 
a common origin. In many GRBs, superimposed on the early-time fast 
decaying X-ray light curve, there are X-ray flares, whose peak intensities 
also decrease with time and whose accompanying $\gamma$-ray emission is 
probably below the detection sensitivity of BAT. Yet, their spectral and 
temporal behaviour is similar to that of the prompt X/$\gamma$ pulses. 
Very often the flaring activity continues into the afterglow phase. 
Late-time flares appear to exhibit different temporal and spectral 
behaviours than early-time flares.

Neither the general trend, nor the frequently complex structure of the 
Swift X-ray data were predicted by (or can be easily accommodated within)  
the standard FB models (see, e.g.~Zhang \& M\'esz\'aros 2004, Piran~2005, 
for reviews). Much earlier confrontations between predictions of the FB 
models and the observations also provided severe contradictions, such as 
the failure to understand the prompt spectrum on grounds of synchrotron 
radiation (e.g.~Ghisellini, Celotti, \& Lazzati 2000), or the `energy 
crisis' in the comparison of the bolometric prompt and AG fluences 
(e.g.~Piran~1999, 2000). We have discussed elsewhere other problems of FB 
models (DD2004, Dar~2005 and references therein), including those related 
to `jet breaks' (e.g.~DDD2002a, Dar~2005, DDD2006), and the a-posteriori 
explanations of the reported detections (GRB 021206: Coburn and 
Boggs~2003, see however Wigger et al.~2004 and Rutledge \& Fox~2004; GRBs 
930131 and GRB 960924: Willis et al.~2005; GRB 041219A: Kalemci et 
al.~2007; McGlynn et al.~2007) of large $\gamma$-ray polarization 
(DDD2007b, and references therein).

The Swift data have challenged the prevailing views on GRBs. Kumar et 
al.~(2007) concluded that the prompt $\gamma$-ray emission cannot be 
produced in shocks, internal or external. Zhang, Liang \& Zhang~(2007) found 
that the fast decay and rapid spectral softening ending the prompt 
emission cannot be explained by high latitude emission. The X-ray and 
optical afterglows of Swift GRBs are very chromatic at early time in 
contrast with the fireball model expectation. Moreover, Curran et 
al.~(2006) have carefully examined Swift data and found that X-ray and 
optical AGs have chromatic breaks which differ significantly from the jet 
break of the blast-wave model of AGs. Burrows and Racusin~(2007) examined 
the XRT light curves of the first $\sim\! 150$ Swift GRBs and reported 
that the expected jet breaks are extremely rare. In particular, Liang et 
al.~(2008) have analyzed the Swift X-ray data for the 179 GRBs detected 
between January 2005 and January 2007 and the optical AGs of 57 pre- and 
post-Swift GRBs. They did not find any burst satisfying all the 
criteria of a jet break.

In spite of the above failures, not all authors are so critical. Some 
posit that the Swift data require only some modifications of the standard 
FB models to accommodate the results (e.g.~Panaitescu et 
al.~2006, Dai et al.~2007, Sato et al.~2007). Others still view the 
situation with faith (e.g.~Covino et al.~2006, 
Panaitescu~2008, Dai et al.~2008, Racusin et al.~2008a).

The situation concerning the CB model is different. The model was based on 
the assumption that LGRBs are produced by highly relativistic 
jets of plasmoids of ordinary matter (Shaviv \& Dar~1995) ejected in 
core-collapse supernova (SN) explosions akin to SN1998bw (Dar \& 
Plaga~1999, Dar \& De R\'ujula 2000).  It successfully described the 
broad-band AGs observed before the Swift era (e.g.~DDD2002a, 
DDD2003a) 
and exposed the consistent photometric evidence for a LGRB/SN association 
in all nearby GRBs (DDD2002a, DD2004 and references therein) long before 
GRB 030329. In the case of GRB 030329 the first $\sim\!6$ days of AG data 
were described by the CB model precisely enough to extrapolate them to 
predict even the date in which its associated SN would be bright enough to 
be detected spectroscopically (DDD2003c). General acceptance of the 
GRB-SN association waited until the spectroscopic discovery of SN2003dh, 
coincident with GRB 030329 (Hjorth et al.~2003, Stanek et al.~2003), and 
other spectroscopically-proven  
associations, e.g.~GRB030213/SN2003lw (Malesani et al.~2004), 
GRB021211/SN2002lt (Della Valle et al.~2003), XRF060218/SN2006aj (Campana 
et al.~2006b, Pian et al.~2006, Mazzali et al.~2006) and XRF080109/SN2008D 
(Malesani et al.~2008, Modjaz et al.~2008, Soderberg et al.~2008).

The CB model (DD2004) has been applied successfully to explain all the 
main observed properties of long GRBs and XRFs before the Swift era
(e.g.~Dar 2005 and references therein). The model is summarized in 
$\S$\ref{CBMODEL}. For detailed accounts see, e.g.,~De R\'ujula, 2007a,b. 

In this report we extend and refine our analysis of the temporal and 
spectral behaviour of the $\gamma$-ray, X-ray and optical light curves of 
GRBs during the prompt emission, the rapid-decay phase, and the 
afterglow phase. The observed prompt spectrum in the $\gamma$-ray to X-ray 
domain is the predicted one, which is Compton-dominated in the CB 
model  (DD2004). The observed widths of the $\gamma$-ray and 
X-ray peaks, as well as lag-times between them and their relative 
fluences, are in accordance with the model's predictions, if free-free 
absorption dominates the transparency of the CBs to eV photons in the CBs' 
rest frame. We investigate whether or not the CB model can 
describe all the data in terms of only two emission 
mechanisms: inverse Compton scattering and synchrotron radiation. We shall 
see that this simple picture, explicitly based on the predictions in 
DDD2002a and DD2004, gives a straightforward and successful description 
of the Swift GRB data, at all observed energies and times.

An exploding SN illuminates the progenitor's earlier ejecta, creating a {\it glory}
of scattered and re-emitted light. In the CB model inverse Compton 
scattering (ICS) of glory photons is the origin of the prompt 
$\gamma$/X-ray peaks, as we 
review in $\S$\ref{Inverse}. Each peak is generated by a single CB 
emitted by the `engine', the accreting compact object resulting from a 
core-collapse supernova event. We shall see that ICS correctly describes 
the prompt peaks, extending even into the optical domain in XRFs in which 
the relevant observations are available, such as XRF 060218. The natural 
explanation of the early time flares is the same as that of the stronger 
flares: ICS of glory photons by the electrons of CBs ejected in late 
accretion episodes of fall-back matter on the newly formed central object. 
These CB emissions must correspond to a weakening activity of the engine, 
as the accreting material becomes scarcer.

In the CB model, from the onset of the `plateau' onwards, the X-ray,
optical (DDD2002a) and radio (DDD2003a) afterglows are dominated
by synchrotron radiation (SR), the CB-model predictions for which
are reviewed in $\S$\ref{Synchrotron}. On occasion
these AGs also have transient rebrightenings (`very late' flares),
two notable cases before the Swift era being GRB 970508 (Amati et 
al.~1999, Galama et al.~1998a)
and GRB 030329 (Lipkin et al.~2004). During these 
episodes, the spectrum continues
to coincide with the one predicted on the basis of the 
synchrotron mechanism that dominates the late AGs. These very late flares
are well described by encounters of CBs with density inhomogeneities
in the interstellar medium (DDD2002a, DD2004). 
Very late flares in the {\it XUVONIR} AG may have this origin as well.

In this article we compare the predictions of the CB model and the observed 
X-ray and optical light curves of 33 selected GRBs, which are well sampled 
from very early time until late time, have a relatively long follow-up 
with good statistics and represent well the entire diversity of Swift 
GRBs. These include the brightest of the Swift GRBs (080319B), the GRB 
with the longest measured X-ray emission (060729), a few with canonical 
X-ray light curves (050315, 060526, 061121 and 080320) with and without 
superimposed X-ray flares, GRBs with semi-canonical light curves (060211A, 
061110A, 070220, 080303, 080307, 051021B) and non-canonical light curves 
(061007, 061126, 060206), and some of the allegedly most peculiar GRBs 
(050319, 050820A, 060418, 060607A, 071010A, 061126). We also compare the 
CB model prediction and the observed X-ray light curve of additional 12 
GRBs with the most rapid late-time temporal decay.

In the CB model, LGRBs and XRFs are one and the same, 
the general distinction being that XRFs are viewed at a 
larger angle relative to the direction of the approaching jet of CBs or 
have a relatively small Lorentz factor (DD2004, Dado et 
al.~2004c). Thus we include a Swift XRF of 
particular interest in our analysis: 060218. Its X-ray light curve is 
shown to be the normal X-ray light curve of a GRB viewed far off axis, and 
not the emission from the break-out of a spherical shock wave through the 
stellar envelope. Its optical AG at various frequencies shows, before the 
SN becomes dominant, a series of 
broad peaks between 30 ks and 60 ks after trigger, which we interpreted as 
the optical counterparts of the dominant prompt X-ray peak of this XRF.  
The expressions for an ICS-generated peak at all frequencies allow 
us to predict the positions, magnitudes and pulse shape of these broad 
peaks, a gigantic extrapolation in time, radiated energy and frequency.

After submitting for publication a first version of a comparison between
the CB-model predictions and Swift observations (DDD2007c), we have 
compared many more Swift data with the CB-model predictions, in order to 
further test its ability to predict correctly all the main 
properties of GRB light curves. These included the rapid spectral 
evolution observed during the fast decay of the prompt emission in 
`canonical' GRBs (DDD2008a) and the `missing AG breaks' in the AG of 
several GRBs (DDD2008b). We have also extended the CB model to describe 
short hard bursts (SHBs) and confronted it with the entire data on all 
SHBs with well-measured X-ray and/or optical afterglows (Dado \& Dar 
2008).

Together with the GRBs discussed in this report, we have analyzed 
and published CB model fits to the light curves of more than 100 LGRBs and 
SHBs.  The CB model continued to be completely successful in the confrontation
 of its predictions with the data.

\section{The CB Model}
\label{CBMODEL}

In the CB model (e.g.~DD2004 and references therein)  {\it 
long-duration} GRBs and their AGs are produced by bipolar jets of CBs 
which are ejected (Shaviv \& Dar, 1995, Dar \& Plaga, 1999)
in~{\it ordinary core-collapse} supernova explosions\footnote{Supernovae
associated with GRBs are viewed uncommonly
close to their jet axis, near which
the non-relativistic ejecta from the SN are faster than average. 
The observed initial large velocities of the leading ejecta may,
erroneously in our view, lead to their interpretation  as 
a very special GRB-associated class of super energetic SNe:
{\it `hypernovae'}. Yet, the velocities of their ejecta
 have been observed to decrease within a year or two after the explosion 
(before they have swept a significant amount of circumburst matter) to
a typical 5000-7000 km s$^{-1}$, implying a normal SN kinetic energy 
release of a few times $10^{51}$ erg.}.   
An accretion disk or a torus is 
hypothesized to be produced around the newly formed compact object, either 
by stellar material originally close to the surface of the imploding core 
and left behind by the explosion-generating outgoing shock, or by more 
distant stellar matter falling back after its passage (De R\'ujula~1987). 
As observed in microquasars (e.g.~Mirabel \& Rodriguez~1999, Rodriguez \& 
Mirabel~1999 and references therein), each time part of the accretion disk 
falls abruptly onto the compact object, a pair of CBs made of {\it 
ordinary-matter plasma} with a typical baryonic number, 
$N_{_{\rm B}}\!\sim\! 10^{50},$ are 
emitted with large bulk-motion Lorentz factors, typically 
$\gamma_0\!\sim\! 10^3$, 
in opposite directions along the rotation axis, wherefrom matter has 
already fallen back onto the compact object, due to lack of rotational 
support.

The $\gamma$-rays of a single pulse of a GRB are produced as a CB coasts 
through the SN {\it glory}, the light emitted and scattered by the `wind' ---the 
ejecta puffed by the progenitor star continuously or in a succession of 
pre-SN flares--- after being illuminated by the progenitor's pre-supernova 
and SN light. The electrons enclosed in the CB Compton up-scatter 
the photons of the glory to GRB energies. The initial fast expansion of 
the CBs and the increasing transparency of the wind environment as the CBs 
penetrates it, result in the fast rise of GRB pulses. As the CB coasts 
further through the SN glory, the density and temperature decrease 
rapidly. Consequently, the energy of the up-scattered photons is 
continuously shifted to lower energies and their number decreases rapidly. 
Typically, the ensuing fast decline of the prompt emission is taken over, 
within a couple of minutes of observer's time, by a broad-band synchrotron 
emission from swept-in electrons from the wind and the interstellar medium 
(ISM) spiraling in the CB's enclosed magnetic field.

In the CB model there is no clear-cut {\it temporal} distinction between 
prompt and {\it after}-glow signals. There are, however, two rather 
distinct radiation mechanisms: inverse Compton scattering and synchrotron 
radiation. For all cases we have studied, the prompt emission of 
$\gamma$-rays, X-rays and optical light in XRFs is dominated by ICS 
whereas in ordinary GRBs, only the prompt emission of $\gamma$ and X-rays 
is dominated by ICS, while SR dominates the prompt optical emission and the 
broad-band afterglow emission. Usually, the SR takes over the X-ray 
emission during the fast decay of the prompt emission or at the onset of 
the `plateau' phase. Late flares appear to be dominated by SR.

\section
{Inverse Compton Scattering}
\label{Inverse}

\subsection{The spectrum of ICS pulses}

During the initial phase of $\gamma$-ray emission in a GRB,
the Lorentz factor $\gamma$ of a CB stays put at its initial value
$\gamma_0\!=\!{\cal{O}}(10^3)$, for the deceleration induced by 
the collisions with the ISM has not yet had a significant effect
(DDD2002a, DDD2003a).
Let $\theta$ be the observer's angle relative to the direction of motion
of a CB.
The Doppler factor by which light emitted by a CB is boosted in energy
is, 
\begin{equation}
\delta={1\over \gamma\,
(1-\beta\, \cos\theta)}\approx {2\, \gamma\over
1+\gamma^2\, \theta^2}\; ,
\label{delta}
\end{equation}
where the 
approximation is excellent for $\gamma\gg 1$ and $\theta\ll 1$.
The emitted light is forward-collimated into a cone of 
characteristic opening angle $1/\gamma$,
so that the boosted energetic radiation is observable for 
$\theta\!=\!{\cal{O}}(1/\gamma_0)$. This implies that the typical
initial Doppler factor of a GRB is:
$\delta_0\!=\!{\cal{O}}(10^3)$.

The burst environment is very complex, and can only
be roughly approximated. 
After it is ejected, the fast-expanding CB propagates through a cavity
produced by the pre-supernova ejecta, and shortly encounters 
the previously ejected `windy environment', whose density distribution 
is roughly $n(r)\!\propto\! 1/r^2$. The initially fast-expanding CB scatters  
the quasi-isotropic distribution of glory light and the collimated light from the CBs
themselves\footnote{The CB arrives at the windy environment shortly
after its emitted light, well before the scattered photons
could have left the beaming cone, since
$r/2\,c\,\gamma^2\! \ll\! r/\gamma\, c$.}.  
The glory light has a thin thermal bremsstrahlung 
spectrum 
 \begin{equation}
\epsilon\, {dn_\gamma \over d\epsilon} \approx n_\gamma(r)\, 
\left({\epsilon \over k\,T_g }\right)^{-\beta_g}\, e^{-\epsilon/ k\,T_g}, 
\label{thinbrem}
\end{equation}
with $\beta_g\!\sim\!0$ and a temperature that decreases with distance beyond a 
characteristic $r_g$  like
$T_g(r)\! \sim\! T(0)\, r_g^2/(r_g^2\!+\!r^2) $,
with
$k\, T(0)\!\sim\!1$ eV.
The observed energy of a glory photon which was  
scattered by an electron comoving with a 
CB at redshift $z$, is:
\begin{equation}
E={\gamma_0\, \delta_0\, \epsilon \, (1+\cos\theta_{in})\over 1+z}\,,   
\label{ICSboost}
\end{equation}
where $\theta_{in}$ is the angle of incidence of the initial
photon onto the CB, in the SN rest system.
For a quasi-isotropic distribution of glory light,  
$\cos\theta_{in} $  in Eq.~(\ref{ICSboost})
roughly averages to zero.
The predicted time-dependent spectrum of the GRB pulse produced 
by ICS of the glory photons is given by (DD2004): 
\begin{equation}
E\, {dN_\gamma\over dE} \sim \left({E\over E_p(t}\right)^{-\beta_g}\,
 e^{-E/E_p(t)}+ b\,(1-e^{-E/E_p(t)})\, \left({E \over
E_p(t)}\right)^{-p/2}\,,
\label{GRBspec}
\end{equation}
where 
\begin{eqnarray}
E_p(t)&\approx & E_p(0)\, {t_p^2 \over t^2+t_p^2}\,,
\nonumber\\
E_p(0)&\approx & {\gamma_0\, \delta_0 \over 1+z}\,k\, T_g(0),
\label{PeakE}
\end{eqnarray}
with $t_p\!\approx \! (1\!+\!z)\,r_g /c\,\gamma_0\, \delta_0 $,
 the peak time of  $dN_\gamma/ dt$, discussed in the next chapter.

The first term in Eq.~(\ref{GRBspec}), with $\beta_g\!\sim\! 0$, is the 
result of Compton 
scattering by the bulk of the CB's electrons, which are comoving with it.
The second term in Eq.~(\ref{GRBspec}) is induced by 
a very small fraction of
`knocked-on' and Fermi-accelerated electrons, whose initial spectrum
(before Compton and synchrotron cooling) is $dN_e/dE\propto E^{-p}$, 
with $p\approx 2.2$. For $b={\cal{O}}(1)$,
the energy spectrum predicted by the CB model, Eq.~(\ref{GRBspec}),
bears a striking resemblance 
to the Band function (Band et al.~1993) traditionally used to model the 
energy spectra of GRBs, 
but GRBs whose spectral measurements
extended over a much wider energy range than that of BATSE and Swift's BAT,
are better fitted by Eq.~\ref{GRBspec} (e.g.~Wigger et al.~2008).

For many Swift GRBs the spectral observations
do not extend to energies bigger than $E_p(0)$, or the value of $b$
in Eq.~(\ref{GRBspec}) is relatively small, so that the first term 
of the equation provides a very good approximation. But for its time-dependence,
this term coincides with the `cut-off 
power-law' which has also been recently used to model 
GRB spectra. For $b\!\sim\! 0$ and $\beta_g\!\sim \!0$
it yields a peak value of $E^2\, dN/dE$ at 
$E_p(t)$ whose pulse-averaged value is:
\begin{equation}
E_p\approx E_p(t_p)\approx 0.5\, E_p(0) 
\approx 155 \,  {\gamma_0\,\delta_0\over 10^6}\;
{T(0)\over 1\,{\rm eV}}\;{3.2\over 1+z}\, {\rm keV}\,, 
\label{ICSEp}
\end{equation}
where the numerical result was obtained for the pulse shape discussed in 
the next subsection and the indicated typical values,
including the mean redshift $\langle z\rangle\!\approx\!2.2$ of Swift's
long GRBs (Greiner: {\it http://www.mpe.mpg.de/$\sim$jcg/grbgen.html}).
For $b\!=\!1$ and $\beta_g\sim 0$,  
$E_p$ is larger by 50\% than the result of Eq.~(\ref{ICSEp}). 
The predicted spectrum, Eq.~(\ref{GRBspec}), 
and the range of $E_p$
values, Eq.~(\ref{ICSEp}), are in good agreement with the 
observations of BATSE, 
BeppoSAX, Konus-Wind, INTEGRAL, Suzaku and RHESSI, which cover a much 
broader 
energy range than Swift\footnote{Swift data can determine
$E_p$ only when it is 
well within its 15-350 keV detection range. This results in a biased sample 
of GRBs whose {\it measured} $E_p$ is smaller than the average 
over the entire GRB population.}.

In the CB model XRFs are either GRBs with typical values of $\gamma_0$, 
but viewed from angles $\theta \!\gg \!1/\gamma_0$, or GRBs
with smaller $\gamma_0$  (DD2004, DDD2004a). 
Both choices imply a smaller $\delta_0$ in Eq.~(\ref{delta}),
and consequently the softer spectrum and
relatively small $E_p$ that define an XRF,
see Eqs.~(\ref{GRBspec},\ref{PeakE}). XRFs have light curves
with wider and less rugged peaks than GRBs. This follows from
the time dependence of the light curves, which
we discuss next.

\subsection{The light curves of GRB and XRF pulses}

After launch, as the CB propagates in the progenitor's wind on 
its way  to the ISM, its cross section increases, its density and the 
wind's  density decrease and consequently their opacities decrease. 
Let $t$ be the time after launch of a CB as measured by a distant 
observer. 
Approximating the CB geometry by a cylindrical slab with the same radius, 
density and volume, and neglecting multiple scattering and the spread in 
arrival times of ICS photons from the CB which entered it simultaneously, 
their arrival rate is given by: 
\begin{equation}
{dN_\gamma\over dt} = e^{-\tau_{_W}}\, n_g (t)\, \sigma_{_T}\, \pi\, 
R^2(t)\, 
{[1\!-\!e^{-\tau_{_{CB}}}]\over \sigma_a}\, ,  
\label{dNdtslab}
\end{equation}
where  $\tau_{_W}$  
is the opacity of the wind at the CB location,
$\tau_{_{CB}}$ is the effective opacity of the expanding CB 
encountered  by a photon with energy $E'\!=\!(1+z)\, E/\delta_0$
which begins crossing it at a time $t$,
$\sigma_a(E')$ is the photo-absorption cross section at energy $E'$
and $\sigma_{_T}$ is the Thomson cross section. 
The density of the glory photons seen by a CB 
is quasi isotropic and decreases roughly like
$n_g\!\propto\! 1/(r^2\!+\! r_g^2)$,
where $r_g$ is distance where the wind becomes transparent 
(optical thickness $\sim \!1$) to glory photons. 
At an early time, 
$r\!\approx\! c\, \gamma_0\, \delta_0\, t/(1\!+\!z).$ 
Consequently
$n_g\! \propto\! 1/(t^2\!+ \! \Delta t^2)$,
where  $\Delta t\!=\!(1\!+\!z)\, r_g/ c\, \gamma_0\, \delta_0$. Thus
the shape of an ICS pulse produced by a CB is given approximately by
\begin{equation}
E\,{d^2N_\gamma \over dt\, dE}\propto  e^{-\tau_{_W}}\, n_g\, \pi\, 
R^2\, [1\!-\!e^{-R_{tr}^2/R^2}]\, E\,{dN_\gamma\over dE}\, ,
\label{ICSPulse0}
\end{equation}
where  $R_{tr}$ is the radius of the CB at $t\!=\!t_{tr}$,
when $\tau_{_{CB}}\!\approx\!1$, i.e., when it
becomes transparent to the scattered radiation,
and $E\, dN_\gamma/dE$ is given by Eq.~(\ref{GRBspec}). 
The pre-supernova wind from the progenitor star produces a 
density distribution, $n(r)\!=\! n_0\, r_0^2/r^2$ around it, which 
yields $\tau_{_W}\!=\!a(E)/t$ with
$a(E)\!=\!\sigma_a\, n_0\, r_0^2\, (1\!+\!z)/c\,\gamma_0\, 
\delta_0$.
At sufficiently high energies the opacities of the wind and the CBs
are mainly due to Compton scattering and  
$R_{tr}\!\sim\!\sqrt{\sigma_{_T}\, N_B/\pi}$, 
where $\sigma_{_T}$ is the Thomson cross section and 
$N_B$ is the baryon number of the expanding CB.
At low energies, their opacity
is dominated  by free-free absorption
because the CBs and the wind along their trajectory are completely 
ionized. In the CBs' rest frame the glory photons have typical  
energies, $E'\!\ll\!$ keV.  At such low energies, the opacity of CBs
with a uniform  density behaves like  $\tau_{_{CB}}\!\sim\! 
E'^{-3}\,(1\!-\!e^{\!-\!E'\,k/T'})\,G(E')\,R^{-5}\,,$
where $G(E')$ is the quantum mechanical Gaunt factor that 
depends logaritmically on $E'$ (e.g.~Lang~1980 and references therein).
Thus, when the optical thickness of a CB is dominated 
by free-free photo-absorption, its 
transparency radius increases with decreasing energy like
$R_{tr}\!\propto\! E^{-3/5}\!=\! E^{-0.6}$ at $E'\!\gg\!k\,T'$ and
$R_{tr}\!\propto\! E^{-2/5}\!=\! E^{-0.4}$ at $E'\!\ll\!k\,T',$
yielding  $R_{tr} \!\sim \! E^{-0.5\pm0.1}$.

The initially rapid expansion of a CB slows down as it propagates 
through the wind and scatters its particles
(DDD2002, DD2004).  This expansion may be roughly described by 
$R^2\approx R_{cb}^2\, t^2/(t^2\!+\!t_{exp}^2)$, where $R_{cb}$
is the asymptotic radius of the CB and $t_{exp}\!\gg\!t_{tr}$. Thus,
Eq.~(\ref{ICSPulse0}) can be  approximated by
\begin{equation}
 E\,{d^2N_\gamma \over dt\, dE} \propto {
e^{-a/t}\, \Delta t^2\, t^2\over
         (t^2\!+\!\Delta t^2)\, (t^2\!+\!t_{tr}^2)}\,  E\,{dN_\gamma\over 
dE}\, .
\label{ICSP}
\end{equation}
For nearly transparent winds ($a\! \rightarrow\! 0$) and for
$t_{tr}\!\sim\! \Delta t$,  Eq.~(\ref{ICSP}) has an approximate shape
\begin{equation}
 E\,{d^2N_\gamma \over dt\, dE} \propto {\Delta t^2\, t^2  \over
(t^2+\Delta t^2)^2}\,E\,{dN_\gamma\over dE}\, ,
\label{ICSPulse}
\end{equation}
which peaks around $t\!\approx\!\Delta t$.
Except for very early times, this shape is almost
undistinguishable from that of the `Master' formula  of the CB model 
(DD2004):
\begin{equation}
 E\,{d^2N_\gamma \over dt\, dE} \propto e^{-\Delta t^2/t^2 }
\, [1\!-\!e^{-\Delta t^2/t^2} ]\,E\,{dN_\gamma\over dE}\, ,
\label{ICSMaster}
\end{equation}
which also took into account  arrival-time 
effects that depend on the geometry of the CB and the 
observer's viewing  angle, and  was shown to 
describe well the prompt emission pulses of LGRBs  
(DD2004) and  their rapid decay with a fast spectral 
softening (DDD2008a).

At the relatively low X-ray energies covered by Swift and, 
more so, at smaller ones, the first term on the RHS of
Eq.~(\ref{GRBspec}) usually dominates  $E\, dN_\gamma/dE$. Thus
the light curve generated by a sum of ICS pulses 
at a luminosity distance $D_L$ is generally well described by: 
\begin{equation}
E\,{d^2N_\gamma \over dt\, dE} \approx \Sigma_i\, 
A_i\Theta[t\!-\!t_i]\,{\Delta t_i^2\,(t\!-\!t_i)^2 \over
((t\!-\!t_i)^2\!+\!\Delta t_i^2)^2}\, e^{-E/E_{p,i}[t\!-\!t_i]}\, ,
\label{ICSlc}
\end{equation}
where the index `i' denotes the i-th pulse produced by a
CB launched at an observer time $t\!=\!t_i$, or, alternatively,  
\begin{equation}
E\,{d^2N_\gamma \over dt\, dE} \approx \Sigma_i\,
A_i\Theta[t\!-\!t_i]\, e^{-\Delta t_i^2/(t\!-\!t_i)^2}
\, [1\!-\!e^{-\Delta t_i^2/(t\!-\!t_i)^2} ]\, e^{-E/E_{p,i}[t\!-\!t_i]}\,,
\label{ICSlcmaster}
\end{equation}
where $E_{p,i}[t\!-\!t_i]$ is given by Eq.~(\ref{PeakE}) with $t$ replaced 
by $t\!-\!t_i$ and 
\begin{equation}
A_i \approx {c\, n_g(0)\, \pi\, R_{CB}^2\,\gamma_0\, \delta_0^3\, (1+z)
\over 4\, \pi\, D_L^2}\, .
\label{peakparam}
\end{equation}
Thus, in the CB model, each ICS pulse in the GRB light curve 
is effectively described by four parameters, $t_i,\, A_i,\,
\Delta t_i$  and $E_{p,i}(0)$,
which are best fitted to reproduce its observed light curve.
The evolution of its peak energy is then determined.

Setting $t_i=0$, $E_p(t)$  has the approximate form
$E_p(t)\!\approx\! E_p\, t_p^2/(t_p^2\!+\!t^2)\!.$ 
Such an evolution
has been observed in the time-resolved spectrum of well-isolated pulses 
(see, for instance, the insert in Fig.~8 of Mangano et al.~2007),
until the
ICS emission is overtaken by the broad-band
synchrotron emission from the swept-in ISM electrons. Hence,
the temporal behaviour of the  separate ICS peaks  
is given by: 
\begin{equation}
E\, {d^2N_\gamma\over dt\,dE}(E,t)
\propto {t^2/\Delta t^2  \over(1+t^2/\Delta t^2)^2} 
e^{-E\, t^2/\, E_p\,t_p^2} \approx F(E\,t^2),
\label{law}
\end{equation}
to which we shall refer as the `$E\,t^2$~{\it law'}.
A simple consequence of this law is that unabsorbed ICS peaks have 
approximately identical shape at different energies when 
their $E\,d^2 N_\gamma/dt\,dE$ is plotted as a
function of $E\, t^2$. 

A few other trivial but important consequences of Eq.~(\ref{law}) 
for unabsorbed GRB peaks at $E \,\lsim\, E_p$ are the following: 
\begin{itemize}

\item{}
The peak time of a pulse is at
\begin{equation}
t_p=t_i\!+\!\Delta t_i\,.
\label{peaktime}
\end{equation}

\item{}
The full width at half maximum (FWHM) of a  pulse is
\begin{equation}
{\rm FWHM}\!\approx\! 2\, \Delta t_i,
\label{fwhm}
\end{equation}
and it extends from $t\!\approx\!t_i\!+\! 0.41\, \Delta t_i $ to 
$t\!\approx\!t_i\!+\! 2.41\,\Delta t_i$.

\item{}
The rise time (RT) from half peak 
value to peak value satisfies
\begin{equation}
{\rm RT\approx 0.30\,FWHM},
\label{ratio}
\end{equation}
independent of energy. This result agrees with the empirical
relation that was inferred by Kocevski et al.~(2003)
from bright BATSE GRBs, 
${\rm RT\!\approx\! (0.32\!\pm\! 
0.06)\, FWHM}$.

\item{}
The FWHM increases 
with decreasing energy approximately like a power-law:
\begin{equation}
{\rm FWHM}(E)\sim E^{-0.5}\, .
\label{widthrelation}
\end{equation}
This relation is consistent with the empirical relation
${\rm FWHM}(E) \propto E^{\!-\!0.42\!\pm\! 0.06}$,
satisfied by BATSE GRBs (Fenimore et al.~2003).

\item{}
The onset-time, $t_i$, of a pulse is simultaneous at all energies.
But the peak times $t_p$ at different energies differ, the 
lower-energy ones `lagging'
behind the higher-energy ones: 
\begin{equation}
t_p-t_i \propto E^{-0.5}\, .
\label{lagtime}
\end{equation}

\item{}
The time-averaged value of $E_p(t)$ for GRB peaks, which follows from 
Eq.~(\ref{PeakE}), satisfies: 
  \begin{equation}
E_p= E_p(0)/2=E_p(t_p)\, .
\label{Epeak}
\end{equation}

\end{itemize}

\subsection{X-ray `flares' and $\gamma$-ray pulses}

In more than 50\% of the GRBs observed by Swift, the X-ray light curve, 
during the prompt GRB and its early AG phase, 
shows flares superimposed on a smooth background. 
In the CB model, an X-ray `flare' coincident in time with a $\gamma$-ray
pulse is simply its low-energy tail. Both are
due ICS of photons 
in the thin-bremsstrahlung glory. The glory's
photons incident on the CB at small $\epsilon_i$ or $1\!+\!\cos\theta_i$
result in X-ray or softer up-scattered energies, see Eq.~(\ref{ICSboost}).
The harder and less collinear photons result in $\gamma$-rays.
The light curve and spectral evolution of an ICS X-ray flare
are given approximately by Eq.~(\ref{ICSlc}).
Its width is related to that of the accompanying $\gamma$-ray pulse  
as in  Eq.~(\ref{fwhm}). Relative to its $\gamma$-ray
counterpart, an X-ray flare is wider and  its peak time `lags'
after the peak time of the $\gamma$-ray pulse.
The X-ray flares during a GRB are well 
separated only if the $\gamma$-pulses are sufficiently spaced.

In the CB model, X-ray flares without an accompanying
detectable $\gamma$-ray emission can be of two kinds.
They can be ICS flares produced  by CBs 
ejected with relatively small  Lorentz factors
and/or relatively large viewing angles  (Dado et al.~2004).
Such CBs may be ejected in accretion episodes both 
during the prompt GRB and in delayed accretion episodes 
onto the newly formed central object in core collapse 
SNe (De R\'ujula 1987).
ICS flares satisfy the $E\,t^2$-law and exhibit a rapid softening 
during their fast decline phase 
which is well described by Eqs.~(\ref{GRBspec},\ref{PeakE}),
as shown in detail in DDD2008b. 

We shall see in our case studies that very often, during the rapidly
decreasing phase of the prompt emission, there are `mini X-ray flares'
which show this rapid spectral softening. 
As the accretion
material is consumed, one may expect  the `engine' to have
a few progressively-weakening dying pangs. 

Flares can also be due to enhanced synchrotron emission during the passage 
of CBs through over-densities produced by mass ejections from the 
progenitor star or by interstellar winds (DDD2002a, DDD2003a). The synchrotron 
emission from CBs is discussed in the following section.  Late flares 
seem to have the predicted synchrotron spectrum and spectral evolution 
which are different from those of ICS flares.

\section{Synchrotron radiation}
\label{Synchrotron} 

A second mechanism besides ICS, which generates radiation from a CB,
is synchrotron radiation (SR). A CB encounters matter in its voyage 
through the interstellar medium (ISM), effectively
ionized by the high-energy radiation of the very same CB.
This continuous collision with the medium decelerates the CB in
a characteristic fashion, and results in a gradual steepening of the
light curves of their emitted synchrotron radiation 
(DDD2002a). In $\S$\ref{Deceleration},
we review the calculation of $\gamma(t)$, the CB's diminishing
Lorentz factor. We have assumed and tested observationally, 
via its CB-model consequences,
that the impinging ISM generates within
the CB a turbulent magnetic field\footnote{`First principle' numerical
simulations of two plasmas merging at a relative relativistic
Lorentz factor (Frederiksen et al.~2003, 2004, Nishikawa et al.~2003)
do not generate the desired shocks, but do generate turbulently
moving magnetic fields.}, in approximate energy equipartition
with the energy of the intercepted ISM (DDD2002a, DDD2003a). 
In this field, the intercepted
electrons emit synchrotron radiation. This radiation, isotropic in the 
CB's rest frame,
is Doppler boosted and collimated around the direction of motion 
into a cone of characteristic opening angle $\theta(t)\sim 1/\gamma(t)$. 
In $\S$\ref{SyncSpec} we summarize the predictions of the synchrotron 
radiation's dependence on time and frequency
(DDD2002a, DDD2003a).

\subsection{The deceleration of a CB}
\label{Deceleration}

As it ploughs through the ionized ISM, a CB 
gathers and scatters its constituent ions, mainly protons. These encounters
 are `collisionless' since, at about the time it becomes transparent to 
radiation, a CB also becomes `transparent' to hadronic interactions
(DD2004). The scattered and re-emitted 
protons exert an inward pressure on the CB, countering its expansion.
In the approximation of isotropic re-emission in the CB's 
rest frame and a constant ISM density $n\!\sim\!n_e\!\sim\!n_p$, 
one finds that within minutes of observer's time $t$, a typical CB
of baryon number $N_B\!\approx 10^{50}$ reaches a roughly
constant `coasting' radius $R\!\sim\!10^{14}$ cm, before it finally 
stops and blows up, after a journey of years of observer's
time. During the coasting phase, and in a constant-density ISM, 
$\gamma(t)$  obeys (DDD2002a, Dado et al.~2006): 
\begin{equation}
({\gamma_0/ \gamma})^{3+\kappa}+
(3-\kappa)\,\theta^2\,\gamma_0^2\,(\gamma_0/\gamma)^{1+\kappa} 
=1\!+\!(3\!-\!\kappa)\,\theta^2\,\gamma_0^2\!+\!t/t_0\,,
\label{decel}
\end{equation}
with
\begin{equation}
t_0={(1\!+\!z)\, N_{_{\rm B}}\over (6\!+\!2\kappa)\,c\, n\,\pi\, R^2\, 
\gamma_0^3}\,,
\label{break}
\end{equation}
where the numerical value $\kappa\!=\!1$, is for the
case in which the ISM particles re-emitted fast by the
CB are a small fraction of the flux of the intercepted ones, and
$\kappa\!=\!0$ in the opposite limit.
In the CB model of cosmic
rays (Dar \& De R\'ujula 2006) the observed spectrum strongly
favours $\kappa\!=\!1$. Thus in all of our fits we use $\kappa\!=\!1$, 
though the results are not decisively sensitive to this choice. 
As can be seen from Eq.~(\ref{decel}), $\gamma$  and $\delta$  
change little as long as $t\!<\! t_b\!=\![1\!+\!\gamma_0^2\, 
\theta^2]^2\, t_0$.
where, in terms of typical CB-model values of $\gamma_0$,
$R$, $N_{_{\rm B}}$ and $n$,
\begin{equation}
t_b= (1300\,{\rm s})\, [1+\gamma_0^2\, \theta^2]^2\,(1+z)
\left[{\gamma_0\over 10^3}\right]^{-3}\,
\left[{n\over 10^{-2}\, {\rm cm}^{-3}}\right]^{-1}
\left[{R\over 10^{14}\,{\rm cm}}\right]^{-2}
\left[{N_{_{\rm B}}\over 10^{50}}\right] \! .
\label{tbreak}
\end{equation}
For $t\!\gg\!t_b$, $\gamma$  and $\delta$ decrease like $t^{-1/4}$. 

The deceleration equation for a non-expanding CB can be integrated 
analytically also for other commonly encountered density profiles, such as 
a step function times $n(r)\! \propto\! 1/r^2$. Such a profile is produced by a 
constantly blowing wind from a massive progenitor star 
prior to the GRB, or from a star formation region, or in an isothermal 
sphere which describes 
quite well the density distribution in galactic bulges, 
galactic halos and elliptical galaxies. 

\subsection{The Synchrotron spectral energy density}
\label{SyncSpec} 

As indicated by first-principle calculations of the relativistic merger of 
two
plasmas (Frederiksen et al.~2004), the ISM ions continuously impinging on
a CB generate
within it turbulent magnetic fields, which we assume to be
in approximate energy equipartition with their energy, 
$B\!\approx\! \sqrt{4\,\pi\, n\, m_p\, c^2}\, \gamma$.
In this field, the intercepted
electrons emit synchrotron radiation. The SR, isotropic in the CB's
rest frame, has a characteristic frequency, $\nu_b(t)$,
the typical frequency radiated by the
electrons that enter a CB at time $t$ with a relative Lorentz
factor $\gamma(t)$. In the observer's frame:
\begin{equation}
\nu_b(t)\simeq  {\nu_0 \over 1+z}\,
{[\gamma(t)]^3\, \delta(t)\over 10^{12}}\,
\left[{n\over 10^{-2}\;\rm cm^3}\right]^{1/2}
{\rm Hz}.
\label{nub}
\end{equation}
where $\nu_0\!\simeq\! 3.85\times 10^{16}\, \rm Hz \simeq 160\, eV/h$.
The spectral energy density of the SR
from a single CB at a luminosity distance $D_L$  is given by (DDD2003a):
\begin{equation}
F_\nu \simeq {\eta\,  \pi\, R^2\,n\, m_e\, c^3\,
\gamma(t)^2\, \delta(t)^4\, A(\nu,t)\,
\over 4\,\pi\, D_L^2\,\nu_b(t)}\;{p-2\over p-1}\;
\left[{\nu\over\nu_b(t)}\right]^{-1/2}\,
\left[1 + {\nu\over\nu_b(t)}\right]^{-(p-1)/2}\,,
\label{Fnu}
\end{equation}
where $p\!\sim\! 2.2$ is the typical spectral index\footnote{The 
normalization
in Eq.~(\ref{nub}) is only correct for $p\!>\!2$, for otherwise the norm
diverges. The cutoffs for the $\nu$ distribution are time-dependent,
dictated by the acceleration and SR coling times of electrons 
and their `Larmor' limit. The discussion of these processes being complex 
(DD2003a, DD2006),
we shall satisfy ourselves here with the statement that for $p\!\leq \!2$
the AG's normalization is not predicted.} of the Fermi-accelerated
electrons, $\eta\!\approx\!1$ is the fraction of the impinging ISM
electron energy that is synchrotron re-radiated by the CB, and $A(\nu, t)$
is the attenuation of photons of observed frequency $\nu$ along the line
of sight through the CB, the host galaxy (HG), the intergalactic medium
(IGM) and the Milky Way (MW):
\begin{equation}
A(\nu, t) = {\rm
exp[-\tau_\nu(CB)\!-\!\tau_\nu(HG)\!-\!\tau_\nu(IGM)\!-\!\tau_\nu(MW)].}
\label{attenuation}
\end{equation}
The opacity $\tau_\nu\rm (CB)$ at very early times, during the
fast-expansion phase of the CB, may strongly depend on time and frequency.
The opacity of the circumburst medium [$\tau_\nu\rm (HG)$ at early times]
is affected by the GRB and could also be $t$- and $\nu$-dependent.  The
opacities $\tau_\nu\rm (HG)$ and $\tau_\nu\rm (IGM)$ should be functions
of $t$ and $\nu$, for the line of sight to the CBs varies during the AG
observations, due to the hyperluminal motion of CBs. 

The dependence of the SR afterglow on the CB's radius, external density 
and extinction, as summarized in Eq.~(\ref{Fnu}), give rise to a 
variety of early-time optical light curves, in contrast to the more 
uniform behaviour of the optical and X-ray SR afterglow at later times.
 
 \subsection{The X-ray afterglow}

The Swift X-ray band is usually above the bend frequency $\nu_b$, 
as given by
Eq.~(\ref{nub}), at all times. It then follows from Eq.~(\ref{Fnu}) that
the spectral energy density of the {\it unabsorbed} 
X-ray afterglow has the form:
\begin{equation}
F_\nu \propto R^2\, n^{(p+2)/4}\,
\gamma^{(3p-2)/2}\, \delta^{(p+6)/2}\,  \nu^{-p/2}=
R^2\, n^{\Gamma/2}\,
\gamma^{3\,\Gamma-4}\, \delta^{\Gamma+2}\, \nu^{-\Gamma+1}\, ,
\label{Fnux}
\end{equation}
where we have used the customary notation
$dN_{\gamma}/dE\!\propto\!E^{-\Gamma}$.

The deceleration of a CB causes a transition of its $\gamma$ and 
$\delta$ values from being approximately 
constant to asymptotic power-law declines. For a constant ambient
density this occurs gradually around $t\!=\!t_b$ and the 
asymptotic decline is $\delta\!\propto\!\gamma\! \propto\! t^{-1/4}$, 
see Eq.(\ref{tbreak}). This induces a gradual {\it bend}  (usually called a ``break")
in the synchrotron AG of a CB from a plateau to an 
asymptotic power-law decay
\begin{equation}
F_\nu \propto t^{-p/2-1/2}\,\nu^{-p/2}= t^{-\Gamma+1/2}\, 
\nu^{-\Gamma+1},
\label{Asymptotic}
\end{equation}
with a power-law in time steeper by half a unit than that in
frequency. For a constant ISM density and in an often used notation,
the asymptotic behaviour $F_\nu(t)\propto t^{-\alpha}\,\nu^{-\beta}\,,$ 
satisfies
\begin{equation}
\alpha=\beta+1/2=p/2+1/2=\Gamma-1/2\, . 
\label{indices} 
\end{equation}

For a fast-falling density beyond a given distance $r_c$, crossed by CBs
exiting density enhancements due to stellar winds or by CBs which escape 
the galactic bulge or disk into the galactic halo, 
$\gamma$ and $\delta$ tend to approximately constant values. Consequently
$r\!-\!r_c$ becomes proportional to $t-t_c\equiv t-t(r_c)$.
As a result, for a density profile such as $n\!\propto\! 1/r^2$ beyond $r_c$,
the unattenuated synchrotron afterglow, as given by  
Eq.~(\ref{Fnux}), approximately tends to
\begin{equation}
F_\nu \propto  n^{(p+2)/4}\,\nu^{-p/2} \propto 
(t-t_c)^{-(p+2)/2}\,\nu^{-p/2} =(t-t_c)^{-\Gamma}\, \nu^{-\Gamma+1}\,,   
\label{Fnurm2}
\end{equation}
and satisfies the asymptotic relation
\begin{equation}
\alpha=\beta+1=\Gamma \approx 2.1\, .
\label{indicesrm2}
\end{equation}
Thus, an unattenuated optical and X-ray AG may steepen at late times to an 
asymptotic decline $\!\sim\! (t-t_c)^{-2.1}$. 
Such an achromatic steepening, which was seen in several late-time 
optical and X-ray AGs of Swift GRBs (see Figs.~\ref{f6},\ref{f7}), may 
have been   misinterpreted as very late 
`jet breaks' (e.g.~Dai et al.~2008, Racusin et al.~2008a).   

All of the afterglows of Swift GRBs which are well sampled at 
late time satisfy one or the other of
the asymptotic relations (\ref{indices})  or (\ref{indicesrm2}) (DDD2008b).

\subsection{Early-time SR}

During the early-time when both $\gamma$ and $\delta$ stay put at their initial values 
$\gamma_0$ and $\delta_0$, Eq.~(\ref{Fnu}) yields a SR light curve
$F_\nu\! \propto\! e^{-\tau_{_W}}\, R^2\, n^{(1+\beta)/2}\,\nu^{-\beta}$.
Since $r\!\propto\!t$, a CB ejected into  
a windy density profile $n\!\propto\!1/r^2$, 
created by the mass ejection from the
progenitor star prior to its SN explosion, emits SR 
with an early-time light curve of the form
\begin{equation}
F_\nu \propto  {e^{-a/t}\,
t^{1-\beta} \over t^2+t_{exp}^2}\, \nu^{-\beta}\, ,
\label{SRP}
\end{equation}
where, for a CB ejected at time $t_i$,  $t$ must be replaced by
$t\!-\!t_i$.

In the $\gamma$-ray and X-ray bands, the SR emission from a CB is usually 
hidden under the prompt ICS emission. But in many GRBs the asymptotic 
exponential decline of the energy flux density of the prompt ICS emission, 
$F_\nu\! \propto\! t^{-2}\, e^{-E\, t^2/E_p\, t_p^2}$, is taken over 
by the slower power-law decline, $F_\nu\! \propto\! t^{-\Gamma}$, of the 
synchrotron emission in the windy $\!\sim\! 1/r^2$ circumburst density 
before the CB reaches the constant ISM density and the AG emission enters 
the plateau 
phase. This is demonstrated in Fig.~\ref{f3} for GRBs 051021B, 060211A, 
061110A, 070220, 080303, 080307. As soon as the light curve is dominated 
by SR, the rapid spectral softening of the ICS-dominated light curve stops 
and the spectrum of the unabsorbed X-ray emission changes to the ordinary 
synchrotron power-law spectrum with the typical power-law index, 
$\beta_X\!\approx\!1.1$ ($\Gamma\!\approx\!2.1$). Unlike the sudden 
change in the spectrum when the prompt ICS emission is taken over by SR 
during the plateau phase, there is no spectral change when SR dominates 
the X-ray light curve already before the plateau phase.

When the windy density profile changes to a constant ISM density, the 
light curve of the early-time 
optical AG, as given by Eq.~(\ref{SRP}), changes to a plateau or a shallow 
decline with a typical SR optical spectrum. After the deceleration 
bend, the temporal decline of the optical AG begins to 
approach that of the X-ray AG because the bend frequency, which decreases 
like $\nu_b\!\propto\!  \sqrt{n}\, \gamma^3\,\delta\,,$ moves below the 
optical band. After the cross-over, $\beta_O(t)\!\approx\! \beta_X\! 
\approx\! 1.1$ and the AG becomes achromatic, approaching the asymptotic 
power-law decay $F_\nu\! \sim\! t^{-\!\beta_X\!-\!1/2}$, and yielding an 
optical AG with a similar early and late power-law decline, 
$F_\nu\!\sim\! t^{-1.6}$.  This behaviour has been observed in several 
very bright GRBs whose optical light curve was discovered early enough and 
was followed until late time, e.g.~in GRBs 990123, 021211, 061007, 061126 
and 080319B (see Fig.~\ref{f2}).
Unlike the prompt $\gamma$-ray and X-ray emission in ordinary GRBs, which 
is dominated by ICS, their prompt optical emission is dominated by 
SR. This is because $F_\nu$[SR] increases with decreasing frequency 
like $\nu^{-\beta_O}$ with $\beta_O\sim 0.5$, whereas the 
prompt ICS emission for $h\, \nu\! \ll\! E_p$ is independent of frequency 
($\beta\!\approx\! 0$). 

The entire diversity of the chromatic early-time 
optical light curves of LGRBs measured with robotic telescopes, 
such as GRBs 030418 (Rykoff et al.~2004), 050319 (Wozniak et al.~2005, 
Quimby et al.~2006a), 050820A (Cenko et al.~2006, Vestrand et al.~2006), 
060418 (Molinari et al.~2007) 060605 (Ferrero et al.~2008), 060607A 
(Molinari et al.~2007, Nysewander et al.~2007, Ziaeepour et al.~2008), 
071010A (Covino et al.~2008), and 061126 (Perley et al.~2008; Gomboc et 
al.~2008), is described well by Eq.~(\ref{SRP}). This is shown in 
Figs.~\ref{f5},\ref{f8}.

The early 
SR obeys $F_\nu\propto \gamma_0^{3\,\beta-1}\,\delta_0^{3+\beta}$. Unlike 
ordinary GRBs with large $\delta_0\!\sim\!\gamma_0$, XRFs are GRBs with 
a relatively small $\delta_0\!\sim\! \gamma_0$ (near-axis XRFs) 
or far off-axis GRBs with $\delta_0\!\ll\! \gamma_0$. 
Consequently, the prompt optical emission in XRFs is also dominated by ICS.
An optical pulse that is dominated by ICS emission is distinguishable from an optical pulse that is
dominated by SR, in shape, spectrum and spectral 
evolution. Optical peaks produced by ICS satisfy the approximate $E\, t^2$ 
law, they are much wider than their $\gamma$-ray and X-ray counterparts 
and, as a result, they are usually blended. In XRFs such as 060218 
and 080109, however, the ICS optical peaks
have a large lagtime and are clearly visible as humps in the light curves at 
different optical wavelengths, see Fig.~\ref{f9}.

\subsection{Chromatic Afterglows}
\label{Chromo}

The early-time light curves of LGRBs are very chromatic because their 
prompt $\gamma$-ray and X-ray emission is dominated by ICS, while their 
optical emission is dominated by SR with entirely different temporal and 
spectral properties. Even in XRFs, where the prompt optical emission is 
also dominated by ICS, the light curves are very chromatic 
because the ICS pulses, which satisfy the `$E\, t^2$ law', are by themselves 
very chromatic, see Figs.~\ref{f9},\ref{f10}.

The afterglow emission in GRBs is dominated by SR at all 
wavelengths. In XRFs the situation is more interesting:
the same statement is not correct, should one adhere to the
traditional definition of AG as anything seen after the
decline of the prompt X- or $\gamma$- signal. This is discussed in
detail in $\S$\ref{XRF} on XRF 060218.
 The observed chromatic behaviour of the AGs results 
from their dependence on the circumburst density, the bend frequency and 
the attenuation of light along the line of sight to the source of the AG.  The most
general behaviour --that takes into account light attenuation inside the CBs 
and in the circumburst environment, CB expansion and density variation as 
summarized in Eq.~(\ref{Fnu}), and the chromatic light curves of 
superimposed flares-- is rather complex and will not be discussed in detail in 
this paper. The behaviour becomes simpler when the CB and circumburst 
medium become transparent to radiation and the fast expansion of the CB 
has slowed down.

For a constant circumburst density, the simplest situation arises when all 
observed frequencies are above the injection bend. Notice that for typical 
reference parameters, $\nu_b(0)$ in Eq.~(\ref{nub}) corresponds to an 
energy well above the {\it UVO} bands. But there are cases where $\delta$ is 
sufficiently small, such as XRFs and GRBs with very small $E_{iso}$, 
and/or where $n$ is very small, such that $\nu_b(0)\!\propto\! \sqrt{n}\, 
\gamma_0^3\, \delta_0$, is already below the {\it UVO} bands. In that case the 
(unattenuated) synchrotron spectrum is $F_\nu\!\sim\!\nu^{-p/2}$, 
achromatic all the way from the {\it UVO} bands to X-ray 
band at all times, see Eq.~(\ref{Fnu}).

In ordinary GRBs, $\nu_b(0)$ in Eq.~(\ref{nub}) is usually well above the 
{\it UVO} bands, but below the X-ray band. In that case, the unabsorbed 
spectrum of the optical AG evolves in a predicted fashion from 
$F_\nu\!\propto \![\nu/\nu_b(t)]^{\!-\!1/2}$ to 
$F_\nu\!\propto\![\nu/\nu_b(t)]^{\!-\!p/2}$, the behaviour of the unabsorbed X-ray 
AG. Many cases of this very specific chromatic evolution have been studied 
in DDD2003a. The success of their CB-model description corroborates the 
assumption that the CB's inner magnetic-field intensity $B(t)$, of which 
$\nu_b(t)$ is a function, is approximately determined by the equipartition 
hypothesis.

In a constant density, the CBs begin to decelerate rapidly around the time $t_b$ of 
Eq.~(\ref{tbreak}). Thus, the bend frequency, $\nu_b(t)\!\propto\! 
\gamma^3\, \delta$, declines rapidly with time beyond $t_b$ and crosses 
below the optical band. According to Eqs.~(\ref{tbreak}, \ref{Fnu}), $F_\nu 
$ steepens around $t_b$ to an asymptotic achromatic power-law decline, 
$F_\nu\!\propto\! \nu^{-\beta}\,t^{-\alpha}$ with $\alpha\! \approx\!\beta\! 
+\!1/2\! =\!\Gamma\!-\!1/2$, all the way from X-rays to the {\it UVO} bands. 
This smooth CB-deceleration bend in the AG of canonical GRBs, beyond 
which the $XUVO$ AG becomes achromatic, is not to be confused with the 
achromatic {\it break} predicted in fireball models (Rhoads 1997, 1999). 
The CB-model interpretation of this well understood achromatic bend-time 
(see, e.g.~DDD2003a) is further strengthened by the facts that it is 
observed at the predicted time scale and displays the predicted 
correlations with the prompt GRB emission (DDD 2008b), and that the 
predicted asymptotic relations between the temporal and spectral power-law 
declines beyond it are well satisfied.

A variation of the chromatic behaviour due to
bend-frequency crossing occurs when it happens early enough for
the circumburst density profile to be still dominated by the progenitor's
pre-SN wind emissions. 
At early times, $t\!\ll\! t_b$,
the deceleration of CBs has not significantly affected their motion
and $\gamma(t)$ and $\delta(t)$ are practically constant.
Yet, the observer's
bend frequency, $\nu_b(t)\!\propto\! \sqrt{n(t)}\, [\gamma(t)]^3 \delta(t)$,
decreases with time as $n(t)$ varies.
Keeping track of the $n$-dependence, we concluded (DDD2003a) from 
Eqs.~({\ref{nub},\ref{Fnu}) that the unattenuated
early synchrotron radiation of a CB moving in a windy
density profile, $n\!\sim\! r^{-2}\!\sim\! t^{-2}$, 
is given approximately by
\begin{equation}
F_\nu  \propto n^{(1+\beta)/2}\,\nu^{-\beta}\propto
t^{-(1+\beta)}\,\nu^{-\beta}.
\label{esync}
\end{equation}
In cases for which $\nu_b(0)$ is initially well above the $UVOIR$ bands,
$\beta\!\approx\! 0.5$, and the initial $UVOIR$  behaviour is
$F_\nu \!\propto\! t^{-1.5}\, \nu^{-0.5},$ 
while the X-ray AG, for which 
$\nu_b \!\ll\! \nu$ and $\beta\!\simeq\! 1.1$, behaves like
$ F_\nu \!\propto\! t^{-2.1}\, \nu^{-1.1}$ (DDD2003b). 
If the density continues to decline like $1/r^2$,
then the bend frequency crosses below the $UVOIR$ band, 
and the $XUVOIR$ AG becomes achromatic
with  $\alpha\! \approx\!\beta\! +\!1\!=\!\Gamma\!\sim\! 2.1$.
 
Optical light curves with an early-time power-law decay,
$F_\nu\!\sim\! t^{-1.5}$, 
have been observed, e.g.~in GRB 990123 (Akerlof et al.~1999); GRB 021211 
(Li et al.~2003) and GRB 061126 (Perley et al.~2008). Usually,
the steeper $F_\nu\!\sim\! t^{-(1+\beta_X)}\!\sim\! t^{-2.1}$
early-time decline of the X-ray synchrotron emission is hidden 
under the dominant early-time ICS emission, but in several 
GRBs it is visible as a power-law tail that takes over the initial 
exponential decay of the ICS pulse, see Fig.~\ref{f3}.
This take-over also stops the fast spectral softening 
of the ICS-dominated light curve and changes the soft 
spectrum into the harder SR spectrum.

In the CB model the absorption and extinction in the host galaxy, which 
are frequency dependent, can also be time dependent even far from the 
burst environment. In a day of the highly aberrated observer's time, CBs 
typically move to a few hundred parsecs from their birthplace, a 
region wherein the ionization should have drastically diminished and the 
line of sight to the CBs in the host and IGM has shifted considerably. 
Indeed, Watson et al. (2007) found very different X-ray-to-optical column 
density ratios in GRB afterglows. A strong variation in 
extinction at early times was observed, e.g.~by Perley et al.~(2008) in 
GRB 061126, and by Ferrero et al.~(2008) in GRB 060605. A time-dependent 
IGM absorption was reported by Hao et al.~(2007), but see also Th\"one et 
al.~(2008a). A frequency-dependent extinction and absorption, which change 
 in the host galaxy with the line of sight to the moving CB, can 
change an intrinsically achromatic AG into an observed chromatic AG.

Finally, ICS, which dominates also the prompt optical emission in XRFs,
results in a very specific chromatic 
behaviour: the X-ray light curve is declining rapidly, while the optical 
light curves are stretched by the $E\, t^2$ law and display humps that are 
nothing but the X-ray pulse(s) with their peak time and width stretched by 
the same law. A striking example is shown in Figs.~\ref{f9} and \ref{f10} 
for XRF 060218, discussed in great detail in $\S$\ref{Case}.

Because of the complex chromatic behaviour of GRB afterglows, we have 
limited our discussion to the chromatic properties of the AG of a single 
CB.  The actual situation can be even more convoluted because the AG, 
like the prompt emission, is usually the sum of the contributions from 
many individual CBs ejected at slightly different times with different 
parameters (baryon number, Lorentz factor and emission angle) which have 
or have not merged by the time of the AG phase. For the sake of 
simplicity, brevity and predictivity, we shall assume that the AG from the 
entire ensemble of CBs can be calculated as if it was due to one or at most
two effective CBs during the AG. Despite this simplification, the 
CB model, as we shall show, can reproduce and explain well, within 
observational uncertainties, the entire diversity of the measured light 
curves of the Swift GRBs and their afterglows.

\section{Comparison with observations}
\label{CWO}

To date, Swift has detected over 350 long GRBs, localized them through 
their $\gamma$, X-ray and {\it UVO} emissions and followed most of them until 
they faded into the background.  Beside the Swift observations, there have 
been many prompt optical measurements of Swift GRBs by an increasing 
number of ground-based robotic telescopes, and follow-up measurements by 
other ground-based optical telescopes, including some of the largest ones. 
Incapable of discussing all Swift GRBs, we discuss only a sample of 33 
of them (some $10\%$ of all Swift GRBs), which have  well sampled 
X-ray and optical light curves from early to late time, and which 
represent well the entire diversity of Swift GRBs.

We fit the X-ray light curves reported in the  
Swift/XRT GRB light curve repository:
{\it http://www.swift.ac.uk/xrt\_curves/}; Evans et al.~(2007). 
We used  Eq.~(\ref{ICSlc}) or Eq.~(\ref{ICSlcmaster})
for the early-time ICS contribution, and   Eq.~(\ref{Fnu}) for the  
SR afterglow with its simple explicit limit, 
Eq.~(\ref{SRP}), for the very-early-time synchrotron emission.
The a-priori unknown parameters are 
the number of CBs, their ejection time, baryon number, Lorentz factor and 
viewing angle and the environmental ones along the CBs' 
trajectory, i.e.~the distributions of the glory's light and of the ISM density. 
To demonstrate that the CB model correctly describes all of 
the observed features of the Swift X-ray observations, it suffices to 
include in the fits only the main or the latest few observed pulses or 
flares in the prompt emission. This is because the exponential factor 
in Eq.~(\ref{ICSlc}) suppresses very fast the relative contribution of 
the earlier pulses by the time the data sample the later pulses or flares. 
It also suffices to fix the glory's light and ISM density distribution to 
be the same along the trajectories of all CBs in a given GRB. The pulse 
shapes are assumed to be universal: given by Eq.~(\ref{ICSlc}).  For the 
synchrotron contribution, in most cases it suffices to consider a common 
emission angle $\theta$ and an average initial Lorentz factor $\gamma_0$ 
for the ensemble of CBs.

The ISM density along the CBs' trajectories was generally taken to have
a windy contribution ($\propto\! 1/r^2$) near the ejection site, changing 
later to a constant ISM density. The windy density  
is only relevant in some optical synchrotron-dominated AGs for which 
very early data are available, as discussed in the previous section. 
Density bumps with a density decline $n\propto\! 1/r^2$ were assumed to 
generate X-ray and optical flares in the synchrotron emission at late 
times (DDD2003). Only for a small fraction of cases ($\sim$ 14 out of 350 Swift GRBs) 
the observed late-time decline could not be well fit unless a transition from a 
constant density to a density proportional to $ 1/r^2$ was assumed.

Case by case,
the X-ray and optical light curves were calculated with the same CB 
parameters. The spectral index $p$ of the Fermi-accelerated electrons in 
the CBs was treated as a free parameter. In the CB model it determines 
both the spectral and temporal declines of the AG, as in Eq.~(\ref{Fnu}). 
In cases where the fit was insensitive to $p$ we fixed its value to be the 
canonical one: $p\!=\!2.2$ (DDD2002). The relative normalization of the 
X-ray and optical AGs was generally treated as a free adjustable 
parameter except when both extinction of the optical light and absorption 
of the emitted X-rays could be estimated reliably. In such cases the 
predicted dependence on frequency could be tested. These favourable cases include 
prompt ICS flares and the late-time SR afterglow where both the X-ray 
band and the optical band are above the bend frequency and the SR 
afterglow becomes `achromatic'. No attempt was made to derive the 
environmental parameters, because the use of simplifying assumptions --the 
single-CB approximation and the lack of exact knowledge of the extinction 
of the optical radiation in the host galaxy-- make such attempts potentially
unreliable.

In the following case studies of a representative sample of the entire 
diversity of Swift GRBs, we include GRBs with canonical, non-canonical and 
semi-canonical light curves, with or without superimposed flares, GRBs 
with very chromatic early-time afterglows, GRBs with exceptionally rapidly 
decaying late afterglows and GRBs with very complex and chromatic light 
curves. Special attention is given to XRF 060218. The parameters used in 
the CB-model description of the ICS flares and the synchrotron afterglow 
of all the GRBs and XRFs to be discussed anon are listed in Tables 
\ref{t1} and \ref{t2}.

\subsection{Case studies}
\label{Case} 

\subsubsection{GRBs with a canonical X-ray light curve} 

\noindent
{\bf GRB 060729.} {\it X-ray observations:}
This GRB  is described and discussed in detail in Grupe et al.~(2007).
It was detected and located by Swift at UT 19:12:29 July 29, 2006 
(Grupe et al.~2007). It is one of the brightest Swift GRB in X-rays 
and has the longest follow-up observations in X-rays: 
more than 125 days after burst. The X-ray 
observations were triggered by the detection of a GRB precursor by the 
BAT, which dropped into the background level
within 6 s. Two other major overlapping peaks 
were detected 70 s after trigger and a fourth one around 120 s. The 
end of the fourth peak was seen also by the XRT at the beginning of its 
observations. The XRT detected another flare around 
180 s followed by a rapid decay of the  
light curve by three orders of magnitude before it was 
overtaken by a plateau at $530 \pm 25$ s, which lasted
for $\!\sim\!0.5$ day before it bent into a power-law 
decline. During the fast decline of the prompt emission, the X-ray 
spectrum changed dramatically from a hard spectrum to a 
very soft one. After this phase the unabsorbed spectrum of the X-ray 
afterglow hardened to a power-law with $\beta_X\!\approx \!1.1$ and
remained unchanged during the plateau phase and the late power-law decline.
The complete 
light curve obtained from the 
observations with the BAT (extrapolated to the XRT band), the XRT and 
XMM-Newton is shown in Fig.~\ref{f1}a; a magnified view of the early 
times light curve is shown in Fig.~\ref{f1}b.

\noindent {\bf GRB 060729.} {\it Interpretation:} The CB-model fit to the 
complete X-ray light curve of GRB 060729 is shown in Fig.~\ref{f1}a.  The overall 
good agreement between theory and observations extends over some eight orders of 
magnitude in time and in flux. The spectral evolution of the X-ray emission is 
also in good agreement with the CB-model predictions (see DDD2008b). In the XRT 
0.3 -10 keV band the spectrum of the early time flares and their spectral 
evolution are well described by the exponential cut-off power-law obtained by ICS 
of a thin thermal bremsstrahlung spectrum, Eq.~(\ref{ICSlc}). In particular the 
exponential factor in Eq.~(\ref{ICSlc}) describes well the rapid softening of the 
spectrum as a function of time during the fast decaying phase of the burst 
(DDD2008b). But, as soon as the X-ray afterglow is taken over by the synchrotron 
emission around 325 s its unabsorbed spectrum changes to the much harder SR 
power-law spectrum with $\beta_X\!=\!1.1$. The temporal shape of the AG is best 
fit with electron spectral index $p=2.20$, which implies $\beta_X\!=\!1.1$, in 
good agreement with the observations of the XRT and of XMM-Newton (Grupe et 
al.~2007). The late power law decay satisfies the CB model prediction, 
$\alpha\!=\!\Gamma\!-\!1/2$.

\noindent
{\bf GRB 060729.} {\it Optical observations and a sketch of their
interpretation:}
The ROTSE-IIIa telescope in Australia took a first 5-s image of this GRB, 
starting about a minute after the burst, which showed no afterglow down to 
magnitude 16.6. Some 23 s later, an AG of magnitude 15.7 was clearly 
detected. The AG brightened over several minutes, and faded very slowly 
(Quimby et al.~2006b). In the CB model such a behaviour is expected 
from the early synchrotron emission  in the $UVOIR$ band during and 
shortly after the prompt GRB, see Eq.~(\ref{SRP}). The UVOT followed the 
{\it UVO} emission from 739 s to 20 days after trigger. The VLT in Chile
 obtained spectra, and determined a relatively low redshift of $z\!=\!0.54$ for this 
burst (Th\"one et al.~2006). The light curves of its {\it UVO} AG show a striking 
similarity to the X-ray light curve (Grupe et al.~2007) as predicted by 
the CB model for the optical AG when the bending frequency is below the 
{\it UVO} bands, and the extinction along the lines of sight to the hyperluminal 
CBs stays constant. All in all, the well sampled observations of the $XUVO$ 
light curves of GRB 060729 agree well with the expectations of the CB model.

\noindent {\bf GRB 061121.} {\it Observations.}
This canonical GRB at $z=1.314$ was described and discussed in detail in 
Page et al.~(2007). It is one of the brightest GRBs in X-rays observed to 
date by the XRT. The BAT triggered on a precursor to the main burst,
allowing observations of the latter from 
the optical to $\gamma$-ray bands. Many telescopes, including 
Konus-Wind, XMM-Newton, ROTSE, and the Faulkes Telescope North, also 
observed the burst. Its most intense activity began 60 s after trigger, and consisted 
of three overlapping peaks of increasing brightness, some 63, 69 and 74 s 
after trigger, as one can see in Fig.~\ref{f1}d. The 
$\gamma$-ray emission decayed fast after 75 s and became undetectable by 
the BAT beyond 140 s. The burst was also detected by Konus-Wind. The 
spectrum and its evolution  were well fit with a broken 
power-law. Its `peak' energy appeared to increase during the rise of each 
flare and decreased as their flux decayed. But its isolated main strong 
flare at $\sim\! 74$ s, as in many cases studied before, had a maximum 
$E_p$ at its beginning, which decayed monotonically thereafter. After the 
bright burst, the X-ray emission ---measured by the XRT and later also 
with XMM-Newton--- began to follow the `canonical' decay. Superimposed on 
the initially rapid decay from the major flare are two smaller flares 
around 90 s and 125 s. The rapid decay is taken over by the plateau around 
200 s and gradually breaks into a power-law decline, with an asymptotic 
power-law index $\alpha\!=\!1.59^{+0.09}_{-0.04}\, .$ The AG spectrum was well fit 
by an absorbed power-law with $\Gamma\!=\!2.07\! \pm \!0.06$, which 
slightly hardened after the AG bent down.

\noindent
{\bf GRB 061121.} {\it Interpretation:}
The hard mean spectral index, $\Gamma\!\sim\! 1.3$, and the continuously 
decreasing $E_p$ during the main flare are as predicted by the CB model 
(DD2004). The apparent increase of $E_p$ during the rise time of the 
smaller flares is probably an artifact of overlapping peaks, wherein the 
decay of a previous flare is taken over by a new flare only near its peak 
time. A comparison between the observed complete X-ray light curve (Page 
et al.~2007) and the CB model's fit is shown in Fig.~\ref{f1}c. The 
general trend before the onset of the synchrotron plateau is dominated by 
the main X-ray peak. The two smaller overlapping preceding peaks seen in 
the $\gamma$-ray light curve in Fig.~\ref{f1}d have been included in the 
fit, and so have the two late X-ray flares not intense enough to be seen 
in $\gamma$-rays. Assuming a constant density ISM, the CB model reproduces 
very well the observed light curve over seven orders of magnitude in 
intensity and five in time. The model's best fit to the 
entire X-ray light curve yielded $p\!=\!2.2$, which implies 
$\beta\!=\!1.1$ and 
$\alpha\!=\!1.6$, in good agreement with their observed values. The CB 
model 
also correctly predicts the spectrum and spectral evolution of the X-ray 
emission during the rapid-decline phase of the prompt emission (DDD2008a). 
The observed strong softening of the spectrum during the rapid decline is 
in full agreement with the spectral evolution predicted by 
Eq.~(\ref{ICSlc}). When the plateau phase takes over, the spectral 
power-law index changes to $\beta_X\!=\!1.07\pm 0.06$, remaining in 
agreement with the predicted $\beta\!=\!p/2$. The slight hardening of the 
spectrum at late time to $\beta\!=\!0.87\! \pm\! 0.08$ we have not 
predicted. 
It may be an artifact due to the assumption that the absorbing 
column density is constant during
the AG phase.

\noindent
{\bf GRB 061121.} {\it Optical observations and a sketch of their
interpretation:}
The UVOT detected an optical counterpart in the white filter starting 62 s 
after the trigger, and subsequently in all other filters (optical and UV).  
The UVOT light curve shows a prominent peak around $t\!\sim\!73$ s, 
coincident in time with the blended three main X-ray peaks, followed by a 
plateau phase. The complete optical light curves measured by the Swift 
UVOT and in ground observations with ROTSE, FTN, CTIO and MDM follows 
roughly the canonical X-ray light curve. The peak is well fit 
by SR from expanding CBs in a windy density $\propto\!1/r^2$. The rest 
of the light curve, as the canonical X-ray light curve, is well described by 
SR from a CB decelerating in a constant density with a spectral index 
gradually changing from $\beta\!\sim\!0.5$ above the bend frequency to 
$\beta\!=\!\beta_X\!\sim\! 1 $ below it, well after the AG bends down.

\noindent
{\bf GRB 050315.} {\it Observations:}
This GRB was described and studied in detail by Vaughan et al.~(2006).
It is one of the first Swift GRBs with a well sampled X-ray
light curve from trigger until late time ($\!\sim\! 10$ days).
It was detected and located by BAT at 20:59:42 UT on March 15, 
2005. The BAT light curve comprises two major overlapping peaks separated 
by about 22 seconds. Absorption features in the spectrum of its  optical 
afterglow  obtained  with the Magellan telescope indicated that its 
redshift is $z\!\geq\! 1.949$ (Berger et al.~2006). 
The XRT began observations 80 s after the 
BAT trigger and continued them for 10 days, providing one 
of the best sampled X-ray AGs (Vaughan et al.~2006). 
The extrapolation of the BAT light curve into the XRT band-pass 
showed the X-ray data to be consistent with the tail end of the decaying 
prompt emission. 
The combined light curve showed the canonical behaviour: 
the rapid decline ends $\sim\!300$ s 
after trigger, the plateau lasted for about 
$10^4$ seconds, before it gradually bent into a power-law 
decay. 

\noindent
{\bf GRB 050315.} {\it Interpretation:}
The complete X-ray light curve of this GRB  is compared with the CB-model 
prediction in Fig.~\ref{f1}e. An enlarged view at early times is shown in 
Fig.~\ref{f1}f. Two pulses are used in the fit.  As shown, the 
model reproduces the data well: the exponentially decaying contributions of the 
two pulses describe the changing slope of the fast-decaying phase.  The 
early ICS flares, the decay of the prompt emission and the subsequent 
synchrotron-dominated plateau and gradually bending light curve into a 
power-law decay are well reproduced by the model as shown in 
Figs.~\ref{f1}e,f.

\subsubsection{GRBs with a single power-law afterglow}

\noindent
{\bf GRB 061007.} {\it Broad-band observations:} 
The data were summarized and discussed by 
Mundell et al.~(2007) and Schady et al.~(2007). The Swift BAT detected the event 
on 2006 Oct 07, 10:08:08 UT. The prompt emission was also detected at MeV 
energies by Konus-Wind, Suzaku-WAM and RHESSI. The BAT $\gamma$-ray 
light curve has three peaks with substantial sub-structure, and a small 
fourth peak around 75 s that shows a long exponential decay and a faint 
emission detectable till $\sim\!900$ s. The Swift XRT and UVOT began observing 
80 s after the BAT trigger time and detected a very bright X-ray and optical 
counterpart with a power-law decay identical to that of the soft 
$\gamma$-ray tail, with a temporal slope 
$\alpha_X\!=\!1.6 \pm 0.1$ all the way to at least $10^6$ s with no indication 
of any break. The best-fit spectral index of the unabsorbed X-ray 
afterglow was $\Gamma\!=\!2.1\pm 0.1$. Robotically-triggered 
observations with 
the ground-based telescopes ROTSE and FTS began at 26 s and 137 s after trigger, 
respectively,  and the FTN continued them for 5.5 h. Follow-up 
observations were performed with the VLT and the Magellan-I Baade telescope. 
The spectral indices of the unabsorbed X-ray and unextinct 
optical AGs were found to be the same beyond 200 s.

\noindent
{\bf GRB 061007.} {\it Interpretation:}
In the CB model, afterglows decaying like a single power-law are observed 
when the AG'a bend takes place very early and is hidden under  
the prompt emission, or before the beginning of the observations. In the 
case of a constant ISM density, the decay index and the spectral index are predicted 
to satisfy $\alpha\!=\!\Gamma-1/2$ implying a decay index $\alpha\!=\!1.6 \pm 0.1$. 
The CB-model predictions for the light curves of the X-ray and $R$-band 
AG are shown in Figs.~\ref{f2}a,b. 
The best-fit temporal decay index  is $1.6$, as expected.
Agreement between theory and observations is very good. 
The slight wiggling around the power-law decay follows, in the CB model, from density 
variations along the trajectory of the CBs. The reported broad-band 
spectral index, corrected for extinction and absorption, is $\beta_{OX}\!=\!1.03$.
(Mundell et al.~2007). This is consistent with the CB-model's prediction 
(DDD2002) for a bend frequency below the optical band, as expected
beyond the AG's break. 
 
\noindent
{\bf GRB 061126.} {\it X-ray observations:}
The broad-band observations of this GRB were described and
discussed in detail in the framework of the fireball model 
by Perley et al.~(2008)   
who found their evolution troublesome and by Gomboc
et al.~(2008) who found them intriguing. 
It was a long GRB ($T_{90}=191$ s) dominated by two major 
$\gamma$-ray peaks within the first 40 s, which was followed by a smooth 
power-law decline with a temporal index $\alpha\sim 1.3$. Due to an Earth 
limb constraint Swift slewed to the burst only after 23 min and 
followed its X-ray AG from 26 min to 20 days after burst. Its X-ray AG 
showed roughly a power-law decline with the same power-law index,
$\alpha_X\!\sim\! 1.3$, with marginal evidence for steeper early and very 
late time-declines, with an index $\alpha\!\sim\! 2$. 
The unabsorbed X-ray AG  had a best-fit spectral index  
$\beta_X\!=0.94\!\pm\! 0.05$. 

\noindent
{\bf GRB 061126.} {\it Optical observations:}
In the optical band this was one of the brightest Swift GRBs. Its 
optical emission was detected by RAPTOR during the $\gamma$-ray emission 
21 s after the BAT trigger, and its early decline was followed also by the 
PAIRITEL, NMSU, KAIT, Super Lotis and FTN robotic telescopes. Observations 
continued with several large telescopes. The data were 
summarized and discussed in Perley et al.~(2008) and in Gomboc et al.~(2008). 
The initial power-law decay of the optical AG with $\alpha\!\sim\!1.5$ changed 
to a shallower decline approximately 1000 s after the burst, which 
steepened about 1.5 d after the burst.  The last optical data point 
($0.52\pm 0.05\, \mu$Jy) obtained with Gemini North at 15 d after burst 
(Gomboc et al.~2008) was slightly dimmer than the host galaxy at redshift 
$z\!=\!1.16$ (Perley et al.~2008), whose $R$-magnitude $24.10\pm 0.11$ 
($0.70\pm 0.07\, \mu$Jy)  was measured with the Keck telescope 54 d after burst. 
After correcting for Galactic extinction and estimated extinction in the 
host galaxy, its observed early-time optical emission had 
$\beta\!=\!1.0\pm 0.1$ and showed a strong colour evolution towards 
$\beta_O\!\sim\! 0.4\!$ - $\! 0.5$
around 2000 s. The late-time index was typically $\beta_O\!=\!0.95\pm0.10$, 
consistent with $\beta_O\!=\!\beta_X$.

\noindent
{\bf GRB 061126.} {\it Interpretation:} Swift's XRT detected the X-ray
emission only long after the prompt signal had faded. Its measured light
curve, shown in Fig.~\ref{f2}c, was well fit by SR, with $p\!=\!1.84$,
in a density $n\! \propto\! 1/r^2$, taken
over by a constant at $t\!\sim\!3000$ s. The CB-model
expectation, $\Gamma\!=\!p/2\!+\!1\!=\!1.92$, is consistent with the mean photon
spectral index (corrected for absorption), inferred by Evans et
al.~(2007), $\Gamma\!=\!1.82\pm 0.05$, and by Perley et al.~(2008),
$\Gamma\!=\!2.00\pm 0.07$. The wiggling around the
power-law decay --induced by density variations along the CB's path-- 
we did not model. In the FB model prompt and AG emissions
are SR-dominated and the extrapolation from the $\gamma$- and X-domains 
to the optical band results in a signal much brighter and with a different
light curve than the observed one (Perly et al.~2008).
The changing power law of the declining optical light 
is well described by SR from a CB with the same parameters modeling
the X-ray light curve. A SN contribution akin to
that of SN1998bw placed at the GRB location was added to the CB light curve.
The general behavior is well reproduced by the CB model,
as can be seen from Fig.~\ref{f2}d. The slight wiggling of the 
light curve around its CB model's description and the corresponding changes in
the spectral index $\beta_O$ we attribute, as usual, to small density variations,
which we did not model. The expected variation of the
spectral index from $\beta_O\!\approx\!0.5$ at early time to
$\beta_O\!\approx\! \beta_X$ is supported by the data (Perley et al.~2008). The
evolution of $\beta_O(t)$ at very early time may be due to
variation in light extinction as the CB moves away from the SN.

\noindent
{\bf GRB 080319B} {\it Observations:}
This GRB was detected by the Swift (Racusin et al.~2008b,), INTEGRAL (Beckmann 
et al.~2008) and Konus-Wind (Golenetskii et al.~2008) satellites. 
It lasted  $\sim \! 60$ s. It was the brightest observed
long GRB so far. Three robotic ground 
telescopes detected its extremely intensive optical light emission 
(Karpov et al.~2008, Cwiok et al.~2008, Wozniak et al.~2008) before the 
Swift alert, and saw it brightening to a visual peak magnitude 5.4, 
visible to the naked eye, some 18 s after the start of the burst. Swift 
XRT slewed to the GRB position within 65 s and followed its power-law 
declining X-ray light curve for the first 15 days. Swift's 
prompt alert sent to the world's telescopes triggered many follow-up 
observations including spectral measurements with the VLT (Vreeswijk, et 
al. 2008) and Hobby-Eberly telescope (Cucchiara et al.~2008) 
which determined  the GRB's redshift to be $z\!=\!0.937$. Its 
X-ray light curve is shown in  Fig.~\ref{f2}e. Its
combined $R$- and $V$-band light curve (normalized to the $R$-band), as 
reported in GCNs (see, e.g.,~Bloom et al.~2008 and references therein)
is shown in Fig.~\ref{f2}f.

\noindent
{\bf GRB 080319B.} {\it Interpretation:}
In the CB model the ICS spectrum of the scattered glory's photons is an 
exponential cut-off power-law with a spectral index, $\Gamma\!\approx\!1$, 
cut-off energy $\approx\!E_p$, and a power-law tail, see 
Eq.~(\ref{GRBspec}). The spectral index, $\Gamma\!=\! 1.01\!\pm\!0.02 $ in the 
15 -350 keV range, reported by the Swift BAT team (Racusin et al.~2008b), and  
the Band function fit to the broader 20 keV to 7 MeV energy range, reported 
by the Konus-Wind team (Golenetskii et al.~2008) are in agreement with the
CB model.

In Fig.~\ref{f2}e we compare the X-ray light curve of GRB 080319B, measured 
with the Swift XRT, to its CB-model description, Eq.~(\ref{Fnu}), assuming a 
constant ISM density and a single effective CB. 
The best-fit $p$ is 2.08, yielding an approximate power-law decline 
with $\alpha_X\!=\!1.54$ beyond $t_b$, best fit to 72 s. The description
of the AG is quite good except 
around $4 \times 10^4$ s, where the data are sparse.  As expected (DDD2008a) 
for very luminous GRBs, no AG break is observed. The temporal index, 
$\alpha_X\!=\!\Gamma_X\!-\!1/2\!=1\!.42\! \pm\! 0.07$, predicted from the 
late-time photon spectral index, $\Gamma_X\!=\!1.92\! \pm \!0.07$,
reported by the Swift XRT team (Racusin et al.~2008c) is in 
agreement with the best-fit temporal 
index. At $t\!\sim\!4 \times 10^4$ s, the data lie below the fit. If not a 
statistical fluctuation, this may be due to a failure of the 
constant-density approximation, not surprising at this level of precision. 

GRB 080319B began with a succession of prompt $\gamma$-ray 
pulses, but the XRT observations started too late to
detect their X-ray counterparts, which were seen at 
optical frequencies. Even though the optical pulses are SR-generated, 
their expected time dependence, Eq.~(\ref{SRP}),
is akin to that of the $\gamma$-ray pulses.
The CB-model optical light curve, shown in 
Fig.~\ref{f2}f, was obtained by
fitting each of the three early pulses observed by 
TORTORA (Karpov et al.~2008).  The later-time
AG, described by Eq.~(\ref{Fnu}) for $t_b\,\gsim\, 70$ s,
is essentially a power-law decline, insensitive to
the precise values of the best-fit $\theta\,\gamma_0$ and $t_b$, but
sensitive to $\beta_O$. In the CB model,
the index $\beta$ is $\sim\!0.5$ below and $\sim\!1.1$
above the bend frequency,
which usually crosses the optical band within $t\!\sim\! 1$ day, so that
$\beta_O\!\approx\! \beta_X$ thereafter. Our best fit to the optical AG results in
$\alpha_{O}\!=\!1.40\pm 0.04$, which
implies a late-time  $\beta_{O}\!\approx\! 0.90$, consistent
with the expectation. So far no late-time spectral information is available
to verify it.

When a CB  crosses a density enhancement,
$\nu_b$ increases due the sudden increase in $n$
and the consequently faster CB deceleration. The
bend frequency may then cross the optical band `backwards': from
above it, to below it. Such a spectral evolution may have been observed  
some 5000 s after the onset of the burst (Bloom et al.~2008).
The spectral analysis of the UNLV GRB group (Zhang et al.~2008)
shows a decreased $\beta_X\!=\!0.70\pm 0.05$ around that time.
The expected $\beta_{O}\!\approx\!\beta_X-0.5\!=\!0.2\pm 0.05$
at that time is consistent with the
spectral evolution around 5000 s after burst  reported by Bloom et al.~(2008).

\subsubsection{GRBs with a semi-canonical X-ray light curve}

\noindent
{\bf GRBs 060211A, 061110A, 080307, 051021B, 080303, 070220.} 
{\it Observations and interpretation:} 
These GRBs, detected by the Swift BAT and followed up by its XRT,
have canonical X-ray light curves, but for the fact that their exponentially-declining phase    
at the end of the prompt emission changed into a slower power-law 
decline before it entered the plateau phase. Their X-ray light curves
and their CB-model description
are shown in Figs.~\ref{f3}a-f.
The exponential decline  of the prompt ICS emission, as given by Eq.~(\ref{ICSlc}),
is taken over by the 
SR emission from the CBs, which, in a windy circumburst environment, decays
like a power: $F_\nu \sim t^{-(1+\beta_X)}\, \nu^{-\beta_X}$,
see Eq.~(\ref{SRP}). This takeover by SR is accompanied by 
a sudden hardening of the AG to the ordinary SR spectrum, 
$\sim\!\nu^{-\beta_X}$ with $\beta_X\! \sim\! 1.1$.
The power-law decay changes into the canonical plateau 
when  the CBs enter the constant ISM density.
In the case of GRB 070220, the fast asymptotic decline
was well fit  assuming  an isothermal sphere  density  profile,
$n\propto 1/(r^2\!+\!r_c^2)$, as in the cases shown in 
Figs.~\ref{f6},\ref{f7}.
 
\subsubsection{GRBs with large X-ray flares during the early 
X-ray afterglow}  

\noindent
{\bf GRB 060526.} {\it X-ray observations:} 
This GRB was detected by Swift's BAT at 16:28:30 UT on May 26, 2006 
(Campana et al.~2006a). The XRT began observing the field 73 s after the 
BAT trigger. The burst started with a $\gamma$-ray emission episode 
lasting 18 s. The GRB was thereafter quiet for about 200 s, and then 
emitted two additional pulses which lasted about 50 s and were 
coincident with strong X-ray flares between 220 s and 270 s after 
trigger. The XRT followed the X-ray emission for 6 days until it faded into the 
background. The entire XRT light curve is shown in Fig.~\ref{f4}a. It has 
the canonical behaviour of many Swift GRBs, and two superimposed 
early-time large flares.

\noindent
{\bf GRB 060526.} {\it Optical observations:}
The observations of the optical emission from GRB 0605526 are summarized 
and discussed in Dai et al.~(2007), Khamitov et al.~(2006)
and Th\"one et al.~(2008b). They were 
started as early as 36.2 s after the BAT trigger by the Watcher 40cm 
robotic telescope, in South Africa, which saw the AG 
at a very bright 15th magnitude (French \& Jelinek 2006). 
The UVOT on Swift detected its optical AG 
81 s after trigger (Campana et al.~2006a). The burst was followed 
up with UVOT and ground-based telescopes by several groups. Spectra 
obtained with the Magellan-Clay telescope indicated a redshift of 
$z\!=\!3.21$ (Berger \& Gladders~2006). Its $R$-band light curve obtained 
with the MDM and PROMPT telescopes at Cerro Tololo, amongst others, 
is shown in Fig.~\ref{f4}c (Dai et al.~2007 and references 
therein). It can be seen in Figs.~\ref{f4}a,b that, apart from the 
superimposed large early-time X-ray flares which are not present in the 
optical light curve and the late mini-flares, the X-ray light curve and the 
well-sampled $R$-band data show a roughly achromatic behaviour.

\noindent
{\bf GRB 060526.} {\it Interpretation:} The entire X-ray light curve and 
its CB-model's fit are shown in Fig.~\ref{f4}a. Three ICS pulses were used 
in the fits of the early time emission, although the third pulse may well 
be a superposition of two unresolved ones. The pulse shape and the 
spectral evolution of the last two large flares are typical of ICS 
flares (DDD2008a). Their coincidence in time with the late $\gamma$-ray 
peaks, the absence of corresponding peaks in the optical UVOT light 
curve, and their spectral evolution support their interpretation as part 
of the prompt GRB emission. This GRB's `prompt' emission extends to long 
times partly because of the large redshift of the burst source which 
stretches observer time by the relatively large factor, $z+1\!=\!4.21$. 
A zoom-in on these two ICS X-ray flares is shown in Fig.~\ref{f4}b. The 
decay of the prompt emission is dominated by the decay of the last pulse. 
In Figs.~\ref{f4}a,b,c we show that the early ICS flares, the decay of 
the prompt emission, the subsequent synchrotron-dominated plateau and the
gradual bending into a power-law decline are all well reproduced 
by the CB model. In Fig.~\ref{f4}d we show the theoretical $R$-band light 
curve obtained with the parameters which were fitted to the SR X-rays. 
 Since the bending frequency during the steepening phase is below 
the $R$ band, the temporal decay of the $R$-band light curve practically 
coincides with that of the X-ray one. Both the X-ray and the late-time optical 
light curve are bumpy, which may be caused by mini-flares 
and/or density inhomogeneities, which we have not tried to fit. The 
apparent steeper decay of the optical AG beyond 1 day may be the decline 
following a flare or a transition into the galactic halo with a 
density declining as $1/r^2$.

\noindent 
{\bf GRB 060206.} {\it Observations:}
 This GRB triggered the Swift's BAT on February 6, 2006 at 04:46:53
UT (Morris et al.~2006). Its $\gamma$-ray emission lasted only 
6 s. The XRT started its observations 80 s after the BAT trigger.
Despite its initially poor time sampling,  it detected  
an X-ray decline after 0.5 h and a strong rebrightening after 1 h, 
after which its follow-up was nearly continuous for some 20 days.
The bright optical AG of GRB was detected by Swift at $V=16.7$, 
about 1 minute after the burst.   
RAPTOR   started  observations 
48.1 min after trigger and reported that, after 
an initial fading, the AG rebrightened 1h after burst 
by $\sim\! 1$  magnitude within a couple of minutes 
(Wozniak et al.~2006). Many observatories followed the bright optical AG
(Monfardini et al.~2006, Stanek et al.~2007, and references therein), and 
Fynbo et al.~(2006a) carried out spectral observations to determine its 
large redshift, $z\!=\!4.05$, later confirmed by other groups.
The RAPTOR data clearly shows that the rebrightenning was due to two
flares (Wozniak et al.~2006). Similar `anomalous' rebrightennings  of the
optical AG were seen in some other bursts (Stanek et al.~2007) .

\noindent
{\bf GRB 060206.} {\it Interpretation:}
In Fig.~\ref{f4}a,b we compare the observations of the X-ray and $R$-band 
light curves with the CB-model fits. Superimposed on the plateau phase are 
two strong flares beginning around 1 h after trigger. The coincidence in 
time of the X-ray and optical flares, and the absence of any evidence for 
the typical ICS strong spectral evolution, suggest that these two flares 
are SR flares due to an encounter with a density jump, such as at the boundary 
of a superbubble created by the star formation region.  In Fig.~\ref{f4}c 
we compare the observed light curve of these two flares in the $R$ band 
and their CB-model description via Eq.~(\ref{SRP}). The figures show that the 
agreement is very good and that there is nothing `anomalous' in the X-ray 
and optical data of GRB 060206. Instead, their prominent structures are 
well described and precisely related by their CB-model's understanding in 
terms of SR from late ejections of CBs into the 
circumburst windy environment.

\subsubsection{GRBs with chromatic afterglows}  

\noindent
{\bf GRB 050820A.} {\it Broad-band observations:}
This is one of the Swift GRBs with the best-sampled broad-band data,  
summarized and discussed in detail in Cenko et 
al.~2006. The burst
was detected and observed by Swift and Konus-Wind. Its $\gamma$-ray 
emission was preceded by a soft precursor pulse some 200 s before the main 
burst. The latter lasted some 350 s and consisted of 5 well-separated major 
peaks, with a clear spectral-softening evolution within each peak. 
The main peak observed by Swift during $217\,\rm{s}\!<t\!<\!241$ s, and the 
time-integrated photon spectrum over the entire burst were well fit with a 
cut-off power law with a photon indices $\Gamma\!=\!1.07\pm 0.06$ and 
$\Gamma\!=\!1.12\pm 0.15$,
respectively. 
The Swift XRT began observations 88 s after trigger and 
followed its X-ray emission until 44 days, see 
Fig.~\ref{f5}a. The measured mean photon index of the unabsorbed 
emission in the 0.3-15 keV band during the prompt emission phase was 
$\Gamma\!=\!1.06\pm 0.04$ ($\beta_X\!=\!0.06\pm 0.04$) and 
$\Gamma\!=\!2.06\pm 0.07$ ($\beta_X\!=\!1.06\pm 0.07$) during the 
afterglow phase (Evans et al.~2007).

The prompt optical emission was measured by RAPTOR 
beginning 18 s after trigger. 
The Swift UVOT began observations 80 s after  trigger but became 
inoperable when Swift entered the South Atlantic Anomaly 
approximately 240 s after trigger. The automated Palomar 60-inch 
telescope started observations 206 s after trigger
and followed-up until late time. Later measurements were made by the
Turkish Russian 1.5 m telescope. Late-time images were taken with the 9.2 
m Hobby-Eberly Telescope  and with the Hubble Space Telescope 
until 37 days after burst. The $R$-band light curve is shown in Fig.~\ref{f5}b.
Ignoring host reddening and correcting for 
Galactic extinction  in the burst direction [E (B -V)$=\!0.044$], the fitted 
spectral index (Cenko et al.~2006) in the optical band during the prompt 
emission was $\beta_O\!=\!0.57\pm 0.06$,  steepening to 
$\beta_O\!=\!0.77\pm 0.08$ within the first day. While the optical 
spectrum appeared steeper later on, the poor fit quality precluded the
derivation of a reliable value. In Figs.~\ref{f5}a,b one can see
the very chromatic behaviour of the X-ray- and optical light curves
during the prompt  and AG  phases.

\noindent
{\bf GRB 050820A.} {\it Interpretation:}
The pulse shape of the  prompt-emission $\gamma/$X-ray peaks and their 
spectral index agree well with those predicted by ICS of glory light.
The CB-model fit to the entire XRT light curve 
is shown in Fig.~\ref{f5}a. The early-time light curve 
is well described by the ICS X-ray counterparts of the prompt 
$\gamma$-ray peaks: the very-early-time XRT light curve is  
the tail of the precursor pulse, the next pulse is the 
X-ray counterpart of the first ICS $\gamma$-ray pulse 
around 220 s. The ICS peaks are superimposed on a canonical 
SR afterglow bending down at around 1000 s. 
We interpret the X-ray peak around 5000 s as a flare due to a density bump.
While the prompt $\gamma$-ray and X-ray emission is dominated by 
the ICS of glory light, which yields $\beta\! \sim\! 0$,  
the optical emission is dominated 
by  SR, as in Eq.~(\ref{SRP}), 
with the typical $\beta\!\sim\! 0.5$, as observed. The different
radiation mechanisms are responsible for the 
chromatic behaviour of the prompt emission. 

Although both the X-ray AG and the optical AG are dominated by SR, the 
optical AG evolves differently than the X-ray AG because of its 
dependence on the bend frequency, a function of the ISM density and 
the Lorentz factor of the decelerating CB. Consequently, the early-time 
optical and X-ray AGs are chromatic until the bend frequency crosses well 
below the optical band, after which $\beta_O=\beta_X$. This is shown in 
Figs.~\ref{f5}a,b. The CB-model $R$-band light curve in 
Fig.~\ref{f5}b was calculated with $\beta_O\!=\! 0.77$ and the best-fit 
parameters of the X-ray AG shown in Fig.~\ref{f5}a.
The calculated light curves did not include the late-time flares 
in order not to obscure the chromatic behaviour of the underlying 
smooth AGs.

\noindent
{\bf GRB 060418.} {\it Broad-band observations:}
This GRB was discussed in detail in Molinari et al.~(2007). Its $\gamma$-ray 
emission was detected and observed by the Swift BAT and by Konus-Wind. 
The BAT light curve showed three overlapping peaks at 10, 18 and 27 s 
and a bump which coincided with an X-ray flare at 128 s after trigger. The 
Swift XRT started observing the GRB 78 s after trigger. The XRT 
light curve shows a notable flaring activity superimposed on a smooth 
AG decay. A prominent peak, also visible as a bump in the BAT data, 
was observed at about 128 s after trigger. The REM robotic telescope began 
observing this GRB  64 s after trigger in the $z'JHK$ bands and 
followed it down to the sensitivity limits. The {\it UVONIR} AG was also 
detected by the Swift UVOT, by one of the 16-inch PROMPT telescopes at 
CTIO and by the robotic telescope FRAM 
(part of the Pierre Auger Observatory).  The {\it ONIR} AG was also 
followed up with the 1.3 m telescope at CTIO beginning $\sim$1 h 
post-trigger, and with the PAIRITEL 1.3 m telescope staring 2.53 h after 
 trigger.  The {\it UVONIR} light curves show a very chromatic initial 
behaviour compared to the XRT light curve, see Figs.~\ref{f5}c,d. 
The $NIR$ AG rises until reaching a maximum around 130 s after trigger and 
then gradually changes to a power-law decline shallower than that of the 
X-ray AG, with a weak flare 
superimposed on it at around 5 ks, which roughly coincides in time with a 
strong X-ray flaring activity.

\noindent
{\bf GRB 060418.} {\it Interpretation:}
The XRT light curve was fit by the tail of the prompt ICS 
emission, and an ICS flare around 128 s which was later
taken over by the SR afterglow of a CB moving 
in a constant density ISM (Fig.~\ref{f5}c).
The bend of the SR afterglow
is hidden under the tail of the X-ray flare at 128 s.
The $H$-band light curve, shown 
in Fig.~\ref{f5}d,  was calculated using Eq.~(\ref{SRP})
with an early-time unabsorbed $\beta_O\!=\!0.5$
and an ejection time, $t_i\!=\!26$ s, coincident with 
the start-time of the major $\gamma$-ray peak.   
No attempt was made to model the flaring activity around 5 ks. 

\noindent
{\bf GRB 071010A.} {\it Broad-band observations:}
This GRB at redshift $z\!=\!0.985$ was discussed in detail by 
Covino et al.~(2008). It had a single peak lasting for 6 s, detected by 
the Swift BAT. 
Swift did not slew to this GRB because its automatic slewing  
was temporarily disabled. 
The XRT began observing this GRB only 34 ks after the BAT 
trigger and followed it until 550 ks after trigger. 
The XRT light curve (Fig.~\ref{f5}e)
shows a wide flare  peaking around 60 ks and
followed by a power-law decay with an index $\sim\!1.6\!\pm 0.3$.
The early {\it ONIR}  emission was observed by the 
TAROT, REM and the 2.2 m MPI-ESO telescopes.
Follow-up $NIR$ observations were carried out with Gemini-North,  TNG and the NTT.
The AG was observed a few hours after the GRB
with the Keck-I and Sampurnan telescopes and with  NOT and  VLT.
The {\it ONIR} light curve shows 
an initial rising with a maximum at about 7 min, and 
a smooth decay interrupted by a flare about 0.6 d,
visible in both the {\it ONIR} and in X-rays
(Figs.~\ref{f5}e,f). The {\it ONIR}
spectrum was modeled by a power law with an SMC-like extinction 
law with a best fit  E (B -V) = 0.21. The reported unabsorbed late index was 
$\beta_O\!=\!1.26\pm 0.26 $.

\noindent
{\bf GRB 071010A.} {\it Interpretation:}
The {\it ONIR} light curves correspond to the SR radiation 
from a CB ejected into a windy $1/r^2$ density profile, as 
given by Eq.~(\ref{SRP}),  until taken over by a constant-density 
ISM,  with a standard wide flare superimposed on the AG  
around 0.6 d. The late XRT light curve was calculated with the same 
parameters except for $\beta_X\!=\!1.1$.

\subsubsection{GRBs with very fast-decaying late afterglows}
      
In Figs.~\ref{f6} and \ref{f7} we show the well-sampled XRT light curves of 12 GRBs:
050318, 050326, 050814, 051008, 061019, 060807, 060813, 070306, 070419B, 
070420, 070521 and 080207, with a late decay more rapid than the 
canonical $t^{-1.6}$ decline of the AG of CBs decelerating in a constant-density 
ISM.  In the CB model such a fast decline is produced by a fast-declining 
ISM density or by the tail of a late flare. We have found that all Swift 
GRBs with a well sampled fast-declining X-ray light curve can be 
reproduced by either an asymptotic $n\!\propto\! 1/r^2$ density profile or a 
tail of a late flare, as demonstrated in Figs.~\ref{f6} and \ref{f7}. Such 
an asymptotic density decline is typical of isothermal spheres, for which
$n(r)\!\sim\!n_0/[1\!+\!(r/r_c)^2]$, a fair representation of
the density profile of galactic bulges in spirals, of ellipticals, and of the outskirts
of bumpy density shells created by stellar winds. 
The CB-model prediction is that for $r\! \gg\! r_c$, the AG declines like 
$ F_\nu\! \propto\! t^{-(1+\beta)}\,\nu^{-\beta}$, i.e.~with 
$\alpha\!=\!\beta+1\!=\!\Gamma_{SR}\!\sim\! 2.1$. In some GRBs the 
transition from $\alpha\!=\!\Gamma-1/2\!\sim\!1.6 $ for $r \ll r_c$, 
to $\alpha\!=\!\Gamma\sim 2.1$ for $r \gg r_c$, has probably been 
misidentified as the standard FB-model late achromatic `jet break' 
(e.g.~Dai et al.~2008, Racusin et al.~2008a).

\subsubsection{GRBs with complex chromatic light curves}

\noindent
{\bf GRB 050319.} {\it Observations:}
The Swift BAT, XRT and UVOT observations of this GRB were discussed 
in detail in Cusumano et al.~(2006a) and Mason et al.~(2006). A reanalysis of 
the BAT data showed that its onset was $\sim\!135$ s before the trigger 
time reported by Krimm et al.~(2005). The XRT began its observations 90 s 
after the BAT trigger, continuing them for 28 days (Cusumano et 
al.~2006a). The $\gamma$-ray light curve shows two strong peaks. The X-ray 
light curve had the canonical behaviour: an early fast decline which 
extrapolated well to the low-energy tail of the last prompt $\gamma$-ray 
pulse at around 137 s after the onset of the GRB. After $\sim\!400$ s, the 
fast decline was overtaken by a plateau which gradually bent into a 
power-law decline after $\sim\!10^4$ s.  

\noindent
{\bf GRB 050319.} {\it Interpretation:}
A CB-model fit to the complete X-ray light curve is shown in 
Fig.~\ref{f8}a. Two pulses are used in the early ICS phase. The early ICS 
flares, the decay of the prompt emission and the subsequent 
synchrotron-dominated plateau and gradually bending light curve are well reproduced.
The spectral index of the AG,   
$\beta_X\!=\!0.73\pm 0.05$, and its asymptotic temporal decline index, 
$\alpha\!=\!1.14\pm 0.2$ (Cusumano et al.~2006a),  satisfy well the relation 
$\alpha\!=\!\beta\!+\!1/2$, though they were obtained by correcting only for 
Galactic absorption.

\noindent
{\bf GRB 050319.} {\it UVO observations:}
The UVOT detected an optical counterpart in the initial White filter 
observation, starting 62 s after the trigger, and subsequently in all 
other filters (optical and UV). Swift's UVOT, which followed the typical 
sequence for GRB observations, was able to observe the UVO emission 140 s 
after its detection by the BAT. It was also observed by ground-based 
telescopes RAPTOR (Wozniak et al.~2005), and ROTSE III (Quimby et 
al.~2006a) just 27.1 s after the Swift trigger.  The optical AG was 
followed later with a number of ground-based telescopes (Huang et al.~2006 
and references therein). An absorption redshift, $z\!=\!3.24$, was 
measured with the Nordic Optical Telescope (Jakobsson et 
al.~2006). The afterglow of this GRB is highly chromatic with no apparent 
correlated behaviour between its X-ray and optical emission,
as can be seen from Figs.~\ref{f8}a,b.

\noindent
{\bf GRB 050319.} {\it Interpretation of UVO observations:}
The early-time optical light curve was fit by SR emission from the two 
separate CBs implied by the first two strong $\gamma$-ray peaks. The late 
X-ray and optical AG were calculated with the same deceleration 
parameters. The complex optical light curve is reproduced well, as shown 
in Fig.~\ref{f8}b.

\noindent
{\bf GRB 060605.} {\it Broad-band observations:}
This GRB  at  $z\!=\!3.773$ 
was studied and discussed by Ferrero et al.~(2008). 
It was long and relatively faint,  with 
a duration of about 20 s,  detected by the Swift BAT. 
The BAT light curve showed two overlapping peaks.
The Swift XRT began taking data 93 s after the BAT
trigger and continued for 200 ks. The XRT light curve shows a 
canonical behaviour
with a flare around 265 s after trigger, superimposed 
on a shallow plateau which began at $\sim 200$ s 
and changed into an asymptotic power-law decline beyond 8 ks
(Fig.~\ref{f8}c). The best-fit spectral index of the unabsorbed 
spectrum in the X-ray band 
was $\beta_X\!=\!1.06\!\pm\!0.16$. 
The UVOT, which began observations of the GRB's field 97 s after 
 trigger detected and localized its fading AG.
Follow-up observations in the {\it UVONIR} bands were carried out also with
ROTSEIIIa, which began 48 s after trigger and with the VATT, RTT, TNG   
and the Kitt Peak 2.1 m telescopes. In Fig.~\ref{f8}d we show the 
recalibrated $R_c$-band light curve from these 
observations (Ferrero et al.~2008).
In contrast with the XRT light curve, it shows a chromatic early rise
with a `broken' power-law decay. The {\it XUVONIR} data show a spectral 
evolution at early time from $\beta_{OX}\!=\! 0.8 \pm 0.05$  at 0.07 
d, to $\beta_{OX}\!=\! 1.02 \pm 0.02\!\approx\! \beta_X$  at 0.43 d.

\noindent
\bf GRB 060605.} {\it Interpretation:}
The XRT light curve was fitted with a canonical CB-model X-ray light curve 
(Fig.~\ref{f8}c),  beginning with the tail of the fast decline of the 
prompt ICS emission, and taken over by SR in a 
constant density environment which changes to an $1/r^2$ profile
beyond 8 ks. The 
bump around 250 s was interpreted as a SR flare superimposed on the 
smooth canonical AG. The corresponding $R$-band light curve, shown in 
Fig.~\ref{f8}d, was generated using Eq.~(\ref{SRP}) with the fit 
parameters of the X-ray flare and the canonical $\beta_O\!=\!0.5$ for an 
early optical emission, until it was taken by the SR emission in the 
density profile used in the CB-model description of the late X-ray  AG.

\noindent
{\bf GRB 060607A.} {\it Broad-band observations:}
This GRB  at $z\!=\!3.082$
was studied by Molinari et al.~(2007), Nysewander et al.~(2007) 
and  Ziaeepour et al.~(2008). The Swift XRT began observing it  
73.6 s after the BAT trigger. Its complex X-ray light 
curve, like that of quite a few other
GRBs, was dominated by strong flaring activity. The XRT light curve, shown 
in Fig.~\ref{f8}d,  exhibits 
three early  flares peaking at approximately
97 s, 175 s, and 263 s after trigger and a  continuing
weaker flaring activity superposed on a decaying continuum.
The UVOT began to observe the bright optical AG  75 s after trigger.
The REM telescope began {\it NIR} observations 59 s after
trigger. It detected a brightening smooth light curve which peaked
around  $\!\sim\!155$ s and decayed like a power law  
interrupted by flaring activity beyond 1000 s. The REM followed the decay for 
20 ks down to its sensitivity limit. Four 0.4m PROMPT telescopes began 
observing the AG  44 s after trigger
and measured the {\it UVO} light curves until 20 ks,
which behaved as the {\it NIR} light curve
(Nysewander et al.~2007).  

\noindent {\bf GRB 060607A.} {\it Interpretation:} 
The complex X-ray light curve, shown in 
Fig.~\ref{f8}e, was fit with 6 flares superimposed to the AG of a 
single dominant CB. This fit, which can be improved by splitting the last 
flare into two, is a very rough description $(\chi^2/${\it dof} $=\! 4.9$ 
for 440 {\it dof)}, not a proof of the quality of a prediction. 
Moreover, in cases with such a prominent flaring activity, the mean spectral 
index of the AG data is an average between the typical index of flares, 
$\Gamma\!=\!1.5$, and that of a SR afterglow, $\Gamma\!=\!2$, i.e.~an 
average significantly smaller than that of the SR. Thus, we do 
not expect such a labyrinthine AG to satisfy the CB-model spectral-index 
relations, Eqs.~(\ref{Fnux},\ref{Asymptotic}). The CB model's early {\it UVONIR} 
light curves, shown in Fig.~\ref{f8}f for the $H$ band, is well 
described by the smooth SR afterglow of a single CB moving in a wind 
environment, as given by Eq.~(\ref{SRP}). We did not try to fit the 
weak flaring activity, which is 
probably due to a bumpy environment.

\subsubsection{XRF 060218}
\label{XRF}
\noindent 
{\bf XRF 060218/SN2006aj.} {\it Broad-band observations:} This 
XRF/SN pair provides one of the best testing grounds of theories (De 
R\'ujula 2008) given its proximity, which resulted in very good sampling 
and statistics (see, e.g.~Campana et al.~2006b, Pian et al.~2006, 
Soderberg et al.~2006, Mirabal et al.~2006, Modjaz et al.~2006, Sollerman 
et al.~2006, Ferrero et al.~2006, Kocevski et al.~2007). The XRF was 
detected with the Swift's BAT on February 18, 2006, at 03:34:30 UT 
(Cusumano et al.~2006b). The XRT and UVOT detected the XRF and began 
taking data 152 s after the BAT trigger. Its detection led to 
a precise localization, the determination of its redshift, $z\!=\!0.033$ 
(Mirabal et al.~2006) and the discovery of its association with a 
supernova, SN2006aj (Masetti et al.~2006). The BAT data lasted only 300 s, 
beginning 159 s after trigger, with most of the emission below 50 keV 
(Campana et al.~2006b, Liang et al.~2006).  The total isotropic equivalent 
$\gamma$-ray energy was $E_{\rm iso}\sim 0.8\times 10^{49}$ erg and the 
spectral peak energy, $E_p$, strongly evolved with time from $\!\sim\! 54$ 
keV at the beginning of observations by the BAT down to $\!<\!5$ keV 300 s 
later. The X-ray light curve  was followed up with the XRT until nearly 
$1.1\times 10^6$ s after burst (Campana et al.~2006b). It 
showed the canonical behaviour of X-ray light curves of XRFs and GRBs, 
except that the prompt X-ray emission was stretched in time and lasted 
more than 2000 s. The prompt emission ended with a fast temporal decline 
and a rapid spectral softening (Fig.~\ref{f9}a) that was overtaken around 
10 ks by an ordinary power-law-decaying AG. Follow-up observations with 
the UVOT and ground-based telescopes showed a very chromatic {\it UVONIR} AG 
with a long brightening phase with a peak between 30 and 60 ks, which 
changed into a fast decline and was taken over around 2 d after burst by 
the rising light curve of SN2006aj (Marshall et al.~2006, Campana et 
al.~2006b, Pian et al.~2006, Mirabal et al.~2006, Sollerman et al.~2006, 
Ferrero et al.~2006). Spectral measurements of the the light of SN2006aj 
showed negligible additional extinction (Pian et al.~2006, Guenther et 
al.~2006, Wiersema et al.~2007) beyond the Galactic one, 
E (B -V) = 0.13, along the line of sight.

\noindent 
{\bf XRF060218/SN2006aj.} {\it Interpretation:} 
The spectral energy distribution measured with the Swift BAT and XRT was 
parametrized (e.g.~Campana et al.~2006b, Liang et al.~2006, Butler et 
al.~2007) as the sum of a black-body emission with a time-declining 
temperature from a sphere with time-growing radius, and a cut-off power-law 
with time-dependent amplitude and a constant cutoff energy. 
From this parametrization it was concluded that this event had a 
thermal black-body component in its X-ray spectrum, which cools and shifts 
into the {\it UVO} band as time elapses. This alleged black-body component
was interpreted as the result of a shock's break-out from the stellar 
envelope into the stellar wind of the progenitor star of the core-collapse 
SN2006aj (Campana et al.~2006b, Blustin~2007, Waxman, Meszaros \& 
Campana~2007). From this interpretation, a delay of $\leq 4$ ks between 
the SN and the GRB beginning was concluded. 

The early optical emission from XRF 060218 --the 
first $10^5$ s measured with the UVOT and
interpreted as black-body dominated-- required an intrinsic reddening of 
$\rm{E\,(B -V)}\, = \,0.20\pm0.03$ (assuming a Small Magellanic-Cloud effect) 
in addition to a Galactic reddening of $\rm{E\,(B -V)}\, = \,0.14$ (Campana et al.~2006b, 
Ghisellini et al.~2007) to be consistent with a black-body 
spectrum. Such a host extinction is inconsistent with the 
negligible extra-Galactic one measured from the spectrum 
of SN2006aj by e.g.,~Pian et al.~(2006), Guenther et 
al.~(2006), Wiersema et al.~(2007). With a negligible reddening in the host,
the ratio between the measured fluxes with the V and UVW2 filters 
of the Swift UVOT --de-reddened with the Galactic $\rm{E\,(B -V)}\, = \,0.14$ -- is 
different by nearly a factor 10 from the  
$F_\nu\!\propto\!\nu^2$ behaviour in the Rayleigh-Jeans domain. 
Moreover, the flux ratio between these two bands 
is time-dependent and increases by $\!\sim\!2.5$ between 2 ks and 
20 ks after burst, while it should be constant as long as the optical band 
stays in the Rayleigh-Jeans part of the black-body spectrum.
We conclude that the {\it UVO} emission from XRF 060218 
is not black-body-like. This
is independent of whether it is produced by the same
source which produced the alleged black-body 
component in the prompt X-ray and $\gamma$-ray emission or by another 
 source.

The light curves of XRF 060218/SN2006aj, measured with the 
Swift's UVOT filters, are particularly interesting. Not only they provide evidence 
that the XRF was produced in the explosion of SN2006aj, but, 
together with the BAT and XRT light curves,  they confirm 
the CB-model interpretation of the broad-band emission at all times. Prior 
to the dominance of the associated supernova's radiation, the {\it UVO} light 
curves show wide peaks whose peak-time shifts from $t_{peak}\!\approx\! 
30$ ks at $\lambda\!\sim\!188$ nm to $t_{peak}\!\approx\! 50$ ks at 
$\lambda\!\sim\! 544$ nm, and whose peak-energy flux decreases with 
energy, see Fig.~\ref{f9}b: the lower half of the upper figure. In the CB model 
these are the predicted properties of a single peak generated by a single 
CB as it Compton up-scatters glory's light. The prompt $\gamma$-rays  
and X-rays  of ordinary GRBs are dominated by ICS, while the 
optical emission is dominated by SR. However,  
in low-luminosity XRFs the optical emission is also dominated by ICS 
of glory light. The dominant radiation mechanisms at various times
can actually be identified, using the different spectral and 
temporal shapes of the ICS and SR emissions: while the early unabsorbed SR 
contribution has a spectral energy density $F_\nu \! \propto \! 
\nu^{-0.6}$, ICS has $F_\nu\!\propto\! e^{-E/E_p(t)}$ and  
satisfies the $E\,t^2$ law. 

In order to test whether the prompt X-ray 
peak around 1000 s and the UVOT peaks between 30 and 50 ks belong to 
the same ICS pulse, we have plotted in Fig.~\ref{f9}c the energy fluxes between 
5 ks  and 150 ks
measured with the UVOT  filters, de-reddened for Galactic extinction 
[$\rm {E \,(B -V)} \!=\! 0.14$, Campana 
et al.~(2006b)] and scaled by the $E\, t^2$ law, together with the unabsorbed 
energy flux in the 0.3 -10 keV band of the prompt X-ray pulse which was 
measured with the XRT (Campana et al.~2006b). Each de-reddened
energy flux in the UVOT filters at time $t$ was converted to 
energy flux density using  the UVOT energy
band widths, $  \Delta E \!=\! h\,\nu\, \Delta 
\lambda/\lambda^2$, with $\Delta \lambda$ = 75, 98, 88, 70, 51 and 76 nm 
the FWHM of the V, B, U, UVW1, UVM1 and UVW2  filters of the 
Swift UVOT, with central wavelengths, $\lambda$ = 
544 nm, 439 nm, 345 nm, 251 nm, 217 nm, and 188 nm respectively. 
These energy flux densities were multiplied by the XRT band width 
and plotted at time $(\nu/\nu_x)^{0.5}\,t$, where $h\,\nu_x=5.15$ keV
is the central energy of the 0.3 -10 keV band.  
As can be seen from Fig.~\ref{f9}d, the XRT and UVOT data near their peak 
times satisfy  the $E\, t^2$ (or $\sqrt{\nu}\,t$) law quite well. 
The very large differences between peak times and peak energy fluxes in the 
Swift $XUVO$ bands simply disappear in the scaled-time plot, and
the peaks' shapes coincide.

In Tables~\ref{t3},\ref{t4} and Figs.~\ref{f10}a,b we 
further test the $E\,t^2$ law for the  peak-energy flux (PEF) 
and peak-time in the different UVOT filters. Though these results are
flawless, there remain the small deviations from the $E\, t^2$ law in Fig.~\ref{f9}d,
which may be due to its approximate nature, our rough spectral integrations,
a non-negligible contribution from the SN at a relatively early time 
and/or a significant SR contribution to the UVOT light curves.   
There is a strong indication for the latter possibility:
the spectrum obtained from the de-reddened UVOT data at 
$t\!<\!5$ ks is consistent either with the $E\,t^2$ law 
(Fig.~\ref{f9}d) or with a SR spectrum below the frequency bend, $F_\nu\!\propto\! \nu^{-0.6}$. 
This is demonstrated in Fig.~\ref{f9}c, where we have plotted 
the UVOT de-reddened data of Campana et al.~(2006b) in the form
$\nu^{0.6}\, F_\nu(t)$, which, for $t\!\ll\!(a,t_{exp})$ and $\beta\!=\!0.6$ in Eq.~(\ref{esync}),
should be proportional to $t^{0.4}$. The line in the figure shows that it is.
A black-body shape, $F_\nu\!\propto \! \nu^2$, multiplied by $\nu^{0.6}$ would have, for instance, 
separated the  $V$ and $UVW2$ bands by a factor $\!\sim\!14$, entirely inconsistent 
with Fig.~\ref{f9}c.

From the above relatively model-independent analysis we have 
concluded that the {\it UVO} light curves observed prior to the
dominance of the associated SN, and the early-time X-ray data,
are consistent with ICS of glory light by a jet of CBs breaking out
from SN2006aj, while they are inconsistent with a black-body
radiation from a shock break-out from the stellar envelope
of the progenitor star. But, can the detailed
broad-band observations of this XRF be reproduced by the CB model
in greater detail?
 
Amati et al.~(2006) showed that XRF 060218 complies with the so-called 
`Amati correlation' (Amati 2002) for GRBs and XRFs and concluded that this 
implies that XRF 060218~{\it was not} a GRB viewed far off axis. In the 
CB model the conclusion of the same argument is the opposite one. 
The observed correlation 
between peak and isotropic energies of GRBs~{\it and} XRFs is a prediction 
(Dar \& De R\'ujula~2000, DD2004, Dado, Dar \& De R\'ujula~2007b) 
trivially following from the kinematics of ICS. The fact that XRF 060218 
complies with it corroborates that it~{\it was} a GRB viewed far off axis. 
In this model, the isotropic equivalent $\gamma$-ray energy emission of a 
typical CB is $\approx 0.8 \times 10^{44}\,[\delta_0]^3$ erg (DD2004). 
Thus, the reported $E_{\rm iso}\approx (6.2\pm 0.3)\times 10^{49}$ erg 
implies that the CB which generated the dominant peak of XRF 060218 had 
$\delta_0\!\sim\! 92$. It then follows from Eq.~(\ref{ICSEp}) that its 
measured $E_p\!=\!4.6$ keV implies (for the typical $k\,T(0)\!\sim\! 1$ 
eV) a Lorentz factor $\gamma_0\!\sim\! 103 $, $\gamma_0\, \theta\approx 1.1$ 
and a viewing angle $\theta\!\sim\! 1.08\times 10^{-2}$ rad, an order of magnitude
larger than the typical GRB value $\theta\!\sim\! 1$ mrad.

The best-sampled data set of XRF 060218  
is the XRT 0.3 -10 keV light curve (Campana et al.~2006b). Thus, we 
fit these data first,  with prompt ICS emission plus SR. 
We assume that the prompt ICS emission is dominated by two pulses:
an early one preceding the main pulse, as suggested by the hardness ratio and 
the BAT light curve. The SR contribution was calculated using 
Eqs.~(\ref{decel},\ref{Fnu}) for a constant-density ISM, with the previously-derived
$\!\gamma_0\, \theta\!=\!1.1$, the standard $\beta_X\!=\!1.1$, and a best-fit value for 
$t_b$.  The result is Fig.~\ref{f10}c; the corresponding Swift-XRT hardness ratio
is shown in Fig.~\ref{f10}d (DDD2008a), and the AG parameters are listed in Table~\ref{t1}. 
Next, we use the $E\,t^2$ law for ICS to predict 
the UVOT and BAT light curves. The results are Fig.~\ref{f10}e for
the de-reddened UVOT light curve in the UVW2 filter, and Fig.~\ref{f10}f
for the BAT light curve in the 15-150 keV band.

We conclude that the XRF 060218/SN2006aj pair is in full agreement with 
the predictions of the CB model. The rich structure of its {\it UVO} AG is as 
expected. Its X-ray light curve has the canonical GRB shape (stretched in 
time) and consistent with the observed $E_p$ and $E_{iso}$.
All of these results are explicitly dependent on the fact that XRFs 
are GRBs produced by CBs with smaller Doppler factors, because 
they are viewed at larger angles or have  smaller Lorentz factors. 
The data on this XRF are inconsistent with a black-body component generated 
by a shock break-out through the stellar envelope, or by any other mechanism. 
The start time of the X-ray emission does not constrain the exact time of the 
core's collapse before the launch of the CBs, nor the 
possible ejection of other CBs farther off axis, 
prior\footnote{Intriguingly, Swift detected $\gamma$ rays from the same 
direction over a month earlier on January 17, 2006 (Barbier, et al.~2006).}
to the trigger-time of XRF 060218.

\section{Conclusions and outlook}

\label{outlook}

The rich data on GRBs gathered after the launch of Swift, as interpreted
in the CB model and as we have discussed
here and in recent papers (e.g.~Dado et al.~2006, 2007, 2008a, 2008b) 
has taught us several things:

\begin{itemize}
\item{} 
Two radiation mechanisms, inverse Compton scattering and synchrotron 
radiation, suffice within the CB model to provide a very simple and 
accurate description of long-duration GRBs and XRFs and their afterglows. 
Simple as they are, these two mechanisms 
and the bursts' environments generate 
the rich structure and variety of the light curves at all frequencies and times.

\item{} 
The historical distinction between prompt and afterglow phases is 
replaced by a physical distinction: the relative dominance 
of the Compton or synchrotron mechanisms at different, frequency-dependent 
times.

\item{}
The relatively narrow pulses of the $\gamma$-ray signal, the
somewhat wider prompt flares of X-rays, and the much wider humps
sometimes seen at $UVOIR$ frequencies in XRFs, have a common origin. They 
are generated by inverse Compton scattering.

\item{}
The synchrotron radiation component dominates 
the prompt  optical emission in ordinary GRBs, the broad-band afterglow
in GRBs and XRFs and the late-time flares of both types of events.

\item{}

The early-time XRT and UVOT data on XRF 060218 are inconsistent 
with a black-body emission from a shock break-out through the stellar 
envelope. Instead, they support the CB-model interpretation of ICS of 
glory light by an early jet of CBs from what is later seen as SN2006aj. 
The start time of the X-ray emission does not constrain the exact time 
of the core's collapse before the launch of the CBs, nor the 
possible ejection of other CBs farther off axis. 

\item{} 
Despite its simplicity and approximate nature, the CB model 
continues to provide an extremely successful description of long GRBs and 
XRFs. Its testable predictions, so far, are in complete agreement with the 
main established properties of their prompt emission and of the
afterglow at all times and frequencies.

\end{itemize}

We re-emphasize that the 
results presented in this paper are  based on direct applications
of our previously published explicit predictions.
Our master formulae, Eq.~(\ref{ICSlc}) for ICS and 
Eqs.~(\ref{decel}, \ref{Fnu}, \ref{SRP})
for the synchrotron component describe all the data very
well. But, could they just be very lucky guesses? The general 
properties of the data are predictions. But,
when fitting cases with many flares, are we not `over-parametrizing' 
the results? Finally, the $E\,t^2$ law plays an important role.
Could it also be trivially derived in a different theory?

When their collimated radiation points to the observer, GRBs are the 
brightest sources in the sky. In the context of the CB model and of the 
simplicity of its underlying physics, GRBs are not persistent mysteries, nor
`the biggest of explosions after the Big Bang', nor a constant source of 
surprises, exceptions and new requirements. Instead, they are 
well-understood and can be used as cosmological tools, to study the 
history of the intergalactic medium and of star formation up to large 
redshifts, and to locate SN explosions at a very early stage. As 
interpreted in the CB model, GRBs are not `standard candles', their use in 
`Hubble-like' analises would require further elaboration. The GRB conundra 
have been reduced to just one: `how does a SN manage to sprout mighty 
jets?' The increasingly well-studied ejecta of quasars and microquasars, 
no doubt also fired in catastrophic accretion episodes on compact central 
objects, provides observational hints with which, so far, theory and 
simulations cannot compete.

The CB model underlies a unified theory of high energy astrophysical 
phenomena. The information gathered in our study of GRBs can be used to 
understand, also in very simple terms, other phenomena. The most notable 
is (non-solar) cosmic rays. We allege (Dar et al.~1992, Dar \& Plaga~1999)  
that they are simply the charged ISM particles scattered by CBs, in 
complete analogy with the ICS of light by the same CBs. This results in a 
successful description of the spectra of all primary cosmic-ray nuclei and 
electrons at all observed energies (Dar and De R\'ujula~2006a). The CB 
model also predicts very simply the spectrum of the gamma background 
radiation and explains its directional properties (Dar \& De 
R\'ujula~2001a, 2006b). Other phenomena understood in simple terms include 
the properties of cooling core clusters (Colafrancesco, Dar \& De 
R\'ujula~2003) and of intergalactic magnetic fields (Dar \& De 
R\'ujula~2005). The model may even have a say in `astrobiology' (Dar,
Laor \& Shaviv~1998, Dar \& De R\'ujula~2001b).

{\bf Acknowledgment:} 
A.D. would like to thank the 
Theory Division of CERN for its hospitality during this work.
We would also like to thank S.~Campana, G.~Cusumano, 
P.~Ferrero, D.~Grupe, D.~A.~Kann, K.~L.~Page and S.~Vaughan, for making 
available to us the tabulated data of their published X-ray 
and optical light curves of Swift GRBs and an anonymous referee for
an exceptionally constructive and useful report.

\newpage
\begin{deluxetable}{lllc}
\vskip -2.cm
\tablewidth{0pt}
\tablecaption{CB-model afterglow parameters.}
\tablehead{
\colhead{GRB/XRF} & \colhead{$t_0[{\rm s}]$} & \colhead{$\theta\,\gamma_0 
$}
& \colhead{$p$}
}
\startdata
060729 & (606) & (2.52) &  2.20 \\
061121 & 248 & 1.42 &  2.20 \\
050315 & 12362 & 0.965 & 2.20 \\
061007 &  40   & $\ll 1$ & 2.20\\     
061126  & 142  & 1.08  & 1.84  \\
080319B & 72 & ($\ll 1$) &  2.16 \\
060211A & 64596& 0.54 & 2.11 \\  
061110A & 29402& 0.81 & 2.04 \\ 
080307  & 28893 & 1.00 & 2.13 \\
051021B & 13092 & 0.82 & 2.24\\
080303  & 15196 & 0.79 & 2.15 \\
070220  & 1314  & 0.64 & 2.16 \\
060526  & 1840 & 0.93 & 2.20 \\
060206  & 2570 & 1.035 & 2.20 \\
050820A & 2692 & 1.128 & 2.22 \\
060418  & $<60$ & 1.73 & 2.20 \\
071010A  & 857   & 1.21 & 1.92 \\
050318  & 273   & 1.61 & 2.19  \\
050326  & 379   & 1.28 & 2.16  \\
051008  & 1233  & 1.17 & 2.20  \\
050814  & 7737  & 1.14 & 2.18  \\
061019  & 194   & 2.22 & 2.20  \\    
070306  & 1437 &  1.91 & 2.20  \\
060813  &  273  &  1.60 & 2.20 \\
070521  &  551  &  1.33 & 2.23 \\
080207  & 95    &  0.98 & 1.87 \\  
060807  & 9867  &  1.02 & 2.21 \\
070419B & 1146  &  0.99 & 2.20 \\
070420  &  60   &  2.00 & 2.22 \\
050319  & 73    &  0.92 & 2.20 \\
050319  & 999   &  2.05 & 2.22 \\  
060605  & ($\!<1000$)& (1.00)& 2.20 \\
060607A & (54)& (1.07)& 2.20 \\
060218 & 267 & 1.10 & 1.94 \\
\enddata
\label{t1}
\end{deluxetable}

\begin{deluxetable}{llcccc}
\tablewidth{0pt}
\tablecaption{Time parameters in Eq.~(12) for the two last  
prompt X-ray flares.}
\tablehead{
\colhead{GRB/XRF} & Band 
& \colhead{$t_1$ [s]  } & \colhead{$\Delta t_1$ [s]}
&\colhead{$t_2$ [s]} &
\colhead{$\Delta t_2$ [s]} }
\startdata
060729  &  X & 122    &  6.2   &  153    & 19.1 s     \\
061121  &  X &  52    &  12.4   &  97   &  18.8      \\
050315  &   X & -5    &  6.9   & 16.4    &  5.4      \\
061007  &   X & 23    &  5.5  &          &             \\
061126  &   X & 4.4   &  7.8   &          &             \\
080319  &   X &  37   &  5.0   &          &             \\
060211A &   X &  79   &  30  &          &              \\
061110A &   X &  35   &  54   &          &             \\
080307  &   X &  0   &  373   &          &             \\
051021B &   X &  0   &   67   &          &             \\
080303  &   X &  0   &  114   &          &             \\
070220  &   X &  0   &   75    &          &             \\
060526  &  X & 233    & 16.4   &  272   & 31.6       \\
060206  &  X & 581    & 43.2   & 4187   &  549       \\
050820A &  X & 205    & 29   & 2173    &  2225      \\ 
060418  &  X & 60    & 12   & 118       &  9.8       \\
071010A  &  X & 18990    & 30968   &      &             \\
050814  &  X &           &           &  0   & 2074        \\
070306  &  X & 154   &  20     & 364    &  50        \\
060813  &  X &  37   &  58     &  0     &  246       \\
070521  &  X &  0    &  222     &      &             \\
060807  &  X &  0       & 26     &      &  4635     \\
070419B &  X &  106     & 41      & 134  & 87       \\
070420  &  X &  0       & 54      &      &            \\
050319  &  X &  0       & 54     & 2003   &         \\
060605  &  X &   0       & 83         &  67    &  154   \\
060607A &  X &           &          &      &         \\
060218  &  X  &  0    &  950       &     &        \\   
\enddata
\label{t2}
\end{deluxetable}

\newpage
\begin{deluxetable}{llccccc}
\tablewidth{0pt}
\tablecaption{Peak energy flux (PEF) and peak flux 
density (PFD) of XRF 060218 in the 
Swift UVOT filters, corrected for Galactic reddening E (B -V) = 0.14;
and the PFD predicted, using the $E\,t^2$ law, from   
the XRT unabsorbed PEF in the 0.3 -10 keV 
band.} 
\tablehead{
\colhead{Filter} & \colhead{$\lambda$}      & \colhead{E(center)} &
\colhead{FWHM} & \colhead{PEF}  & \colhead{PFD } & 
\colhead{Predicted PFD}\\ 
\colhead{} & \colhead{[nm]} &\colhead{[eV] } &\colhead{}
&\colhead{${\rm [erg\, cm^{-2}\, s^{-1}]}$} & \colhead{${\rm [\mu\, 
Jansky]}$} & \colhead{${\rm [\mu\, Jansky]}$}  }
\startdata
UVW2  & 188 & 6.60 & 76~nm  & $(2.29\pm 0.23)\times 10^{-12}$ &
$355\pm 36$ & $ 374\pm 135$  \\  
UVM1  & 217 & 5.71 & 51~nm  & $(1.30 \pm 0.10)\times 10^{-12}$ & 
$ 399 \pm 31 $& $ 374\pm 135$  \\  
UVW1 & 251 & 4.94 & 70~nm  & $(1.17 \pm 0.12)\times 10^{-12}$ & 
$ 352\pm 37$& $ 374\pm 135$  \\  
$U$     & 345 & 3.55 & 88~nm  & $(8.89\pm 0.85)\times 10^{-13}$ & 
$406 \pm 44$ & $ 374\pm 135$ \\  
$B$     & 439 & 2.83 & 98~nm  & $(5.99 \pm 0.56)\times 10^{-13}$ & 
$393 \pm 37$ & $ 374\pm 135$  \\  
$V$     & 544 & 2.28 & 75~nm   & $(2.59\pm 0.10) \times 10^{-13}$ & 
$ 340 \pm 34 $ & $ 374\pm 135 $ \\  
\enddata\\
\label{t3}
\end{deluxetable}

\begin{deluxetable}{llcc}
\tablewidth{0pt}
\tablecaption{Peak times of the energy flux of XRF 060218 in the 
Swift XRT and UVOT filters and their expected values from the $E\, t^2$
law.}

\tablehead{
\colhead{Band}   & \colhead{E(eff) [eV]} &
\colhead{Observed $t_{peak}$ [s]} & \colhead{CB Model $t_{peak}$ [s]}
}
\startdata
X     & 5150  & $985 \pm 50$  & $   985\pm 50$     (input)   \\
X     & 3000  & $1,310 \pm 90$  & $ 1,290 \pm 65$            \\
X     & 600   & $2,790 \pm 2,550$  & $ 2,770\pm 150 $         \\
UVW2  & 6.60  & $25,800 \pm 5,000 $  & $ 27,500 \pm 1,400$ \\
UVM1  & 5.71  & $36,208 \pm 8,000 $  & $ 29,600 \pm 1,800$ \\
UVW1  & 4.94  & $41,984 \pm 10,000 $  & $ 32,000 \pm 1,600$ \\
$U$     & 3.59 & $42,864 \pm 10,000 $  & $ 37,500 \pm 1,900$ \\
$B$     & 2.82 & $39,600 \pm 15,000 $  & $ 42,000 \pm 2,100$ \\
$V$     & 2.28 & $47,776 \pm 10,000 $  & $ 47,000 \pm 2,400$ \\

\enddata

\label{t4}
\end{deluxetable}

\newpage
\begin{figure}[]
\centering
\vspace{-1cm}
\vbox{
\hbox{
\epsfig{file=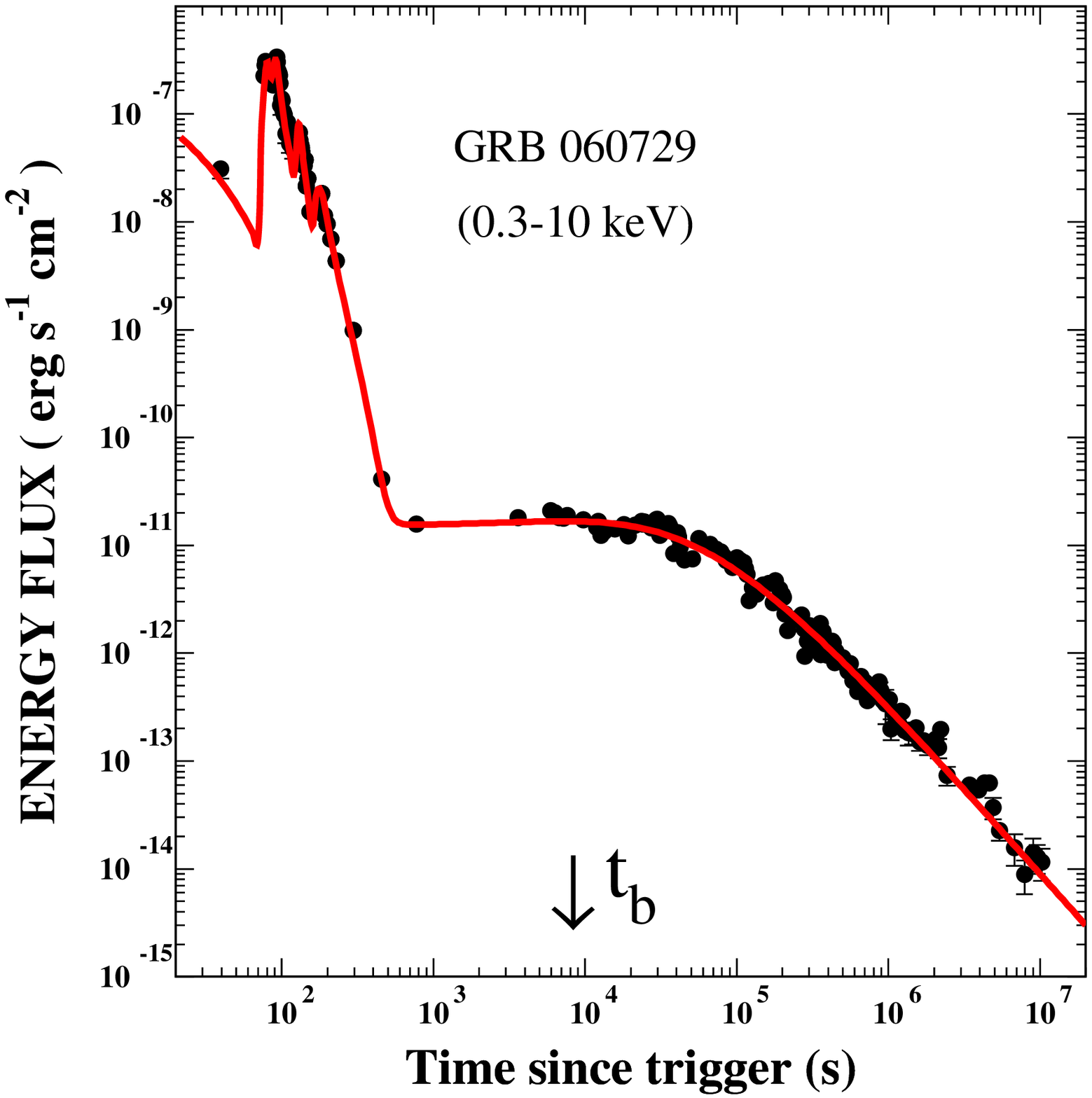,width=8.0cm,height=6.0cm}
\epsfig{file=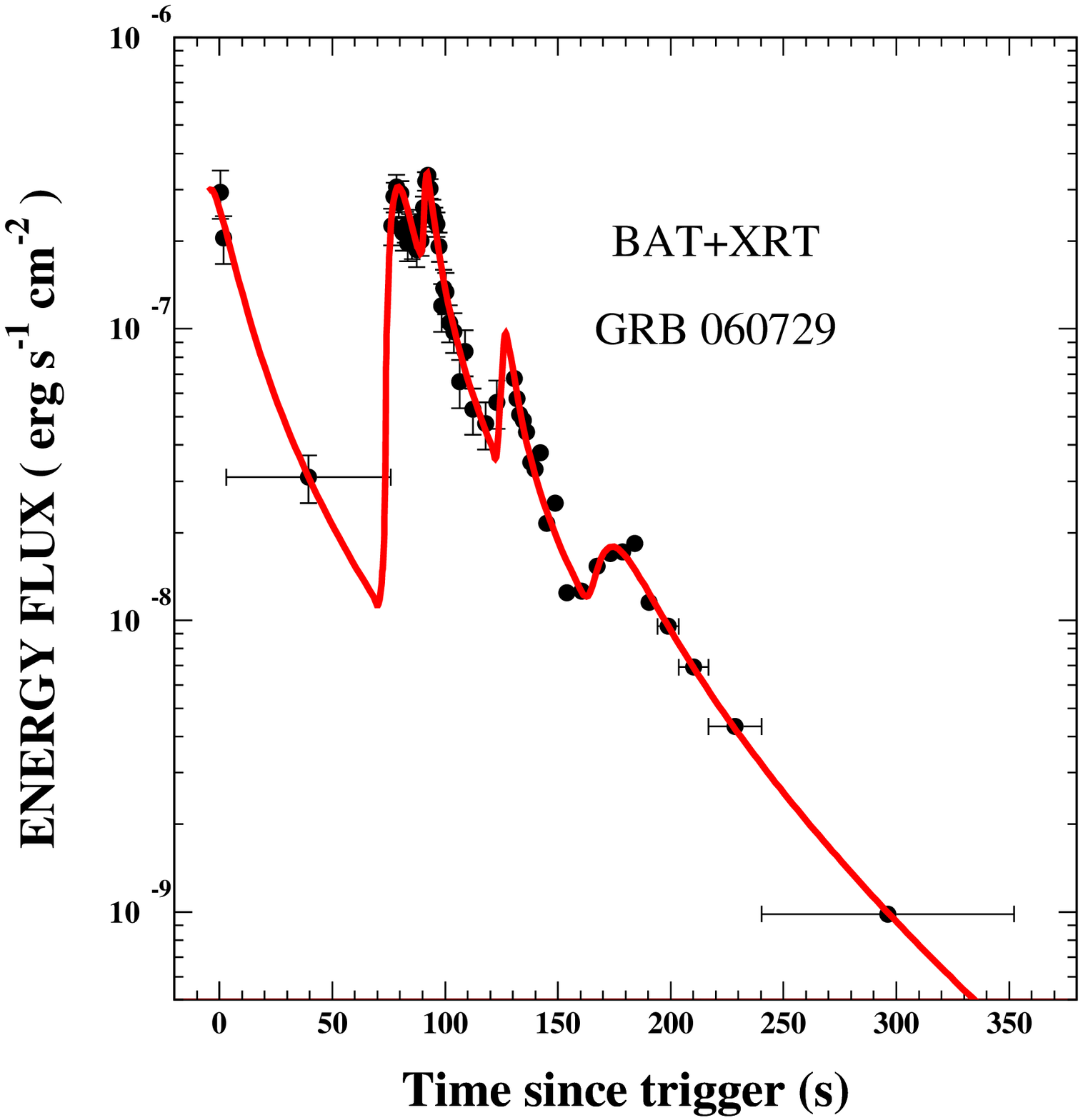,width=8.0cm,height=6.0cm}
}}
\vbox{
\hbox{
\epsfig{file=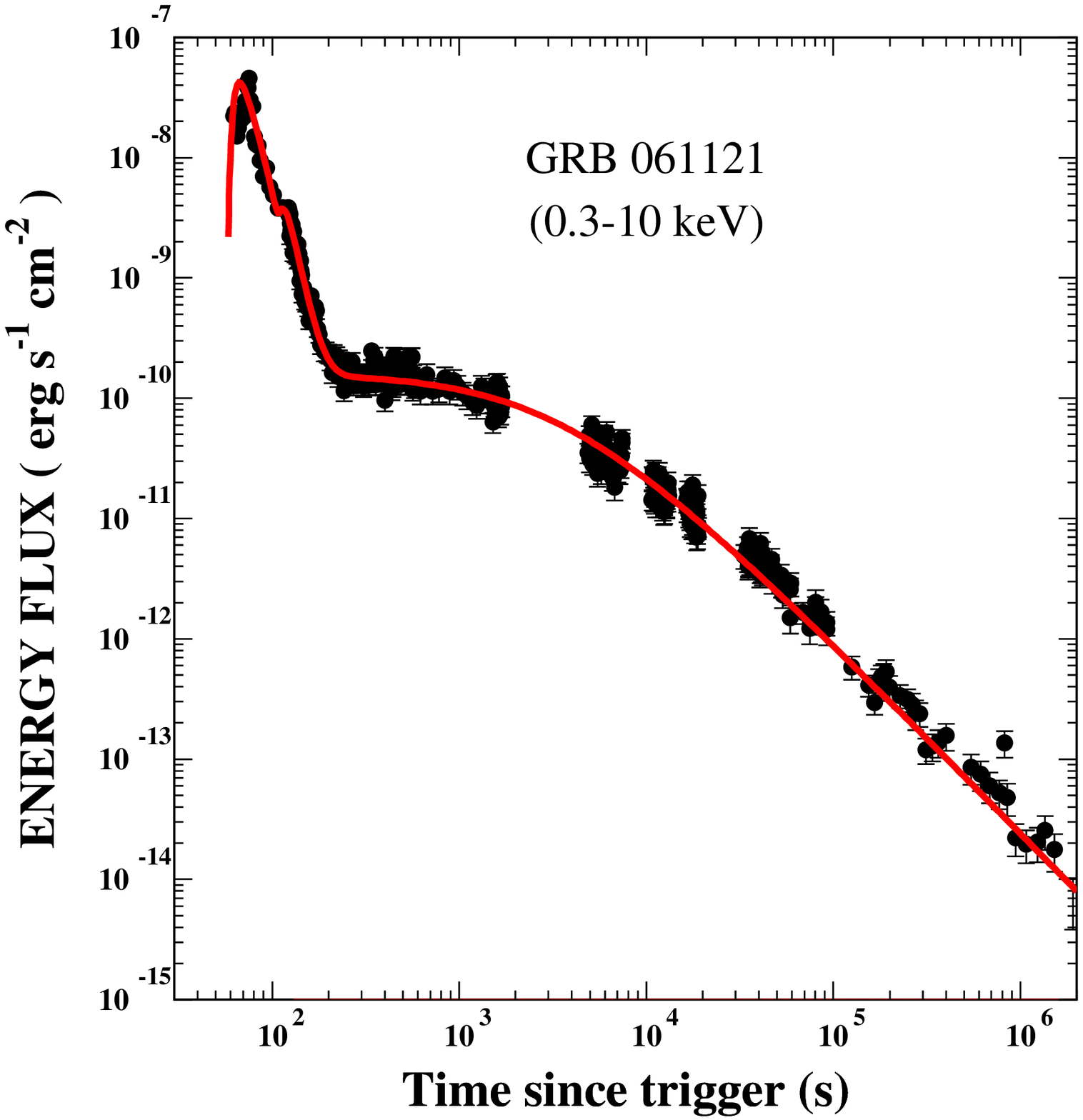,width=8.0cm,height=6.0cm}
\epsfig{file=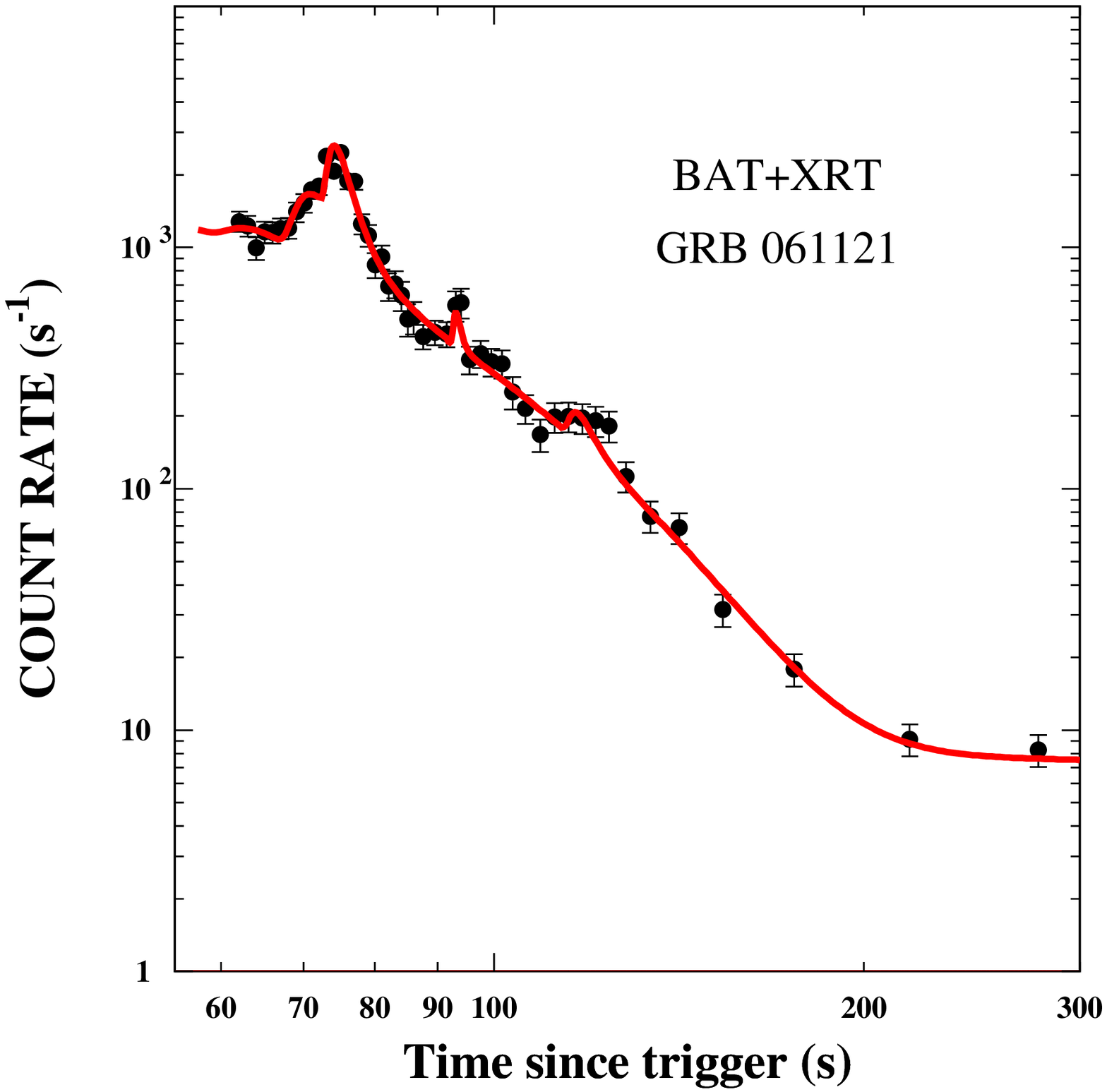,width=8.0cm,height=6.0cm}
}}
\vbox{
\hbox{
\epsfig{file=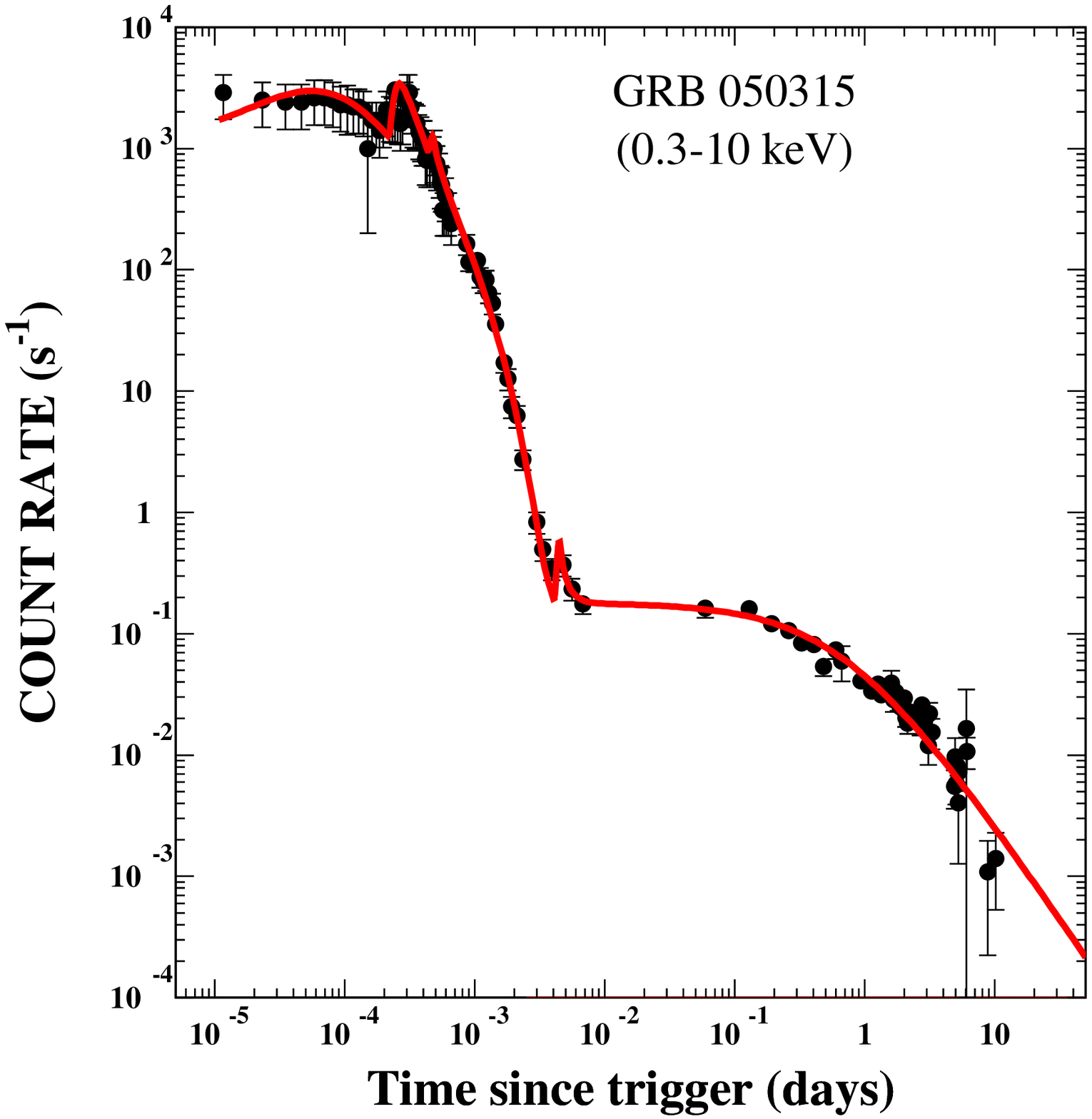,width=8.0cm,height=6cm }
\epsfig{file=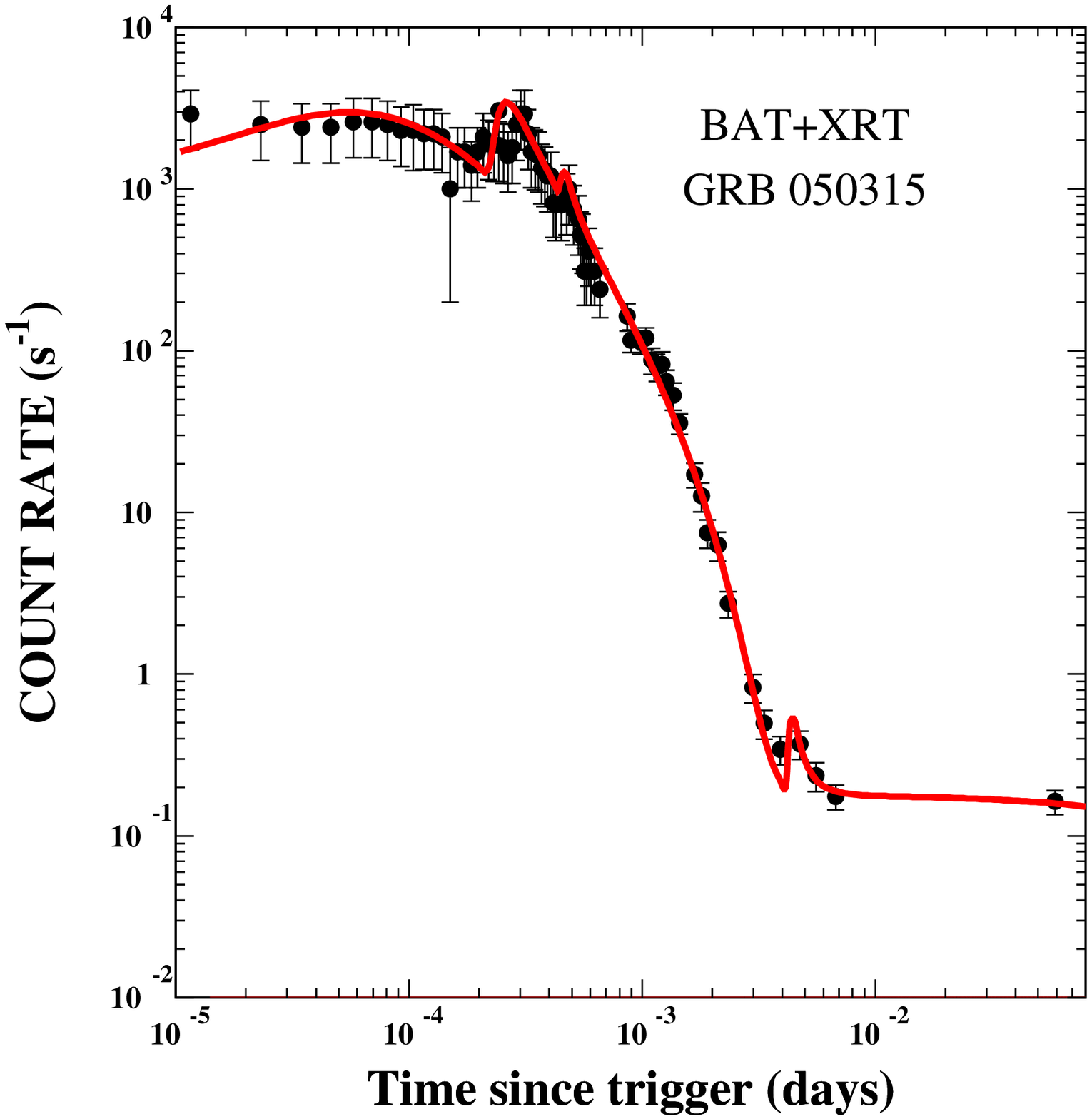,width=8.0cm,height=6cm}
}}
\caption{Comparison between Swift observations of canonical GRB  X-ray 
light curves  and their CB-model description for:
{\bf Top left (a):} GRB 060729.
{\bf Top right (b):}  GRB 060729 at early time. 
{\bf Middle left (c):} GRB 061121.
{\bf Middle right (d):}  GRB 061121 at early time.
{\bf Bottom left (e):}  GRB 050319.
{\bf Bottom right (f):}  GRB 050319 at early time.}
\label{f1}
\end{figure}

\newpage
\begin{figure}[]
\centering
\vspace{-1cm}
\vbox{
\hbox{
\epsfig{file=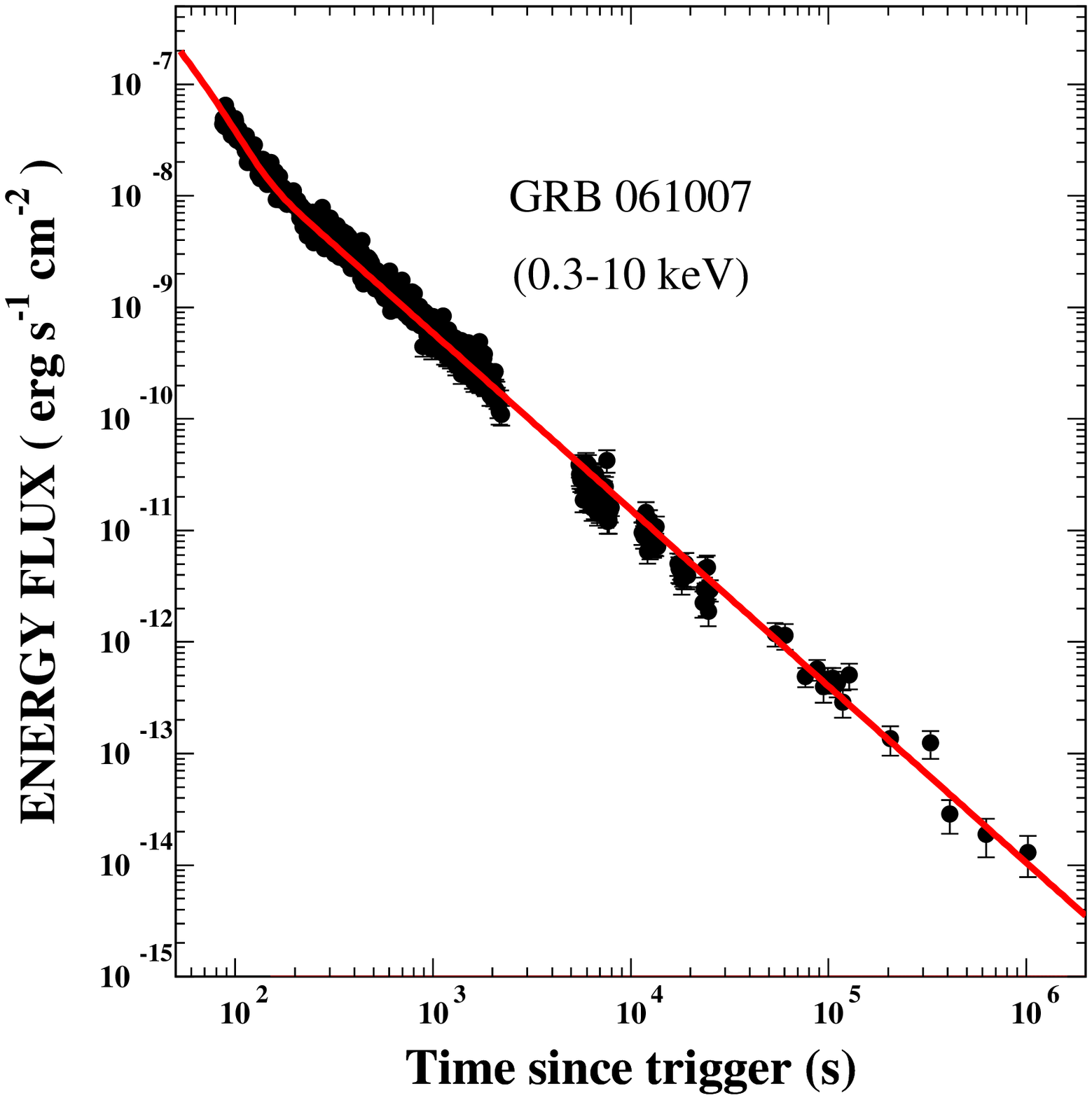,width=8.0cm,height=6.0cm}
\epsfig{file=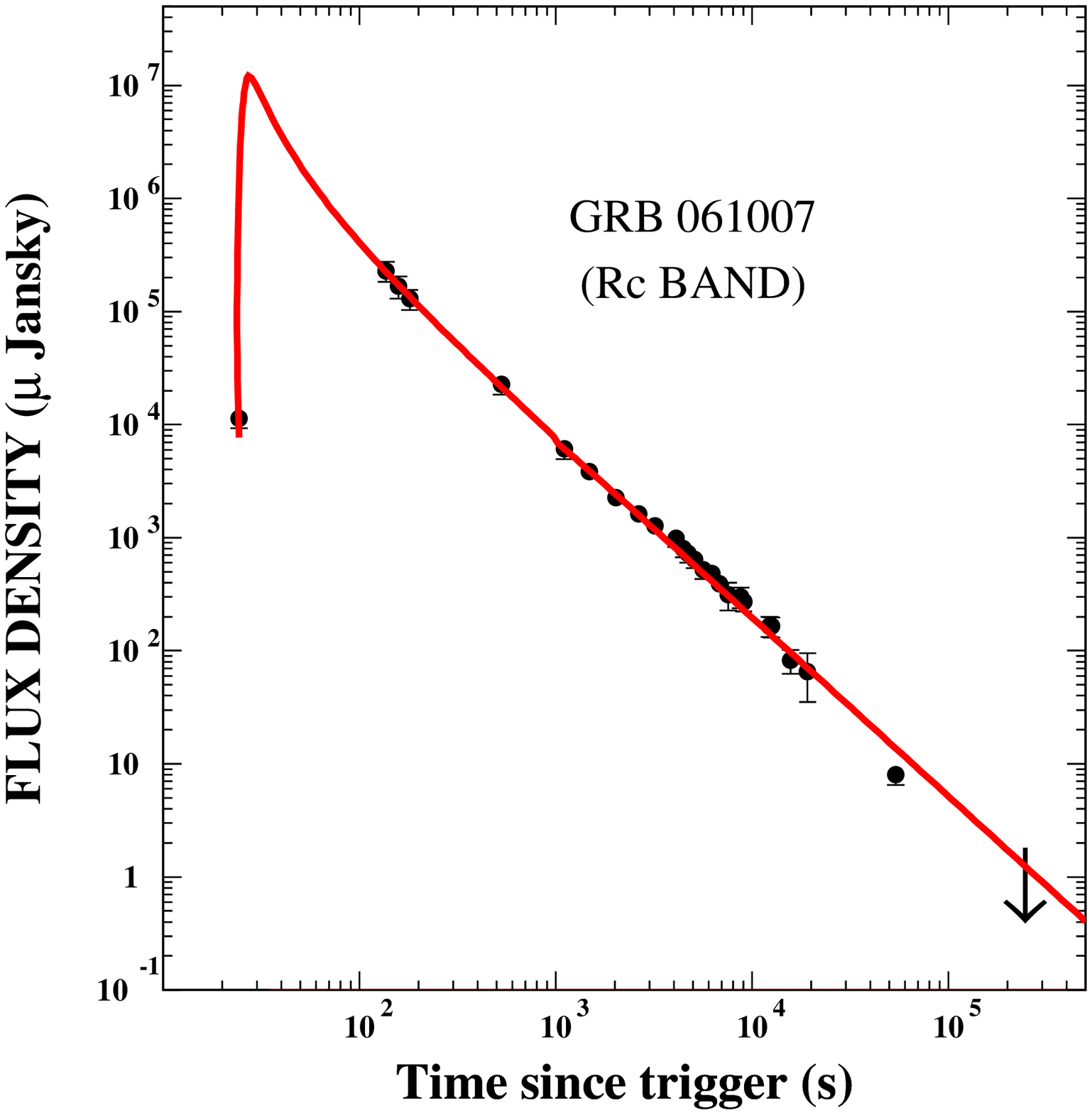,width=8.0cm,height=6.0cm}
}}
\vbox{
\hbox{
\epsfig{file=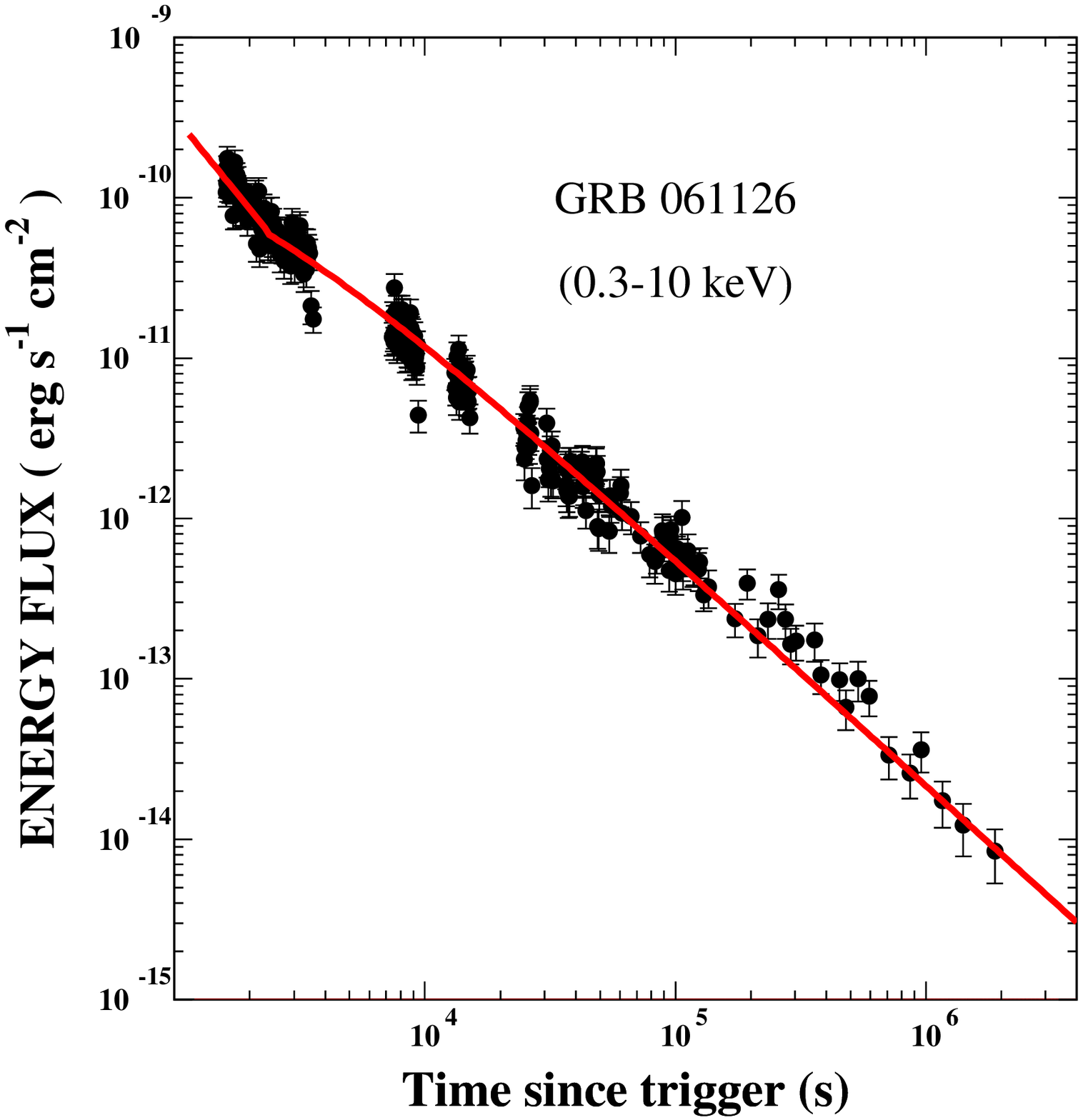,width=8.0cm,height=6.0cm}
\epsfig{file=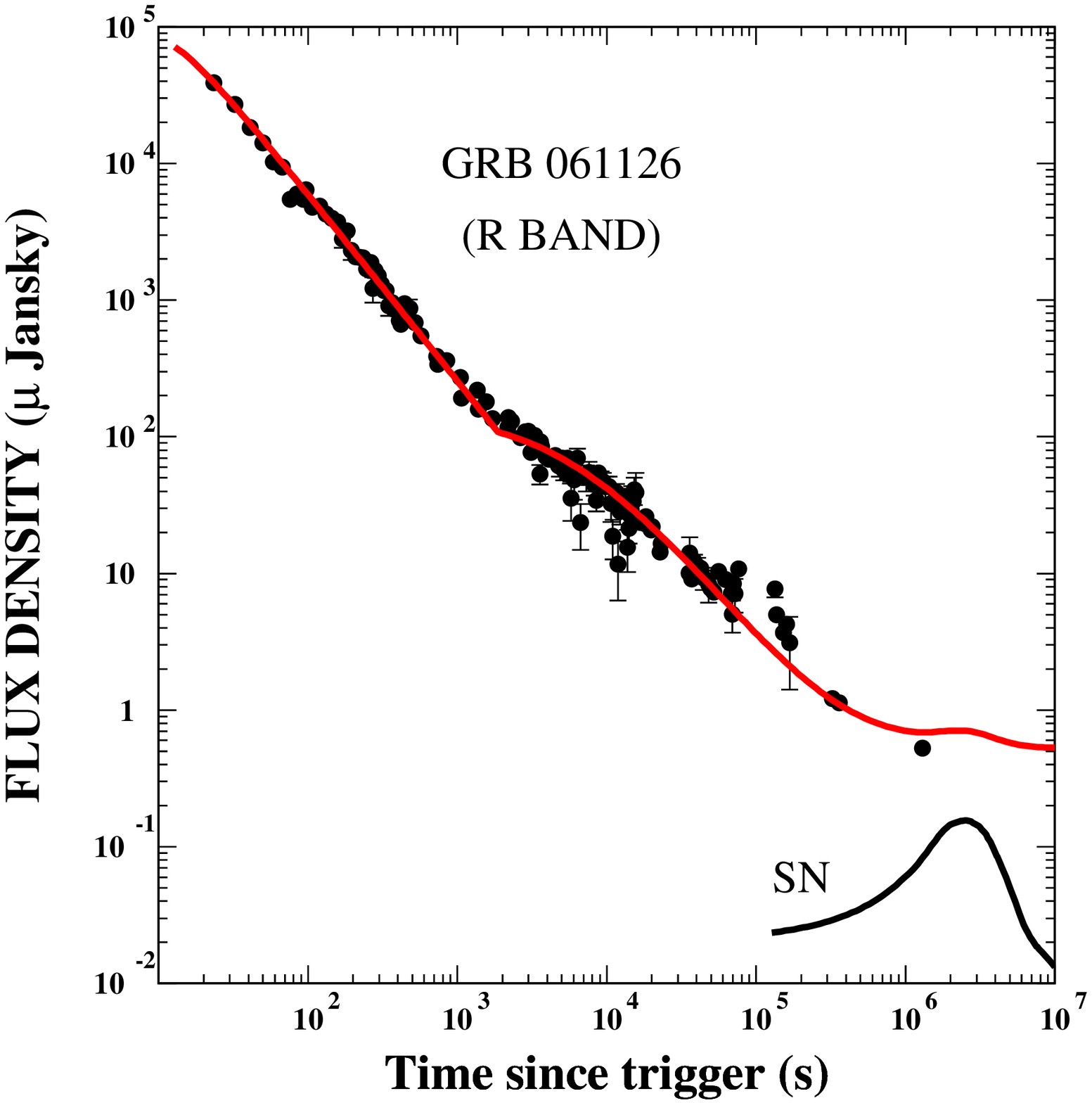,width=8.0cm,height=6.0cm}
}}
\vbox{
\hbox{
\epsfig{file=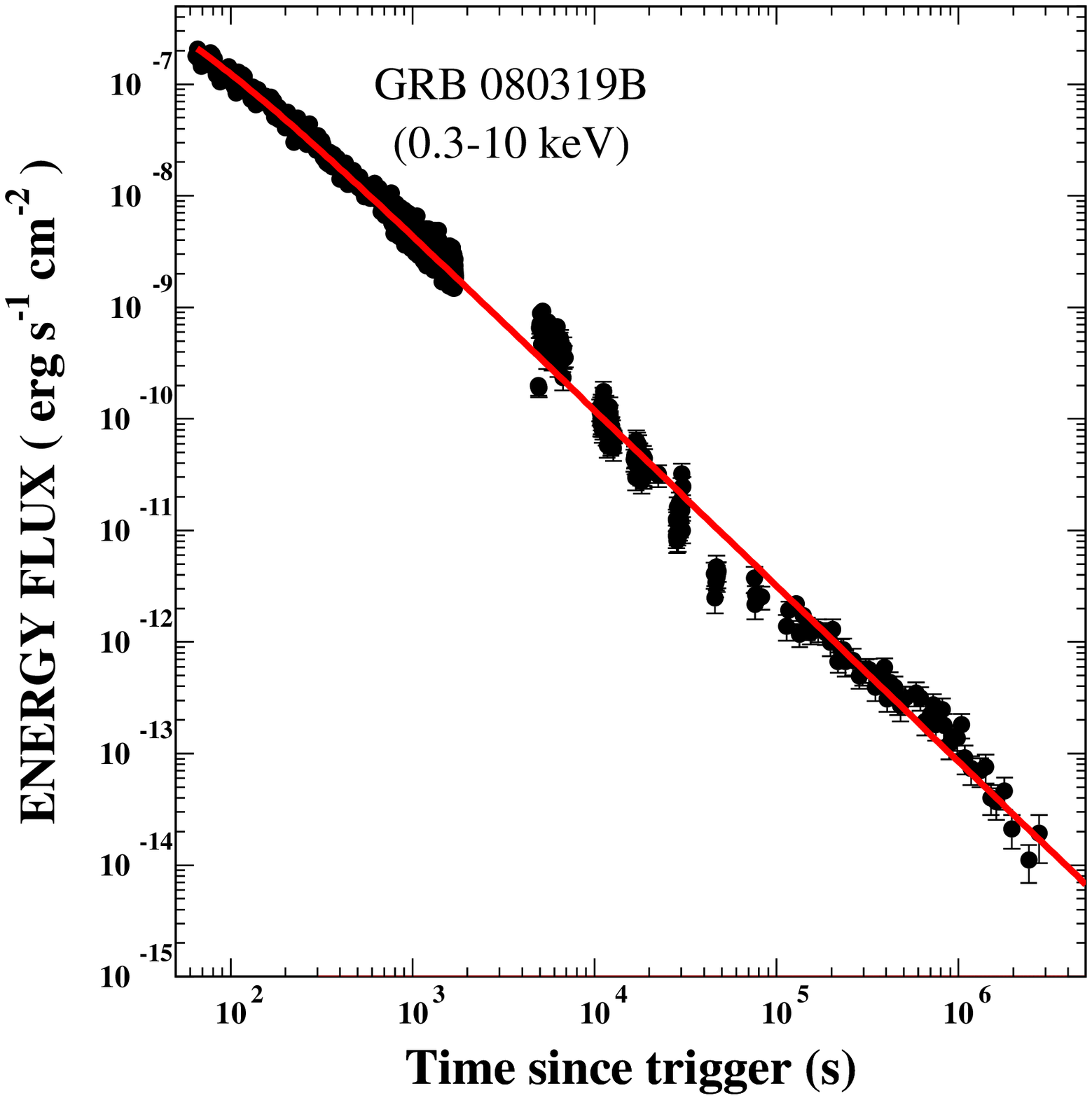,width=8.0cm,height=6cm }
\epsfig{file=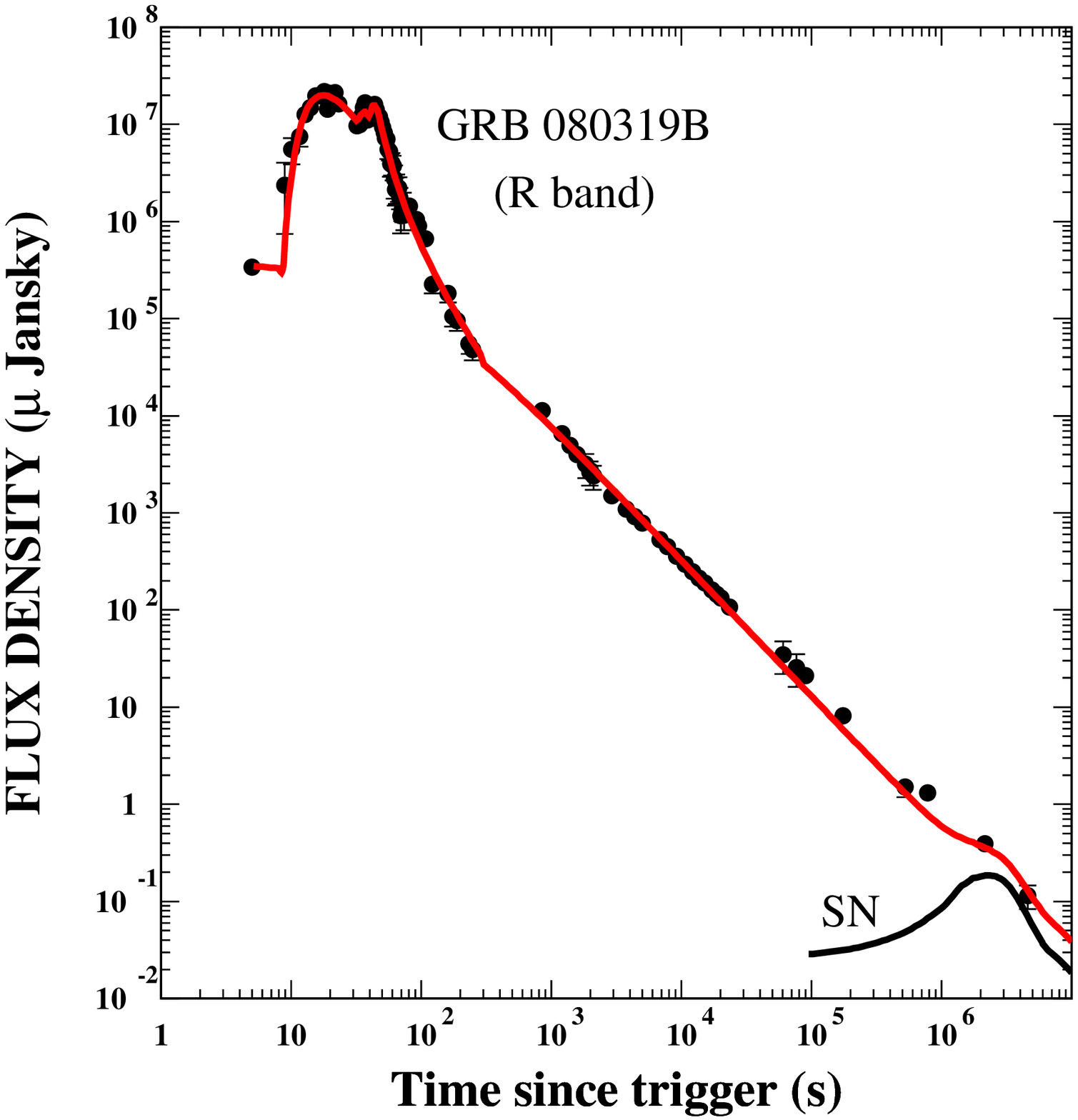,width=8.0cm,height=6cm}
}}
\caption{Comparison between broad-band observations of GRBs with
single-power-law decaying AGs and their CB-model description, for:
{\bf Top left (a):} The X-ray light curve of GRB 061007.
{\bf Top right (b):} The $R$-band light curve of GRB 061007.
{\bf Middle left (c):} The X-ray light curve of GRB 061126.
{\bf Middle right (d):}  The $R$-band light curve of GRB 061126.
{\bf Bottom left (e):} The X-ray light curve of GRB 080319B.
{\bf Bottom right (f):} The $R$-band light curve of GRB 080319B.
Some SN1998bw-like SN contributions are shown.}
\label{f2}
\end{figure}

\begin{figure}[]
\centering
\vspace{-2cm}
\vbox{
\hbox{
\epsfig{file=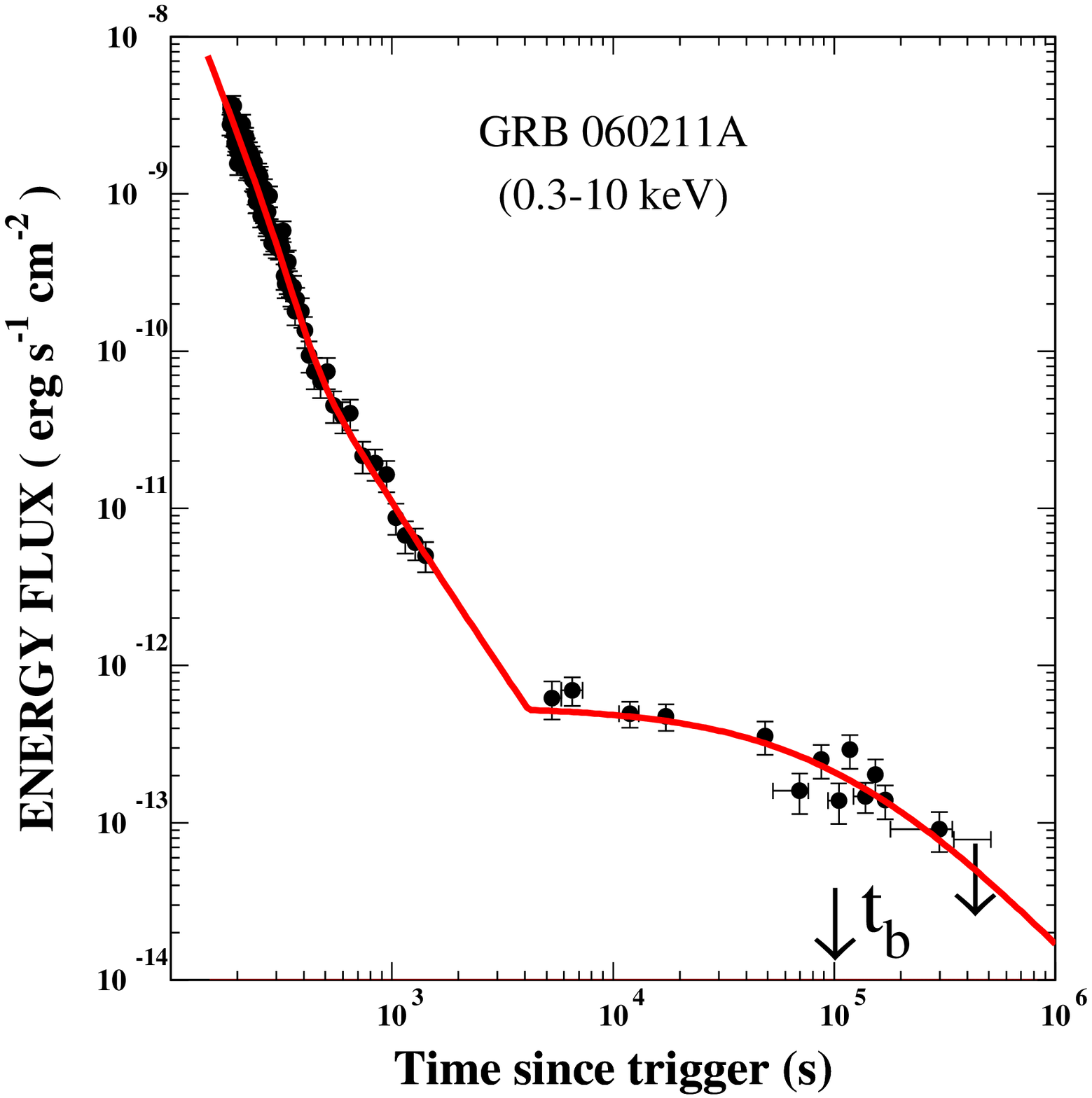,width=8.cm,height=6.cm}
\epsfig{file=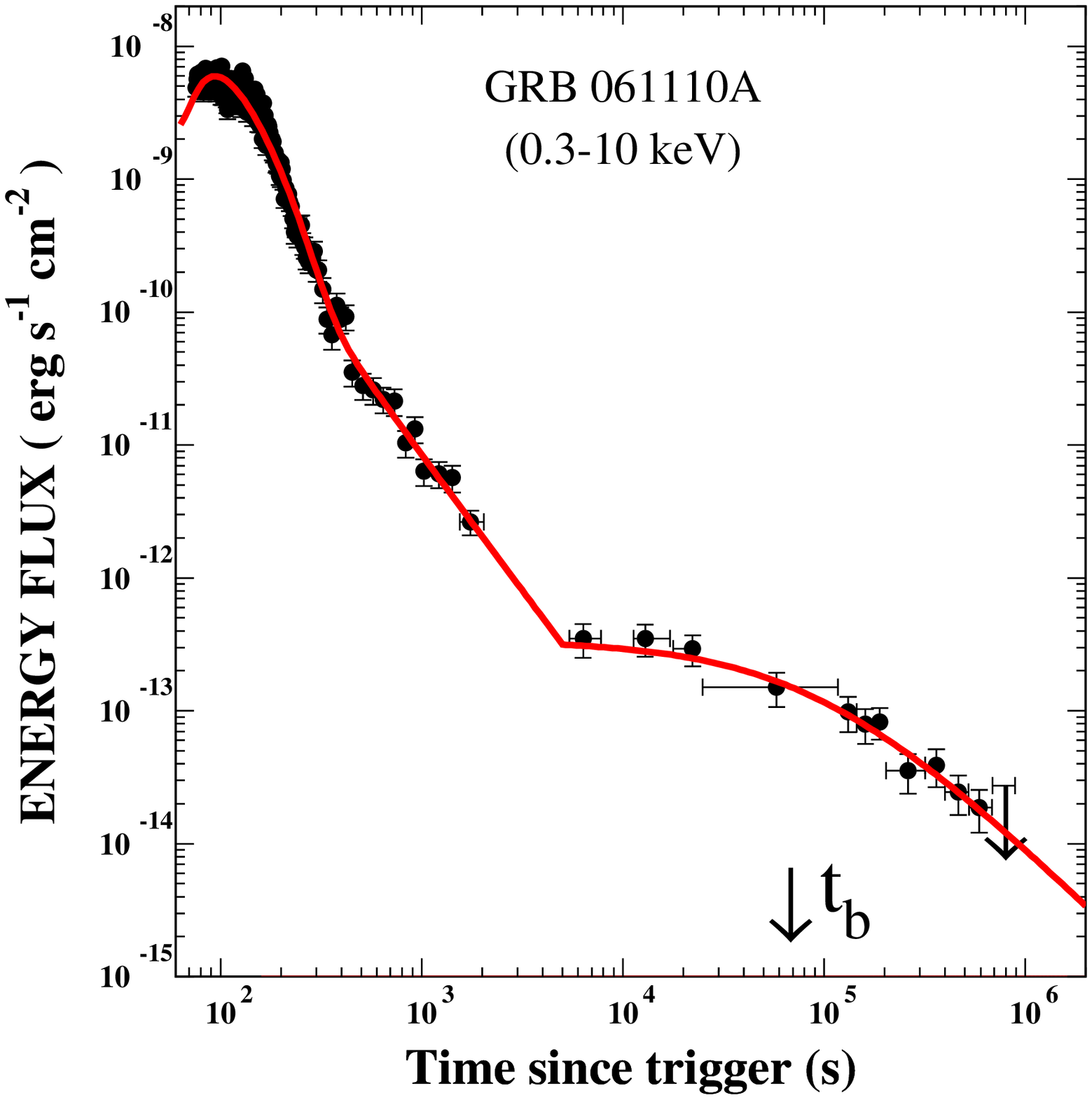,width=8.cm,height=6.cm}
}}
\vbox{
\hbox{
\epsfig{file=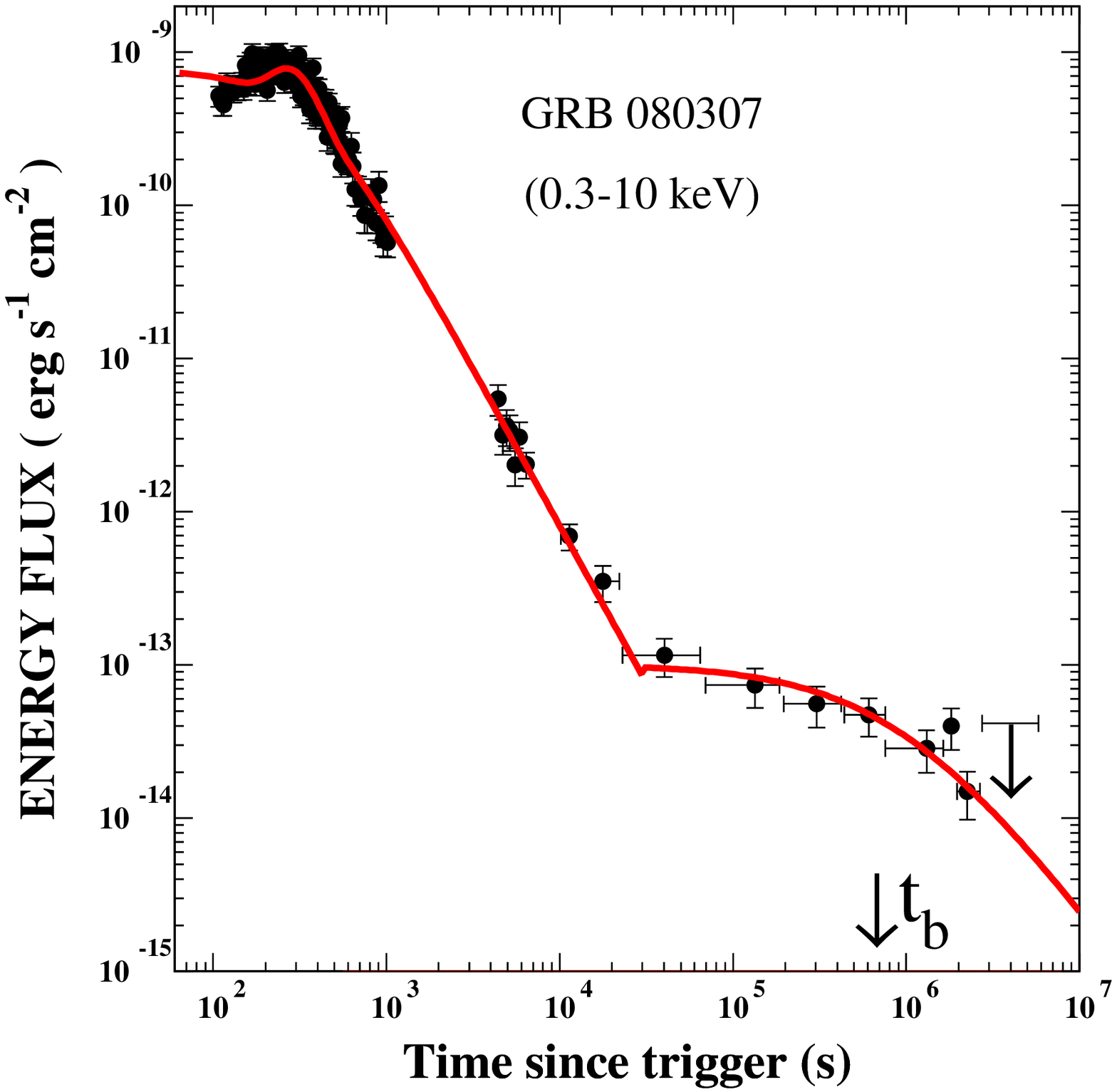,width=8.cm,height=6.cm}
\epsfig{file=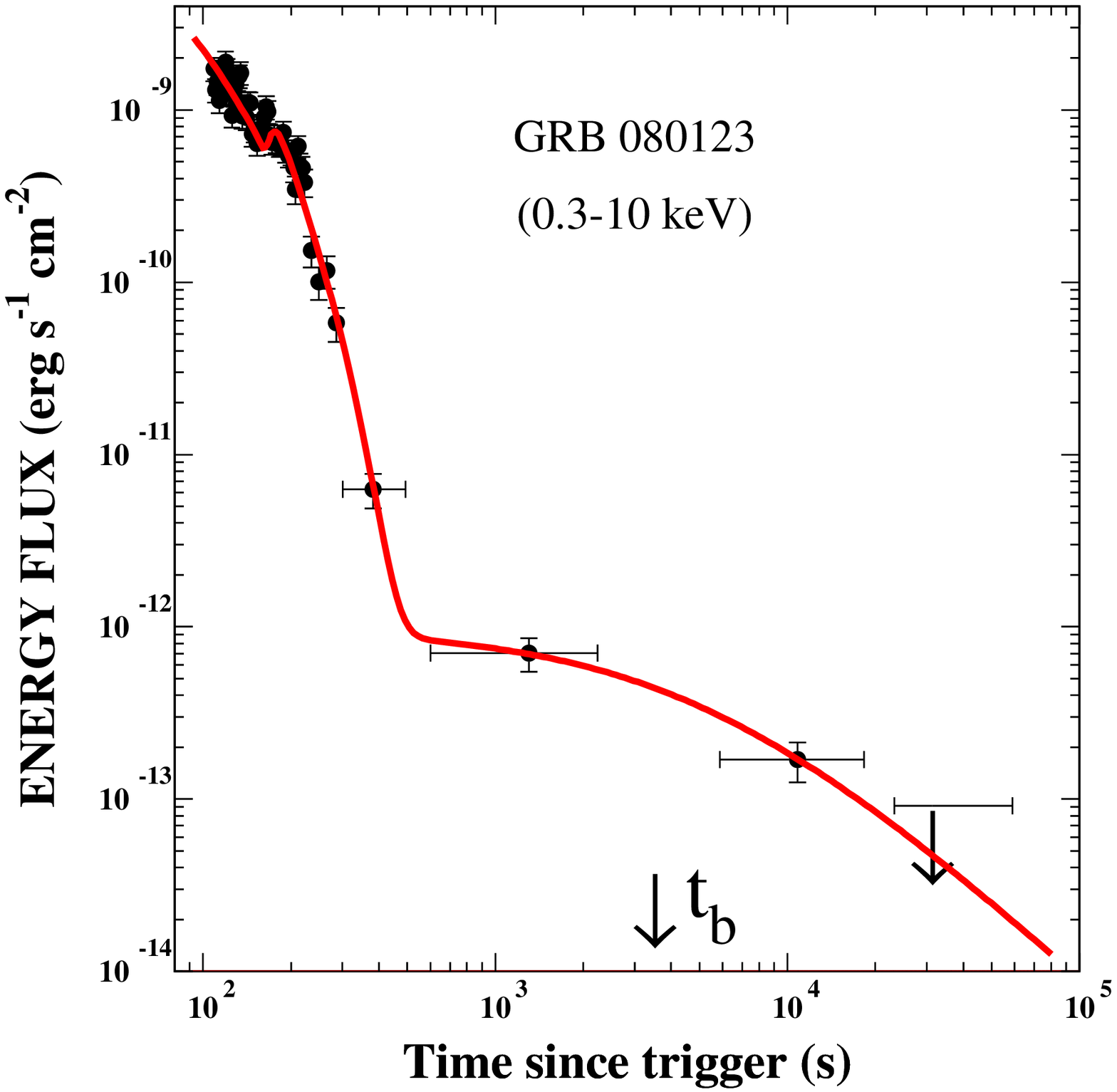,width=8.cm,height=6.cm}
}}
\vbox{
\hbox{
\epsfig{file=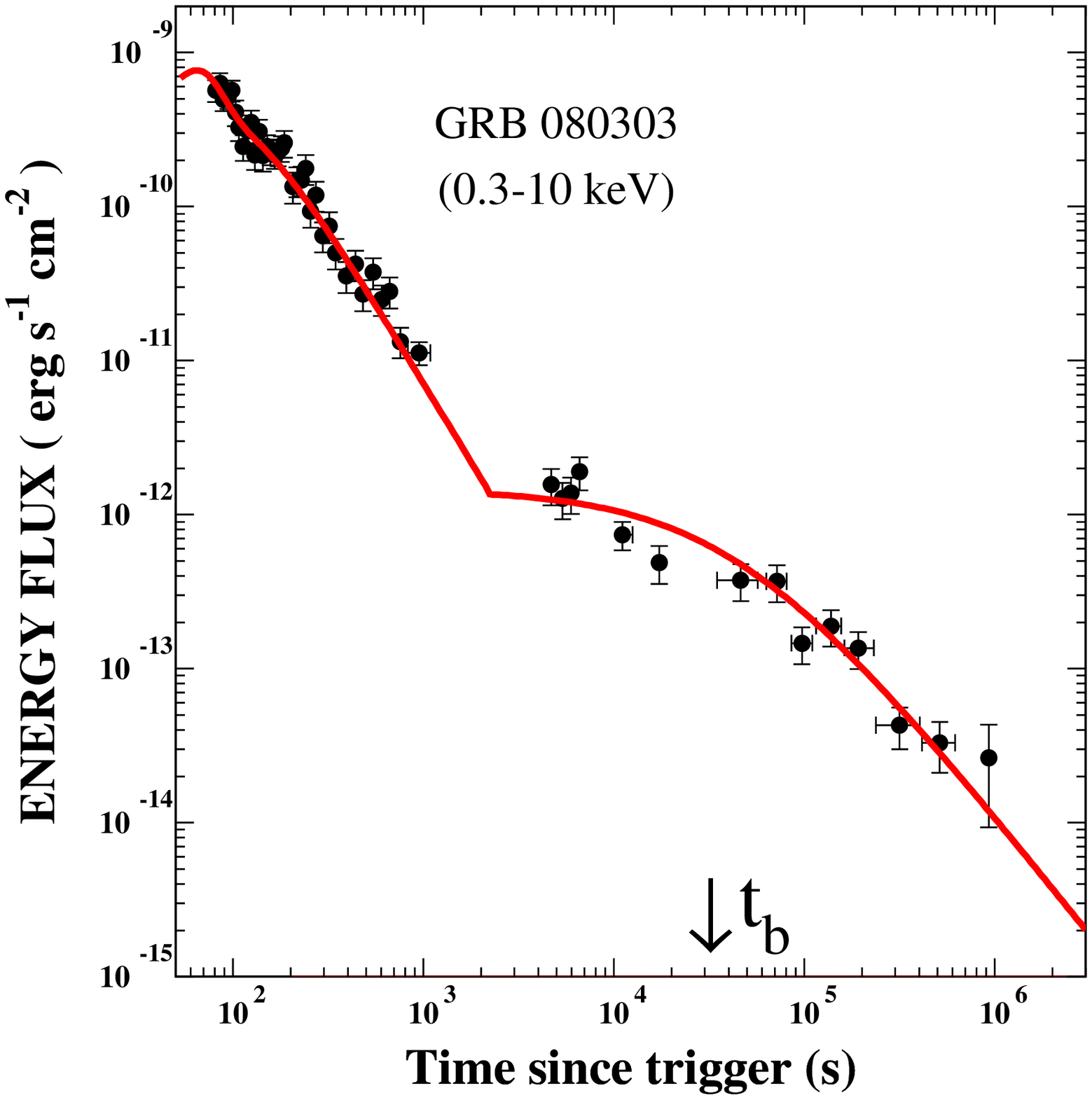,width=8.cm,height=6.cm}
\epsfig{file=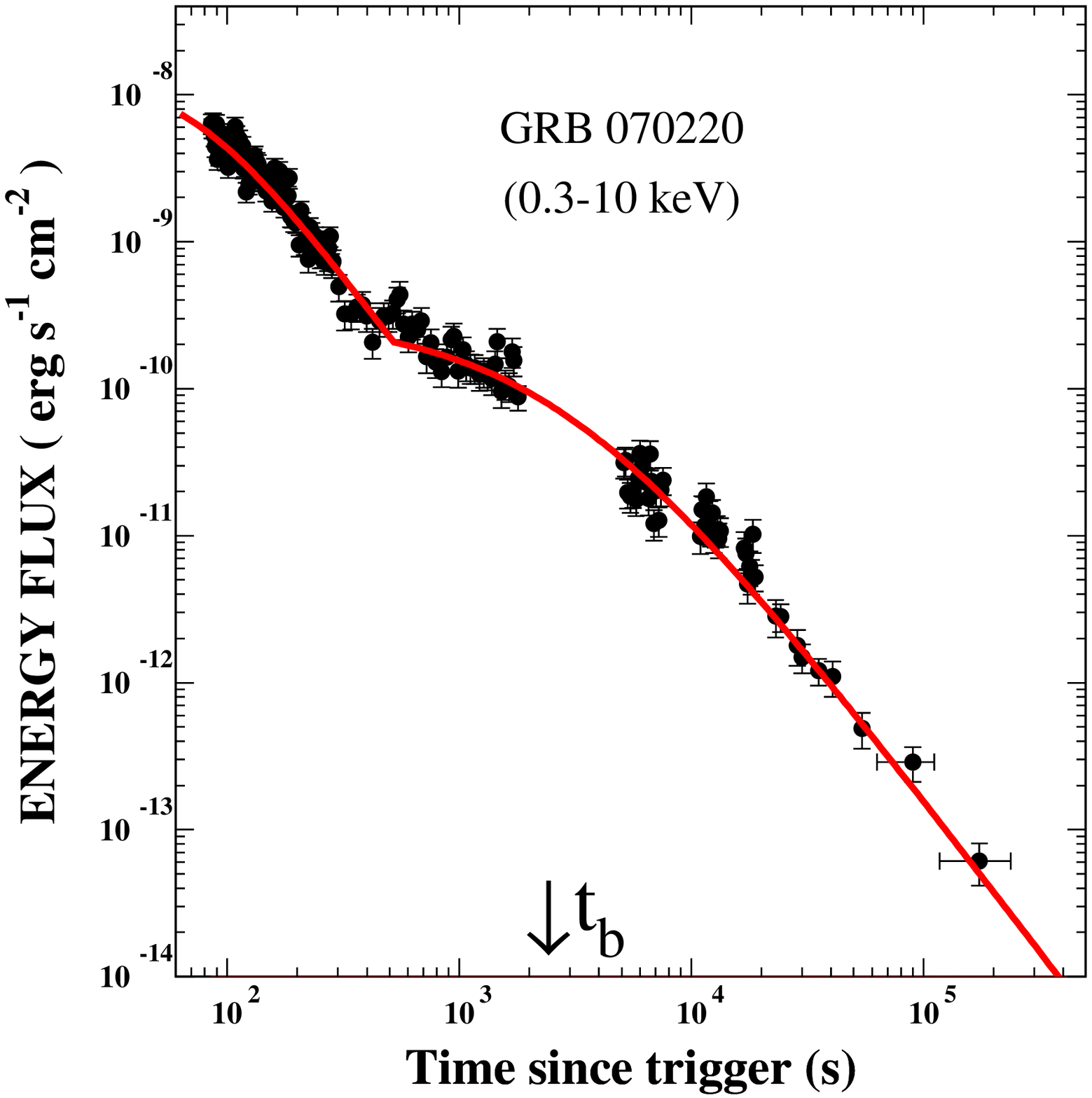,width=8.cm,height=6.cm}
}}
\caption{Comparison between  `semi-canonical' X-ray light curves
of Swift GRBs 
and their CB-model description for:
{\bf Top left (a):} GRB 060211A.
{\bf Top right (b):} GRB 061110A.
{\bf Middle left (c):} GRB 080307.
{\bf Middle right (d):} GRB 051021B.
{\bf Bottom left (e):} GRB 080303.
{\bf Bottom right (f):} GRB 070220.}
\label{f3}
\end{figure}

\newpage
\begin{figure}[]
\centering
\vspace{-2cm}
\vbox{
\hbox{
\epsfig{file=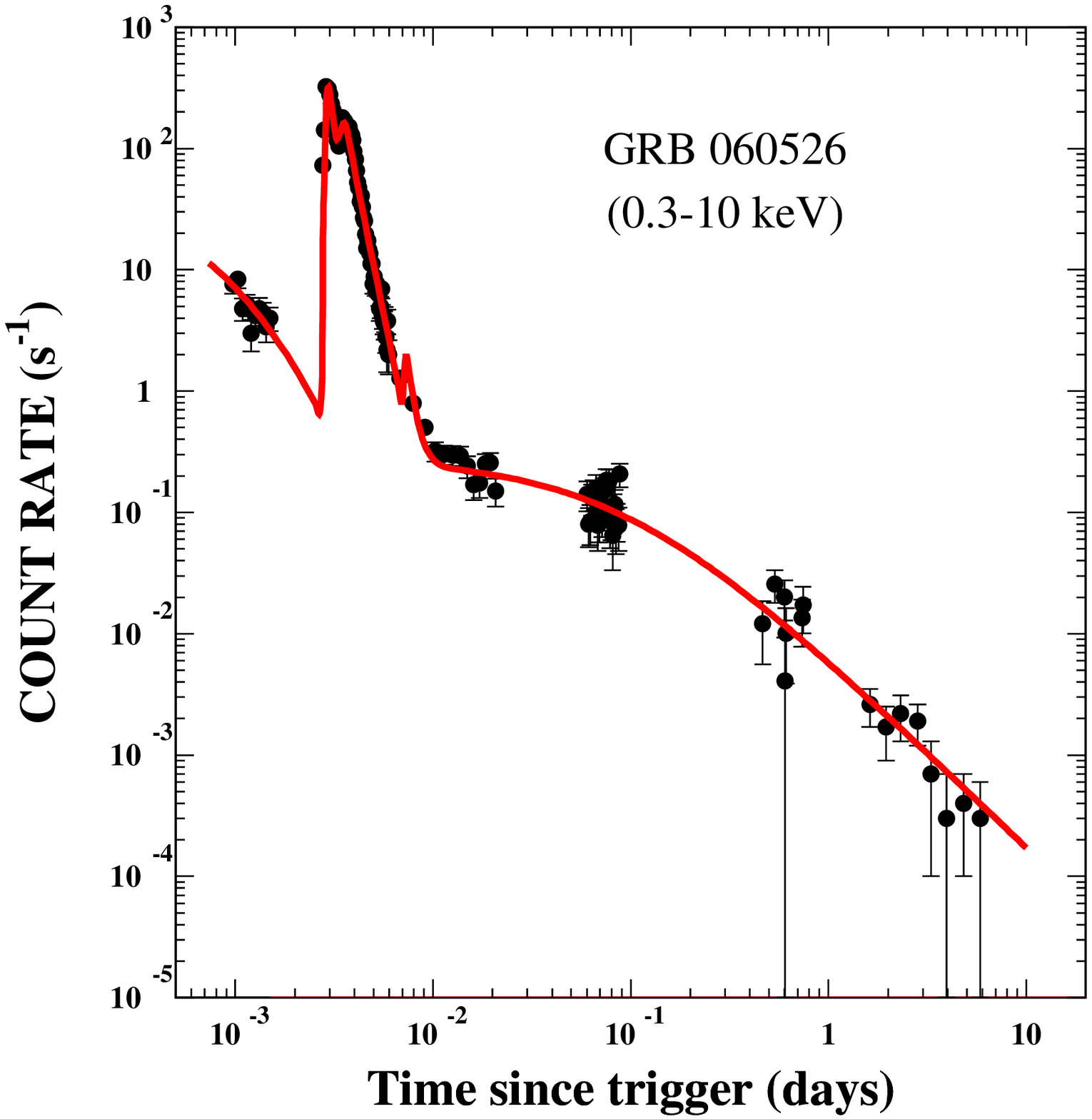,width=8.cm,height=6.cm}
\epsfig{file=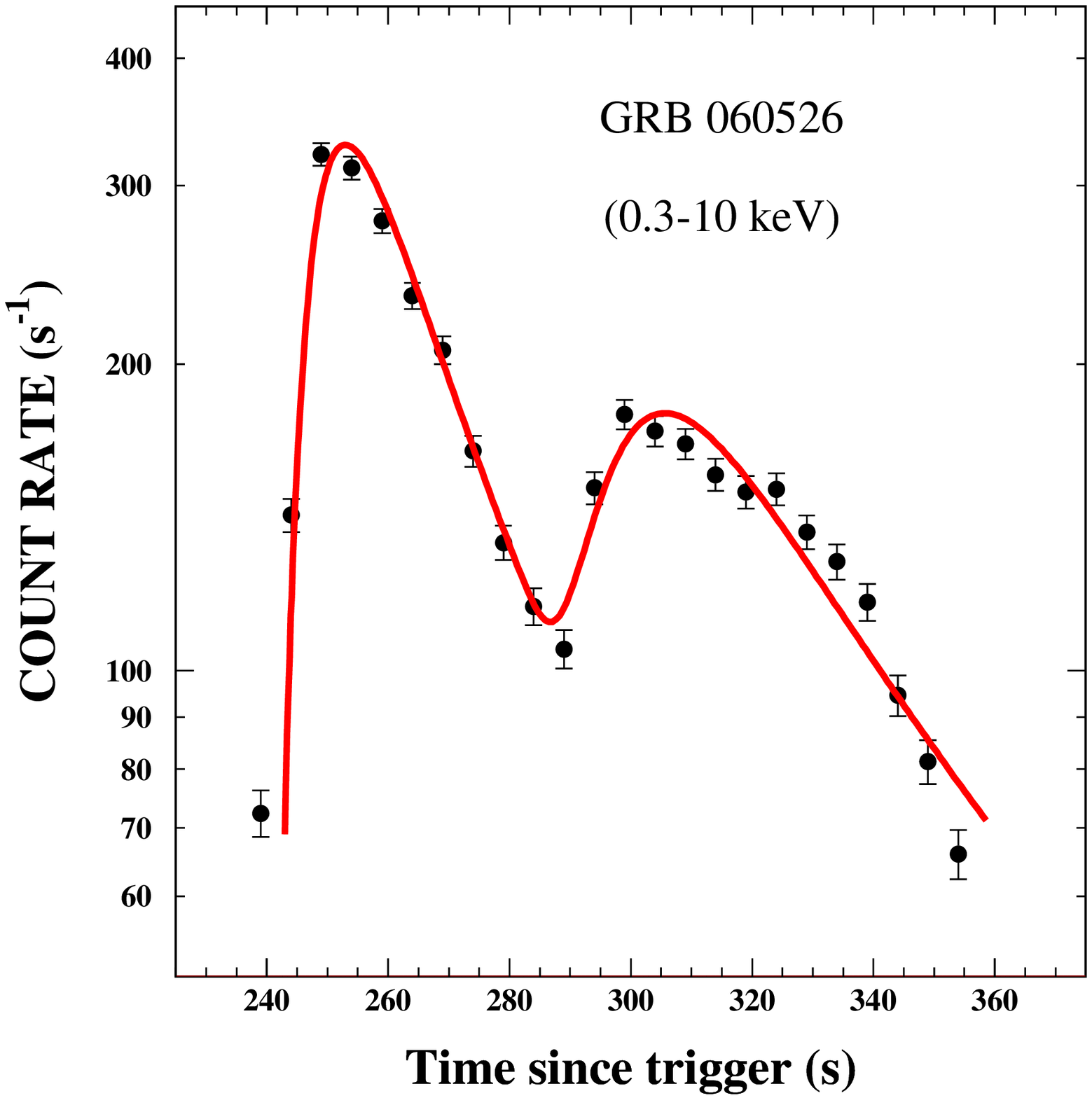,width=8.cm,height=6.cm}
}}
\vbox{
\hbox{
\epsfig{file=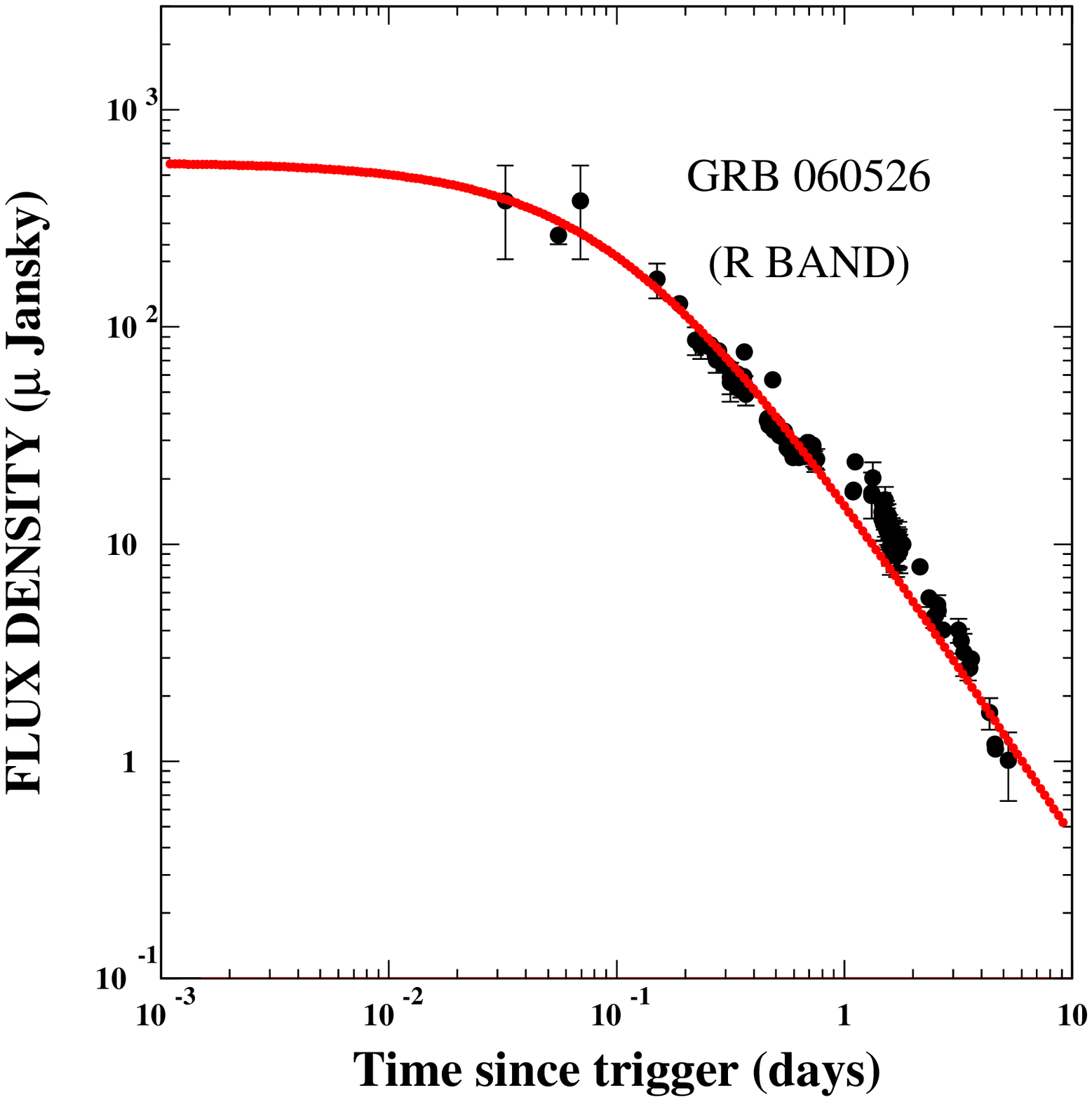,width=8.cm,height=6.cm}
\epsfig{file=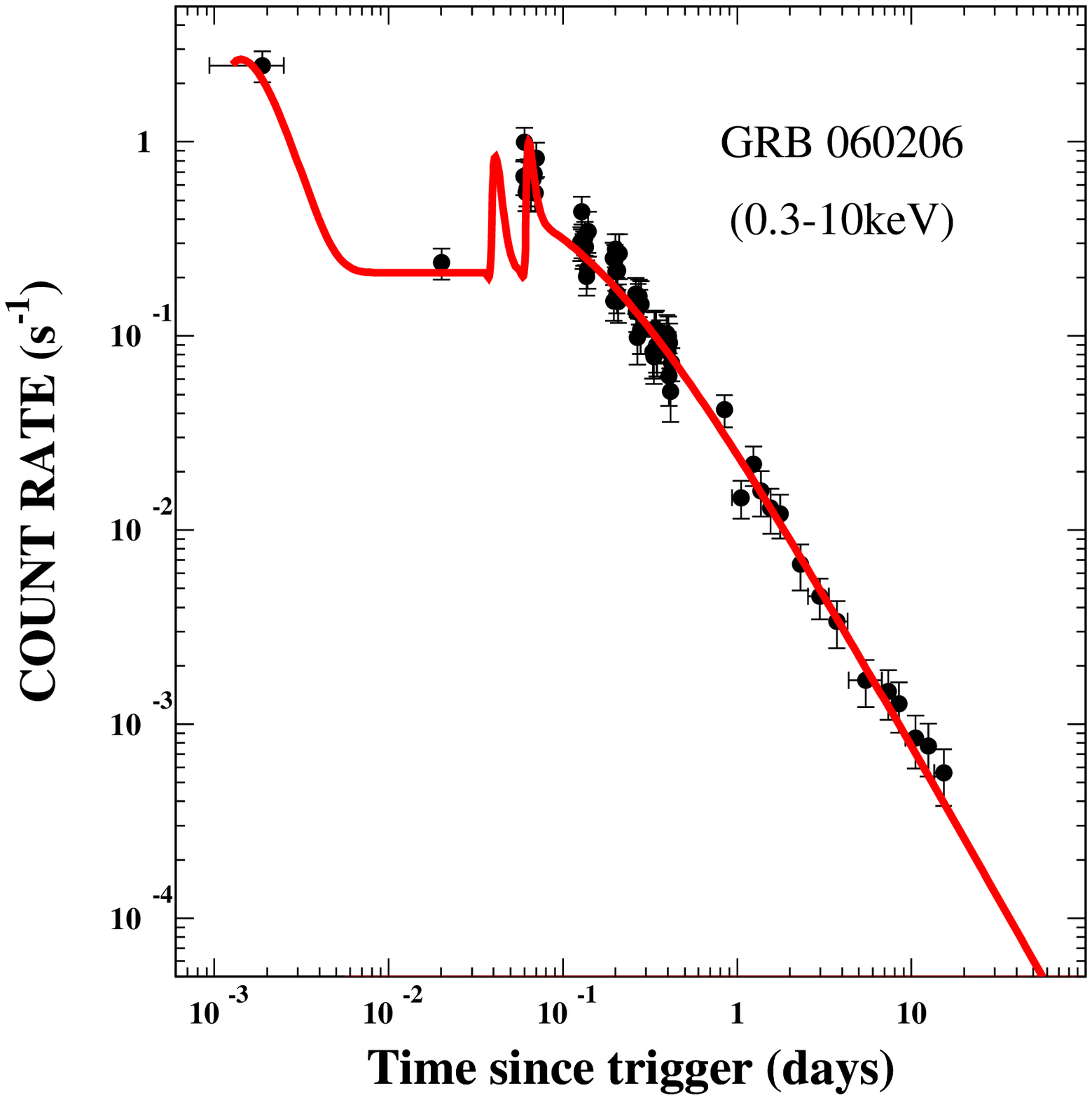,width=8.cm,height=6.cm}
}}
\vbox{
\hbox{
\epsfig{file=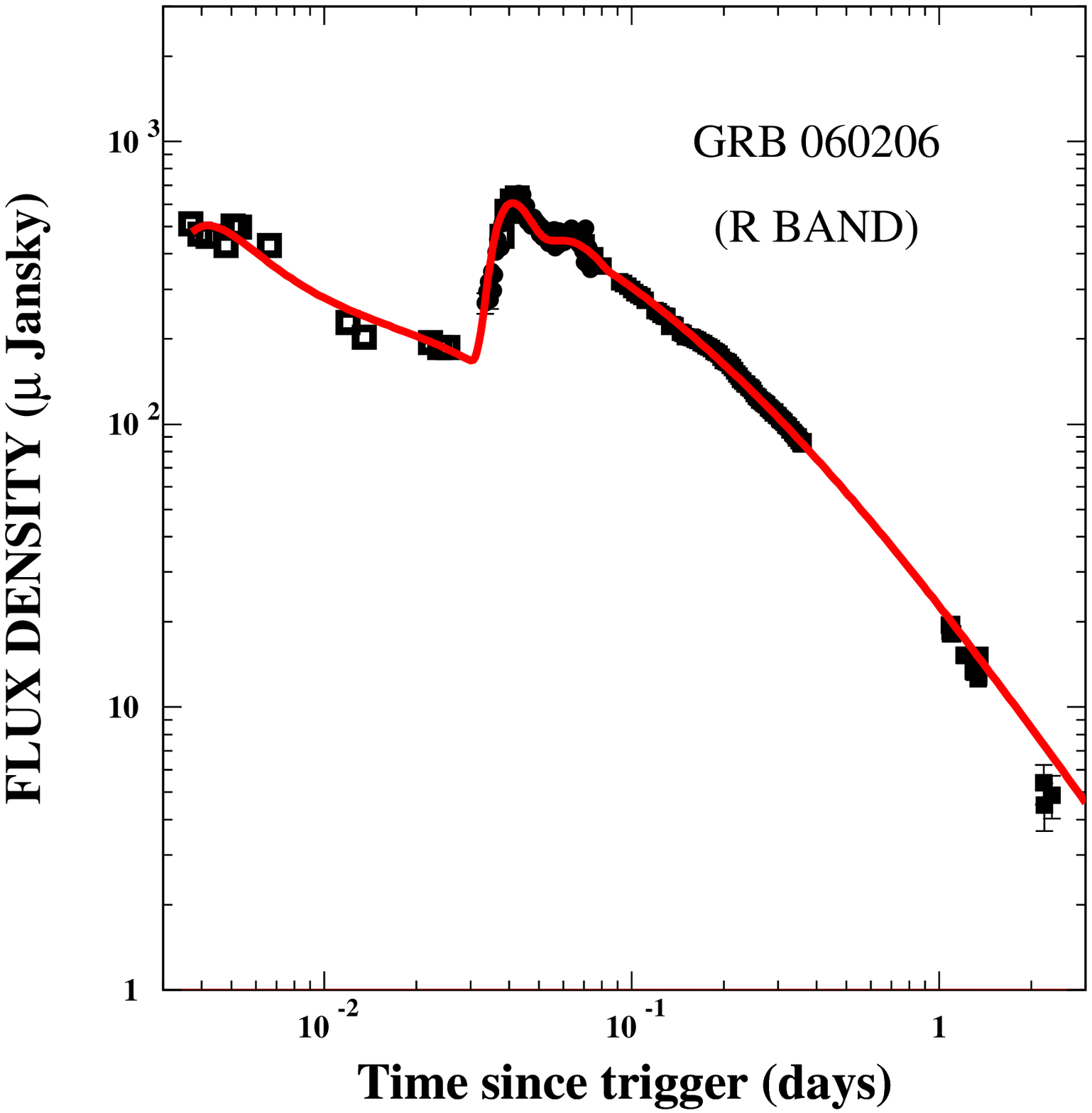,width=8.cm,height=6.cm}
\epsfig{file=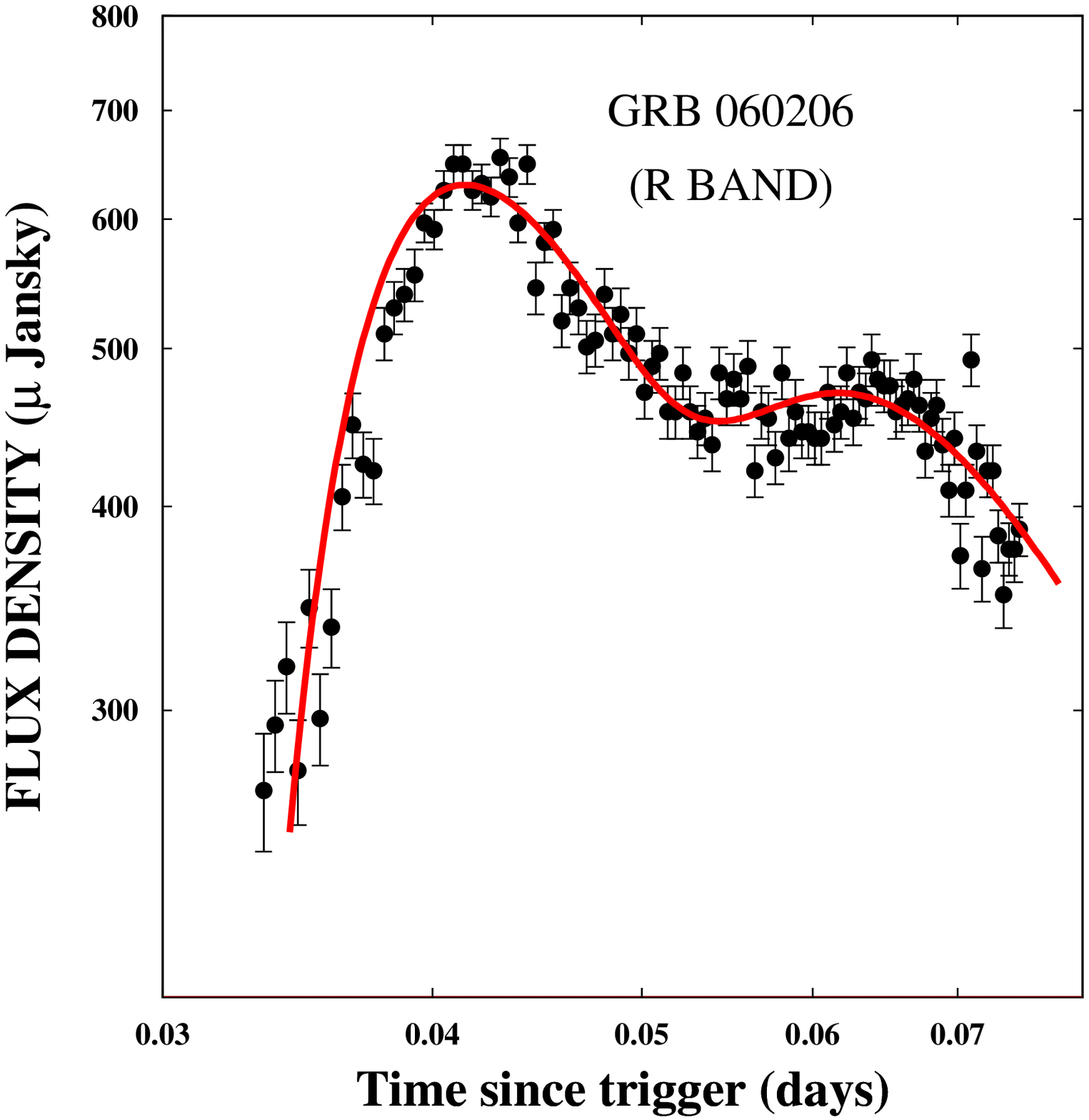,width=8.cm,height=6.cm}
}}
\caption{Comparison between broad-band observations of GRBs with
chromatic early-time afterglow and their CB-model descriptions for:
{\bf Top left (a):} The XRT light curve of GRB 060526.
{\bf Top right (b):} The early X-ray ICS flares of GRB 060526.
{\bf Middle left (c):} The $R$-band light curve of GRB 060526.
{\bf Middle right (d):}  The XRT light curve of GRB 060206.
{\bf Bottom left (e):} The $R$-band light curve of GRB 060206.
{\bf Bottom right (f):} Enlarged view of two-early time 
$R$-band SR flares and their CB-model description.}
\label{f4}
\end{figure}

\newpage
\begin{figure}[]
\centering
\vspace{-2cm}
\vbox{
\hbox{
\epsfig{file=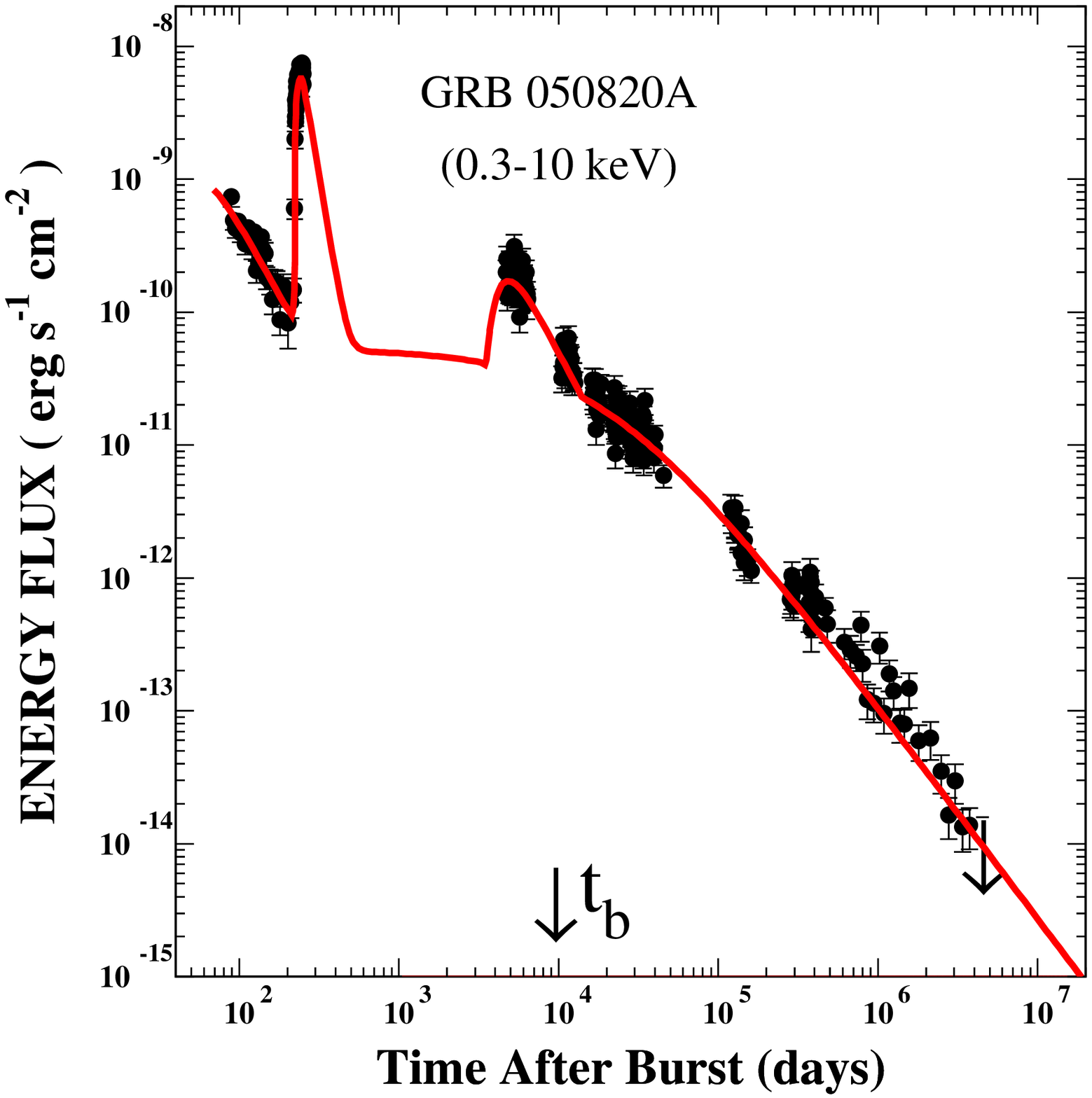,width=8.cm,height=6.cm}
\epsfig{file=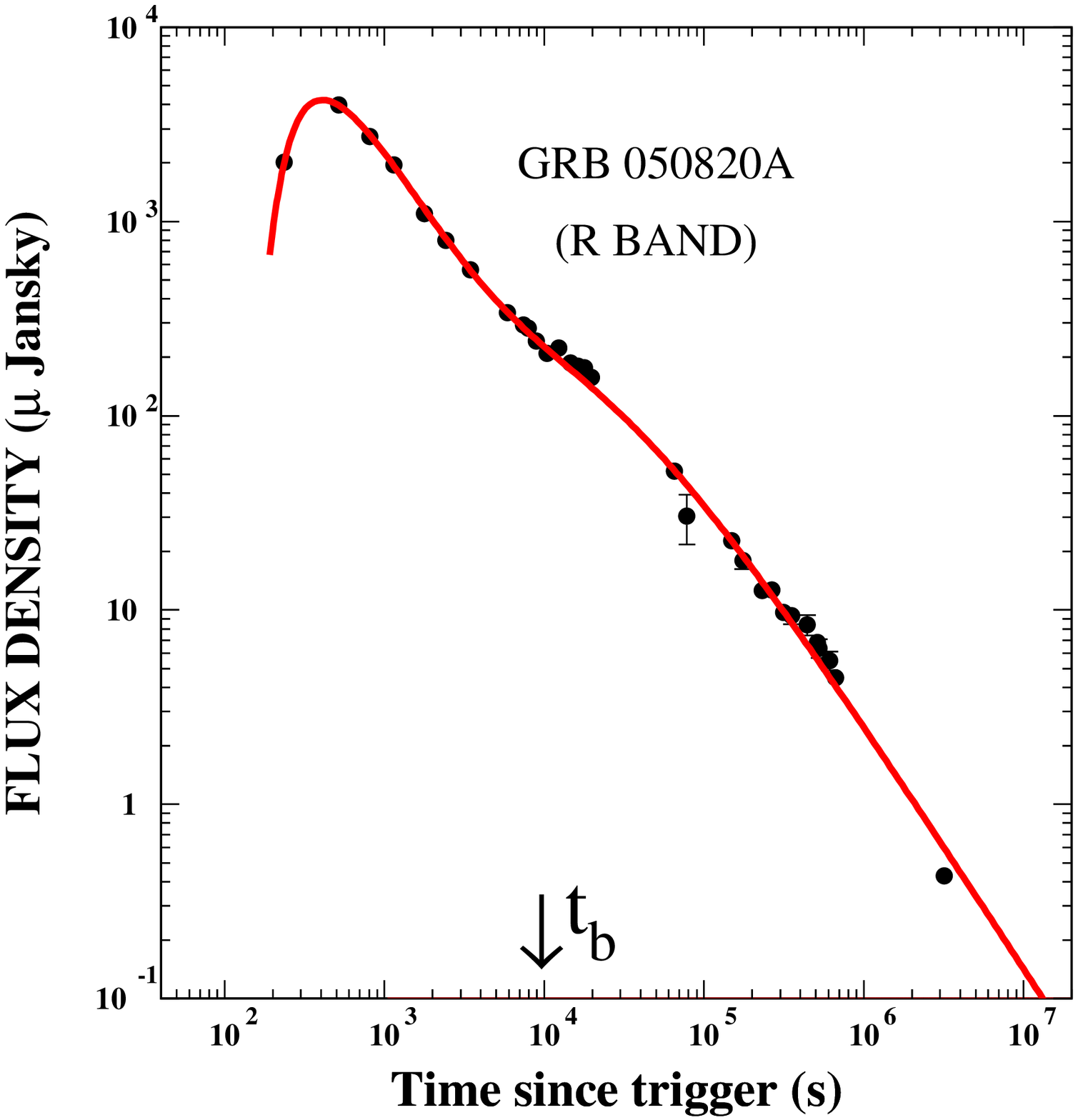,width=8.cm,height=6.cm}
}}
\vbox{
\hbox{
\epsfig{file=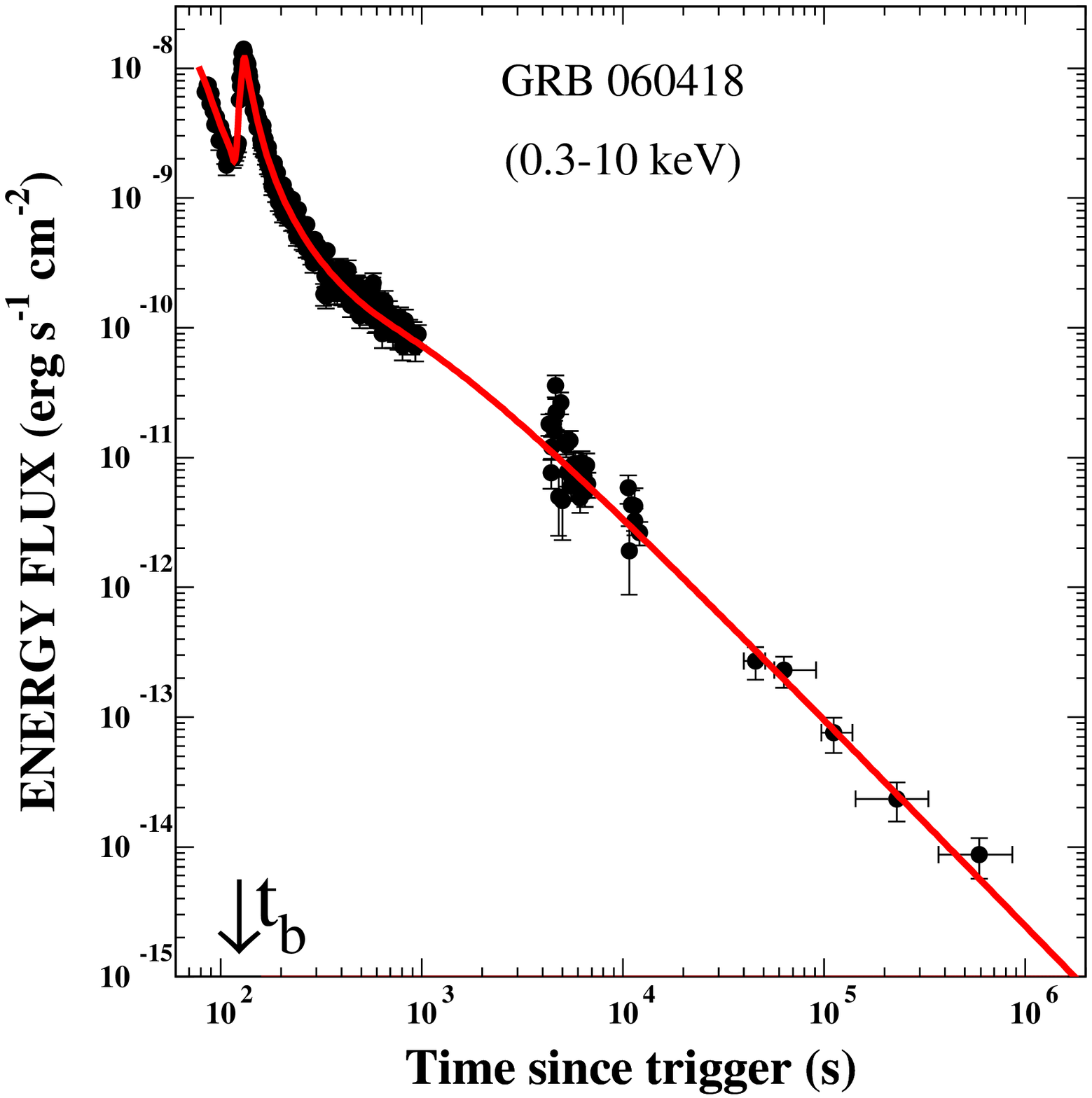,width=8.cm,height=6.cm}
\epsfig{file=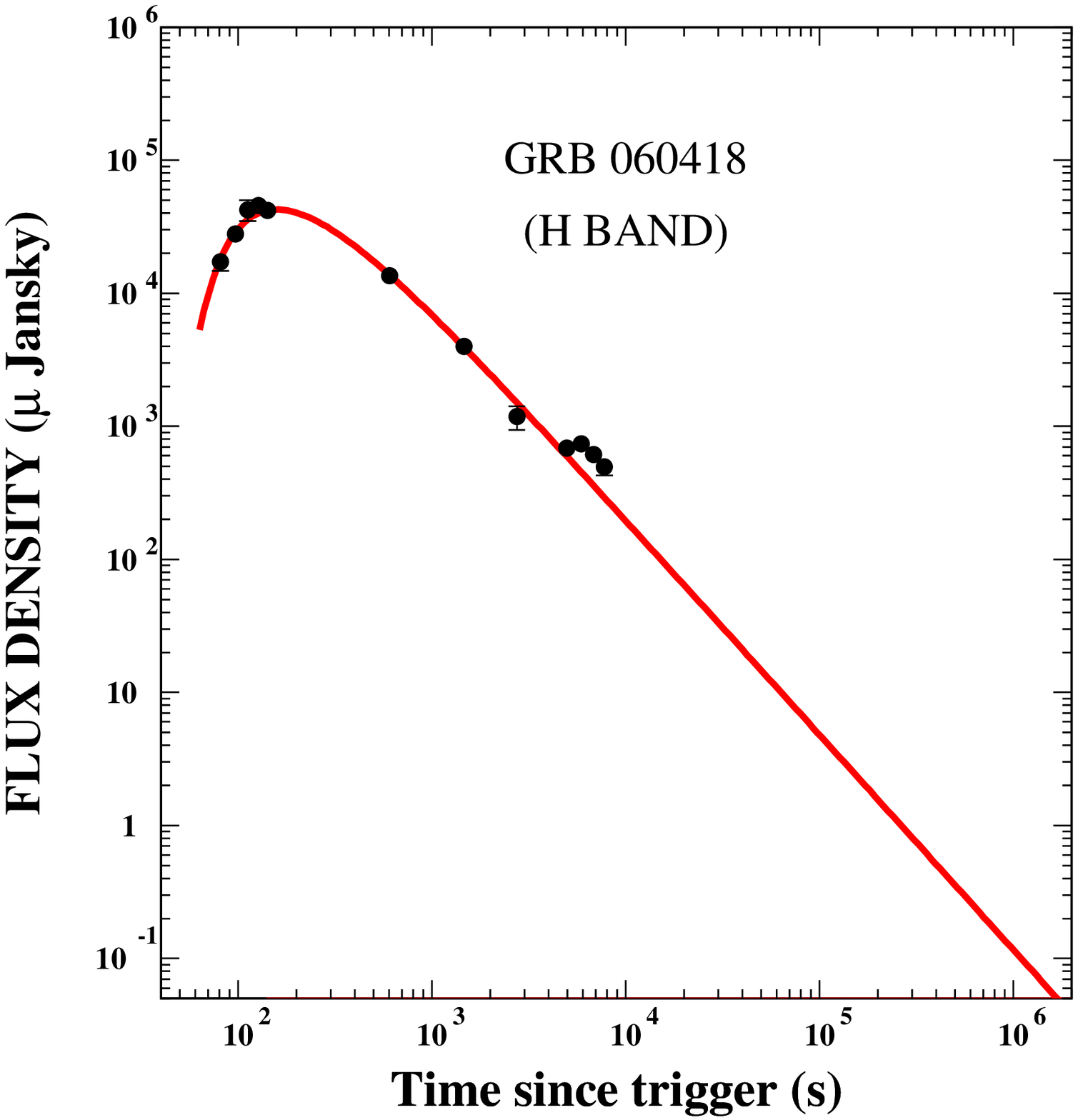,width=8.cm,height=6.cm}
}}
\vbox{
\hbox{
\epsfig{file=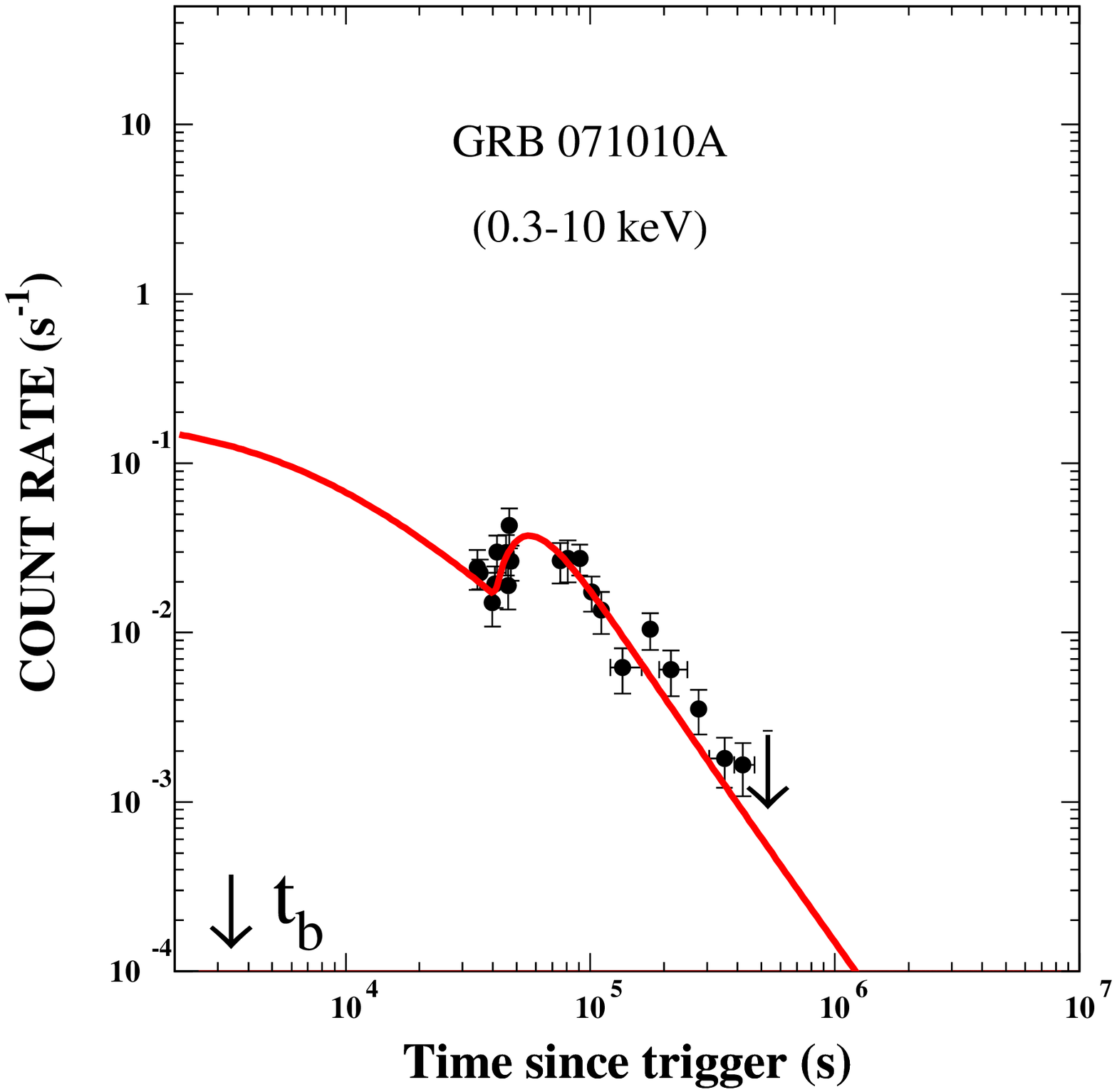,width=8.cm,height=6.cm}
\epsfig{file=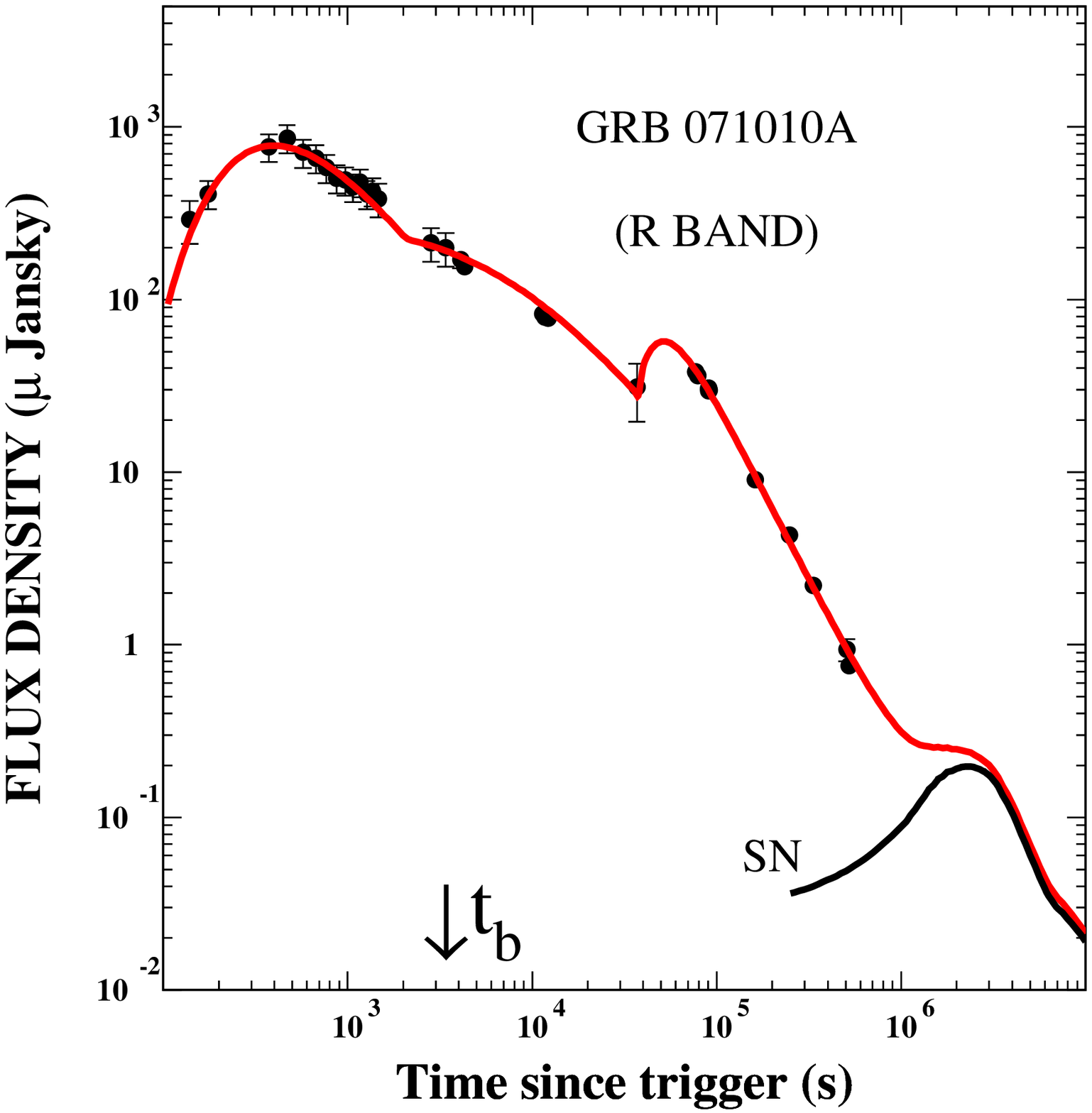,width=8.cm,height=6.cm}
}}
\caption{Comparison between broad-band observations of GRBs with
chromatic early-time afterglow and their CB-model descriptions for:
{\bf Top left (a):} The X-ray light curve of GRB 050820A.
{\bf Top right (b):} The $R$-band light curve of GRB 050820A.
{\bf Middle left (c):} The X-ray light curve of GRB 060418.
{\bf Middle right (d):} The $H$-band light curve of GRB 060418.
{\bf Bottom left (e):} The X-ray light curve of GRB 071010A.
{\bf Bottom right (f):} The $R$-band light curve of GRB 071010A.}
\label{f5}
\end{figure}

\newpage
\begin{figure}[]
\centering
\vspace{-2cm}
\vbox{
\hbox{
\epsfig{file=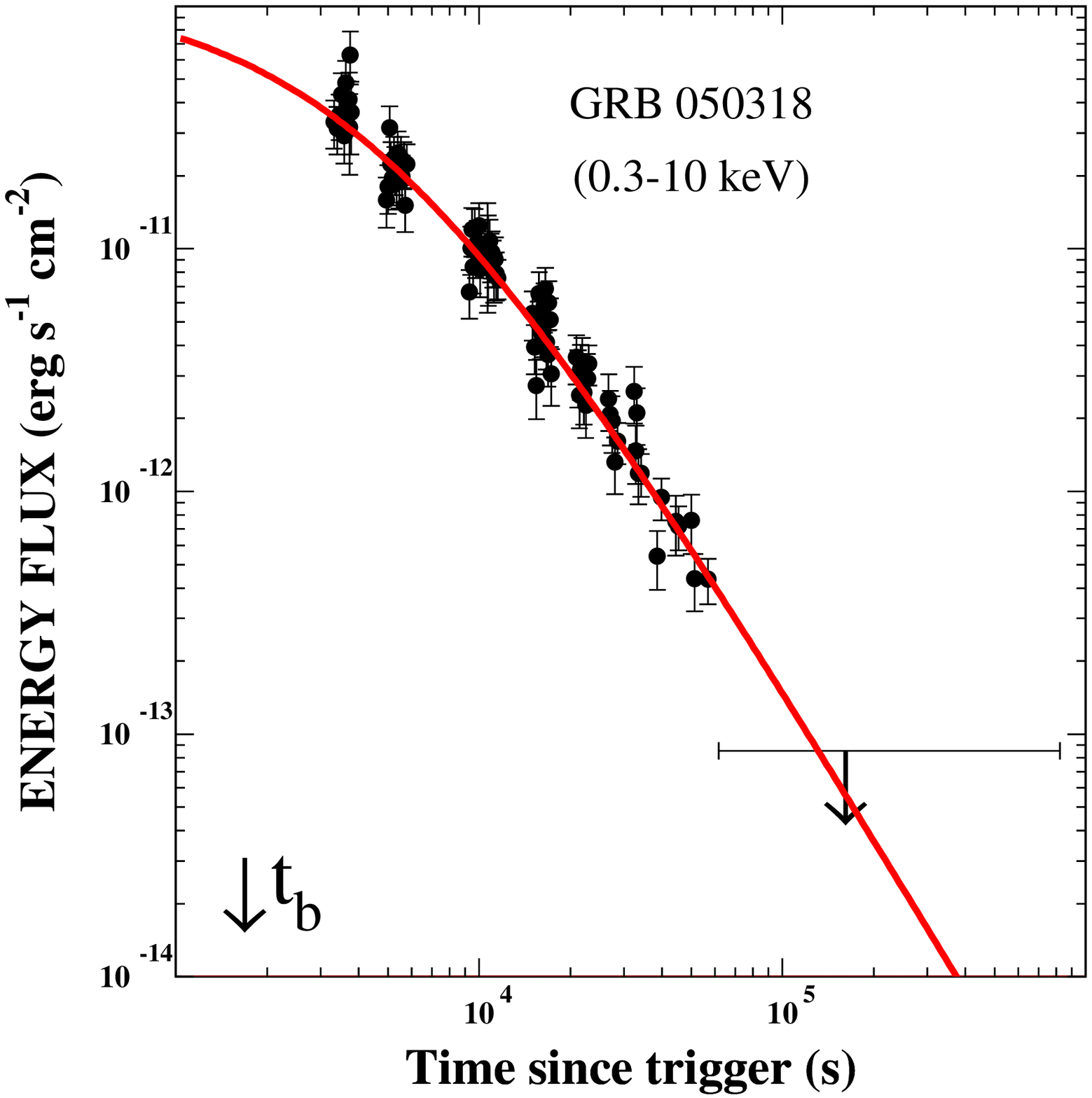,width=8.cm,height=6.cm}
\epsfig{file=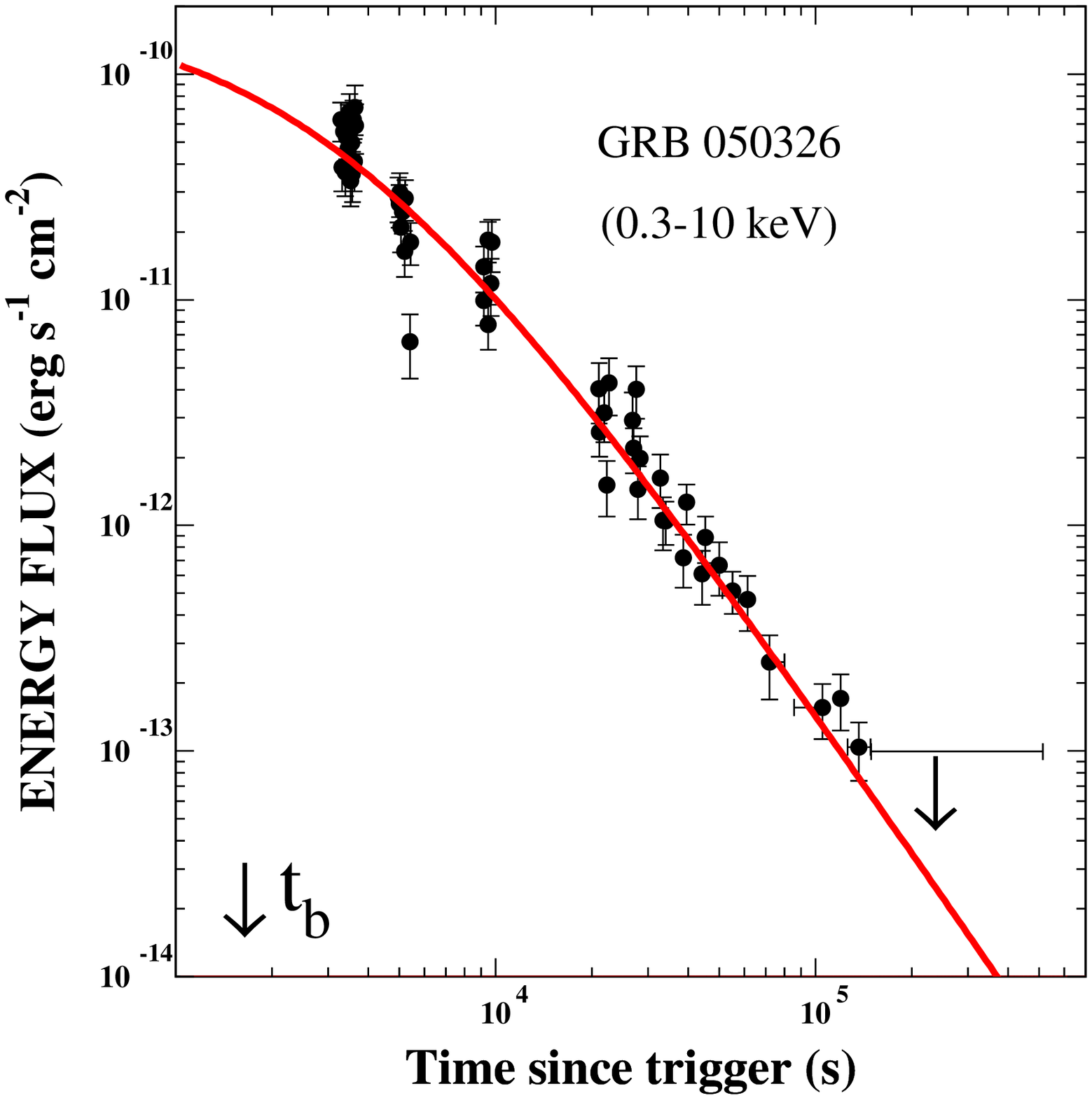,width=8.cm,height=6.cm}
}}
\vbox{
\hbox{
\epsfig{file=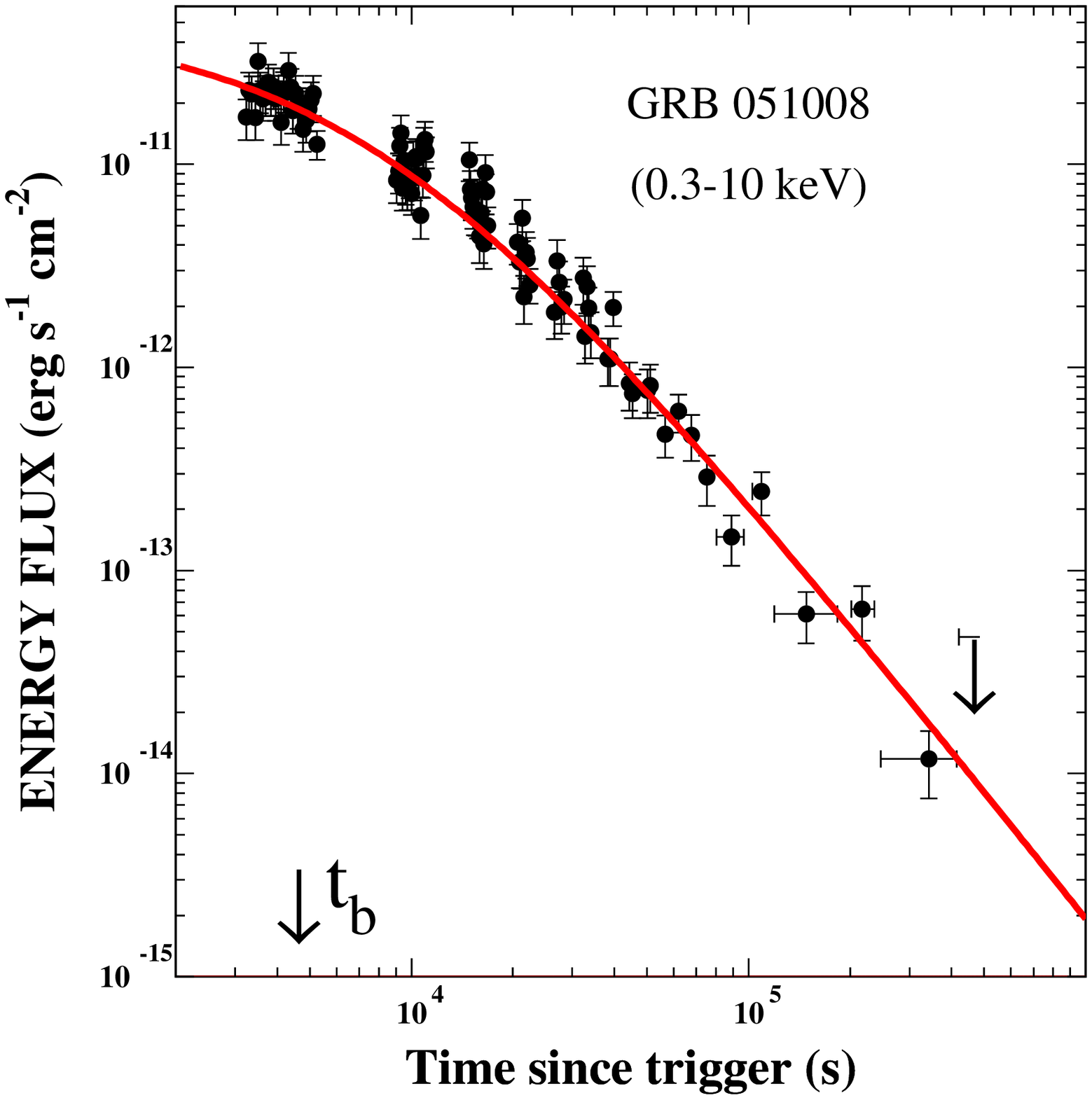,width=8.cm,height=6.cm}
\epsfig{file=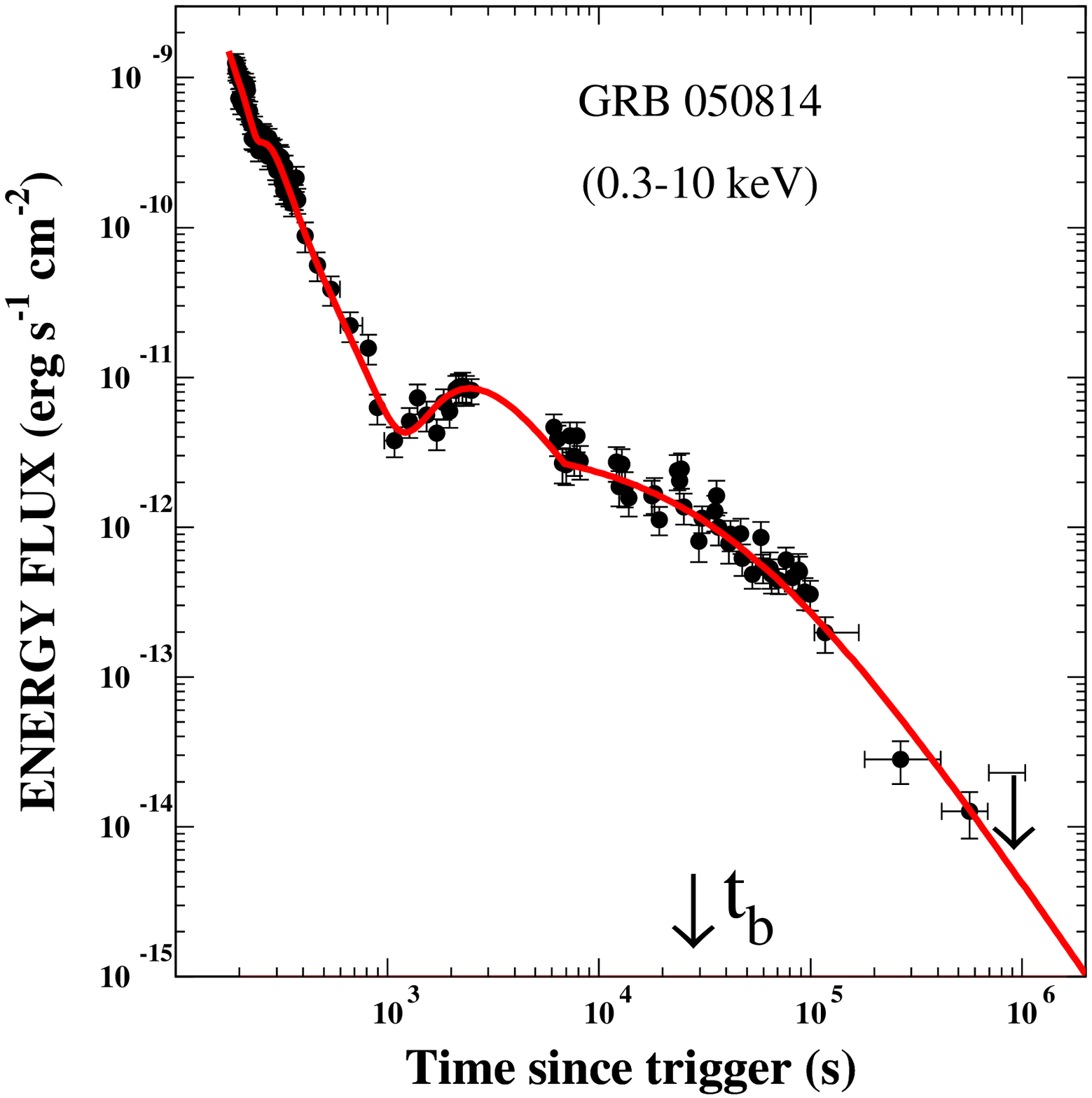,width=8.cm,height=6.cm}
}}
\vbox{
\hbox{
\epsfig{file=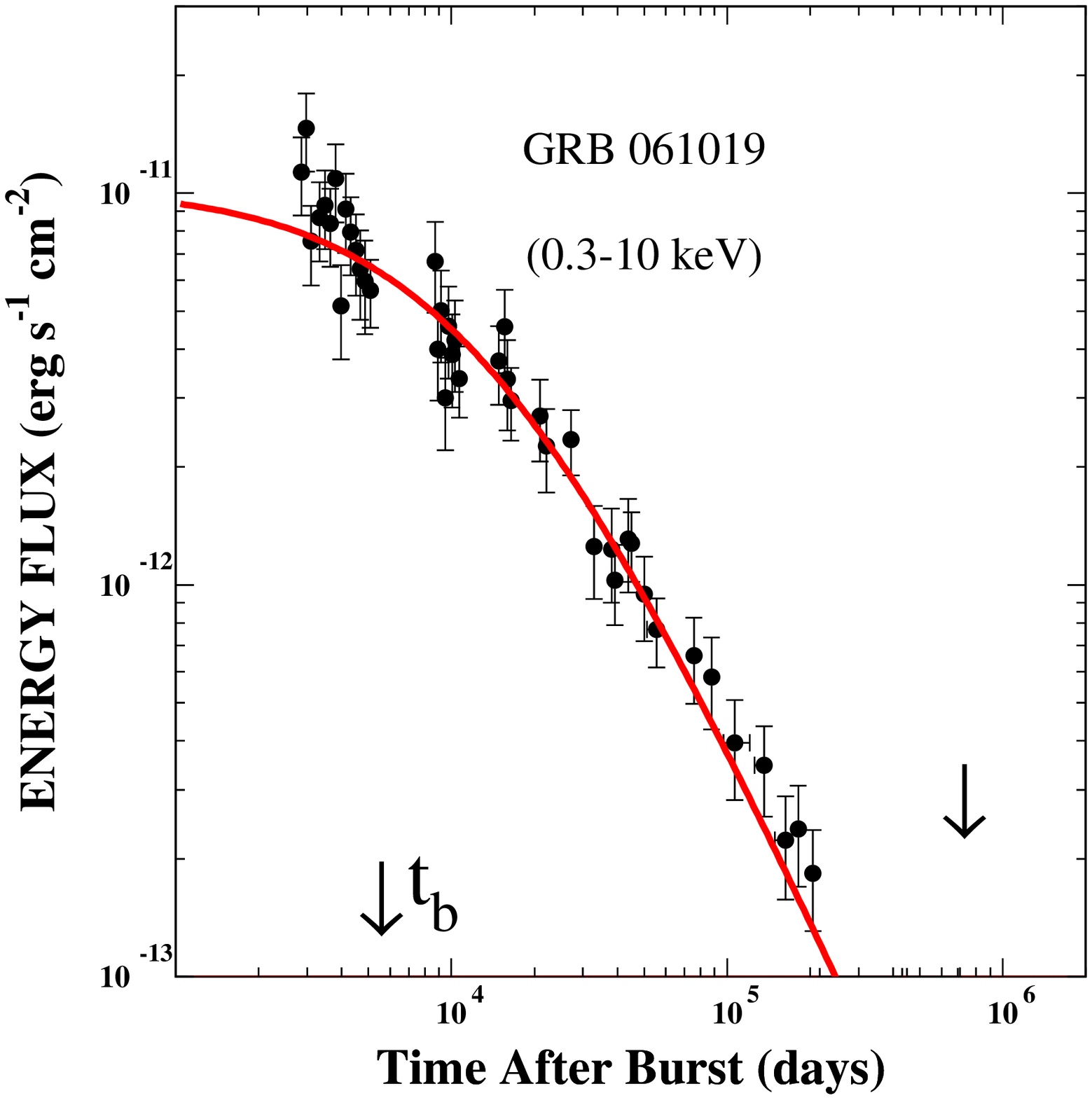,width=8.cm,height=6.cm}
\epsfig{file=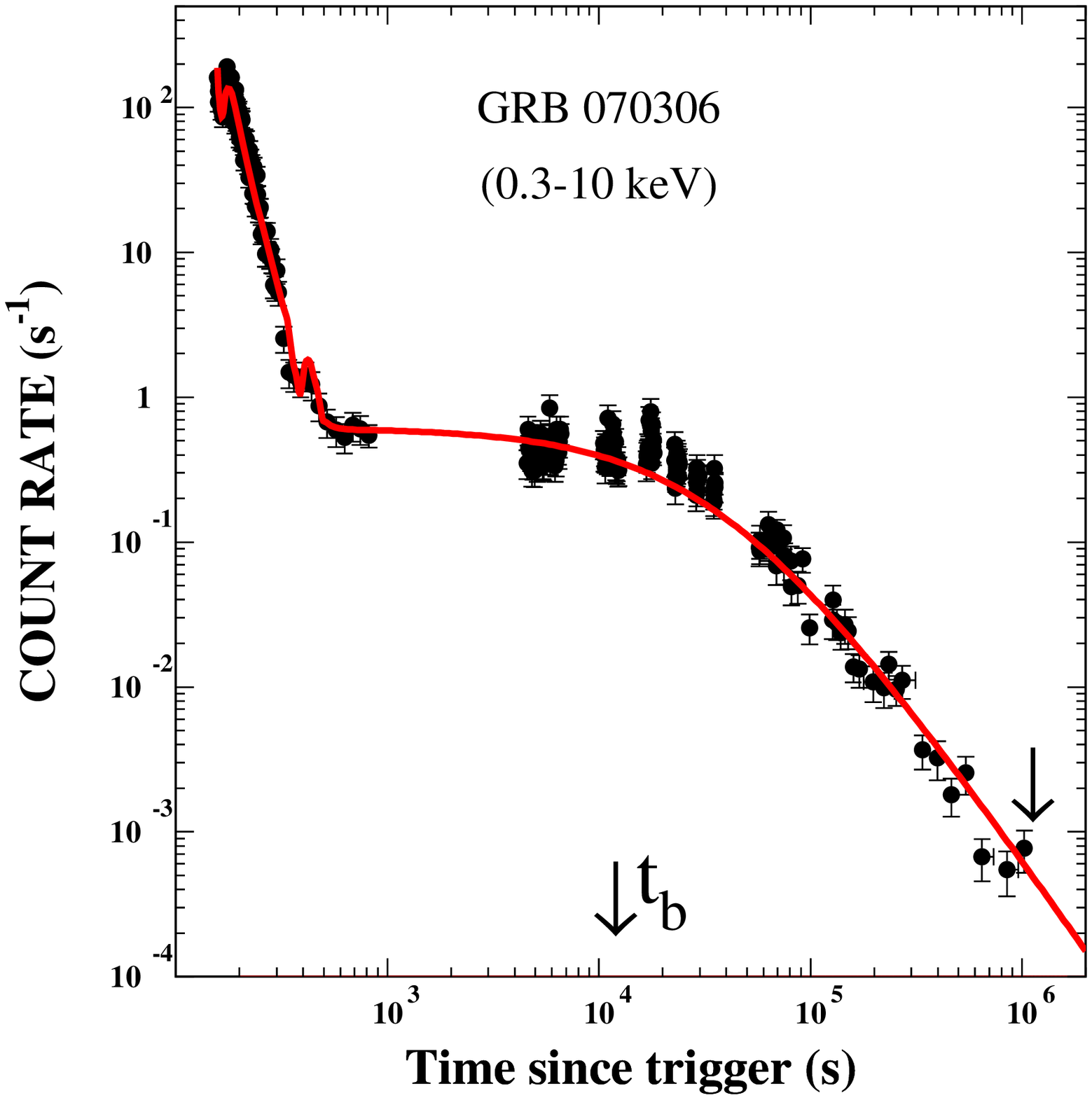,width=8.cm,height=6.cm}
}}
\caption{Comparison between XRT-light curves of Swift GRBs
(Evans et al.~2007) with
late time decay index $\alpha>2$ and their CB-model 
descriptions 
assuming an isothermal-sphere density profile, for: 
{\bf Top left (a):}  GRB 050318.
{\bf Top right (b):} GRB 050326.
{\bf Middle left (c):} GRB 051008.
 {\bf Middle right (d):} GRB 050814.
{\bf Bottom left (e):} GRB 061019.
{\bf Bottom right (f):} GRB 070306.}
\label{f6}
\end{figure}

\newpage
\begin{figure}[]
\centering
\vspace{-2cm}
\vbox{
\hbox{
\epsfig{file=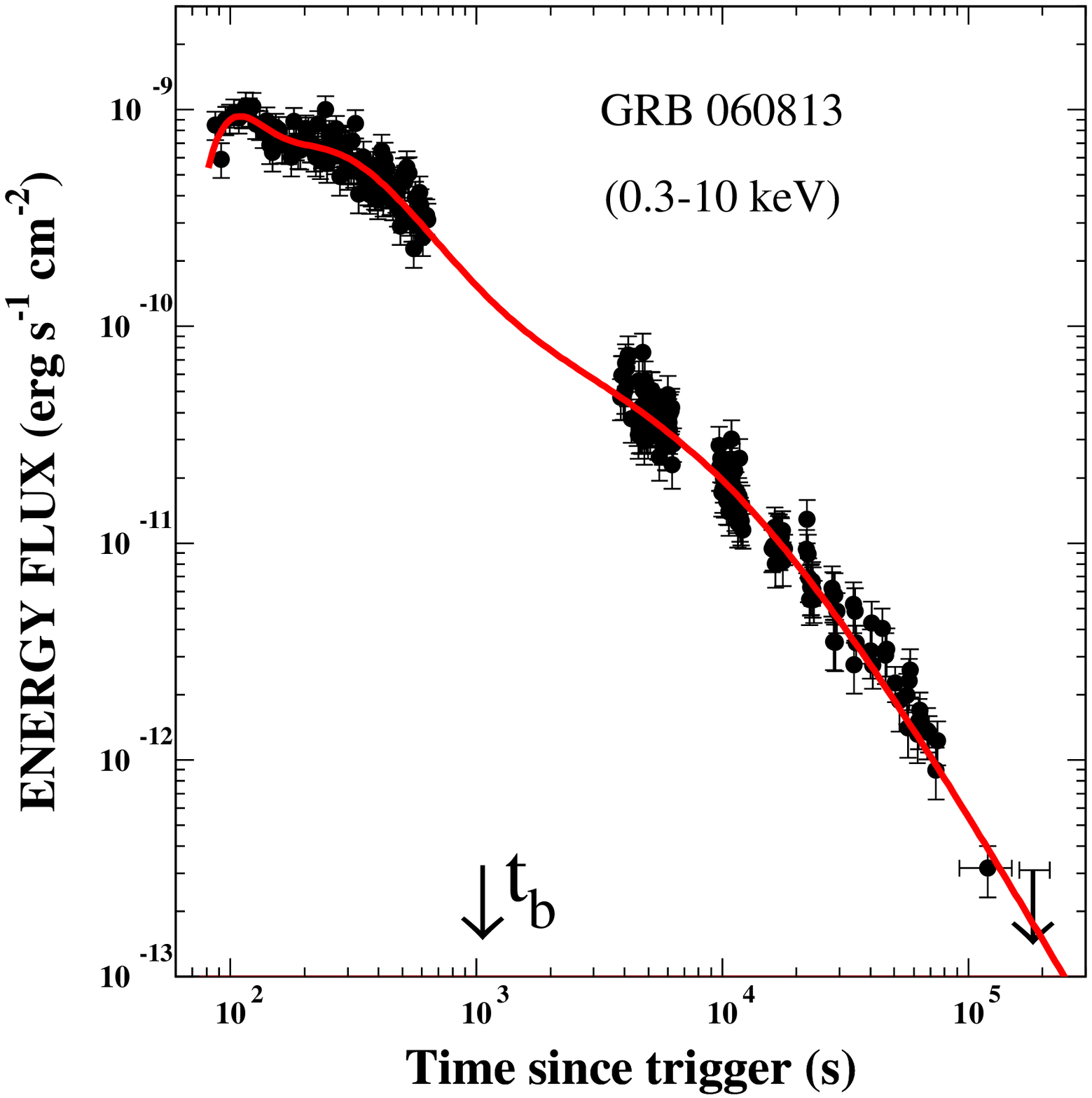,width=8.cm,height=6.cm}
\epsfig{file=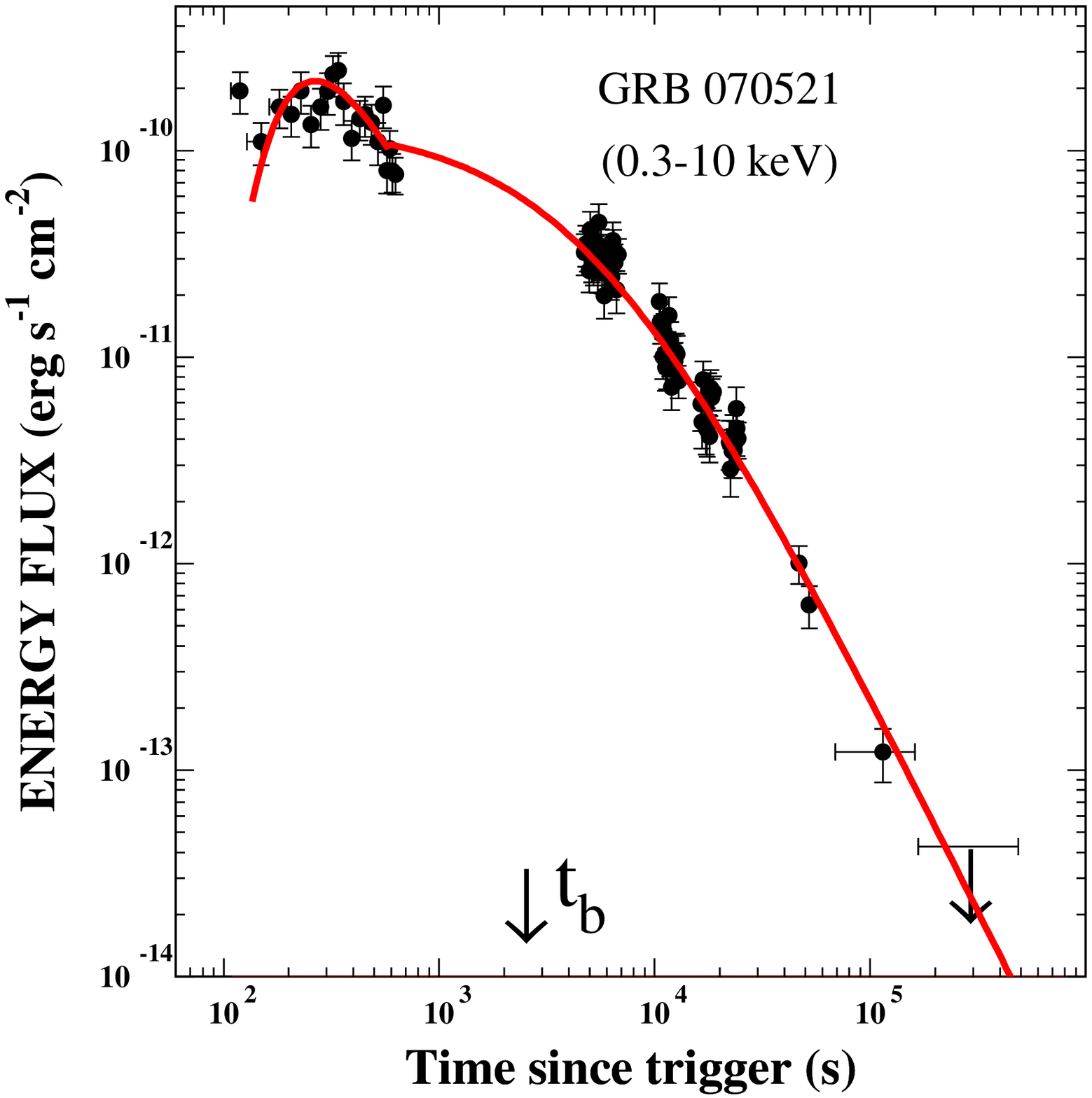,width=8.cm,height=6.cm}
}}
\vbox{
\hbox{
\epsfig{file=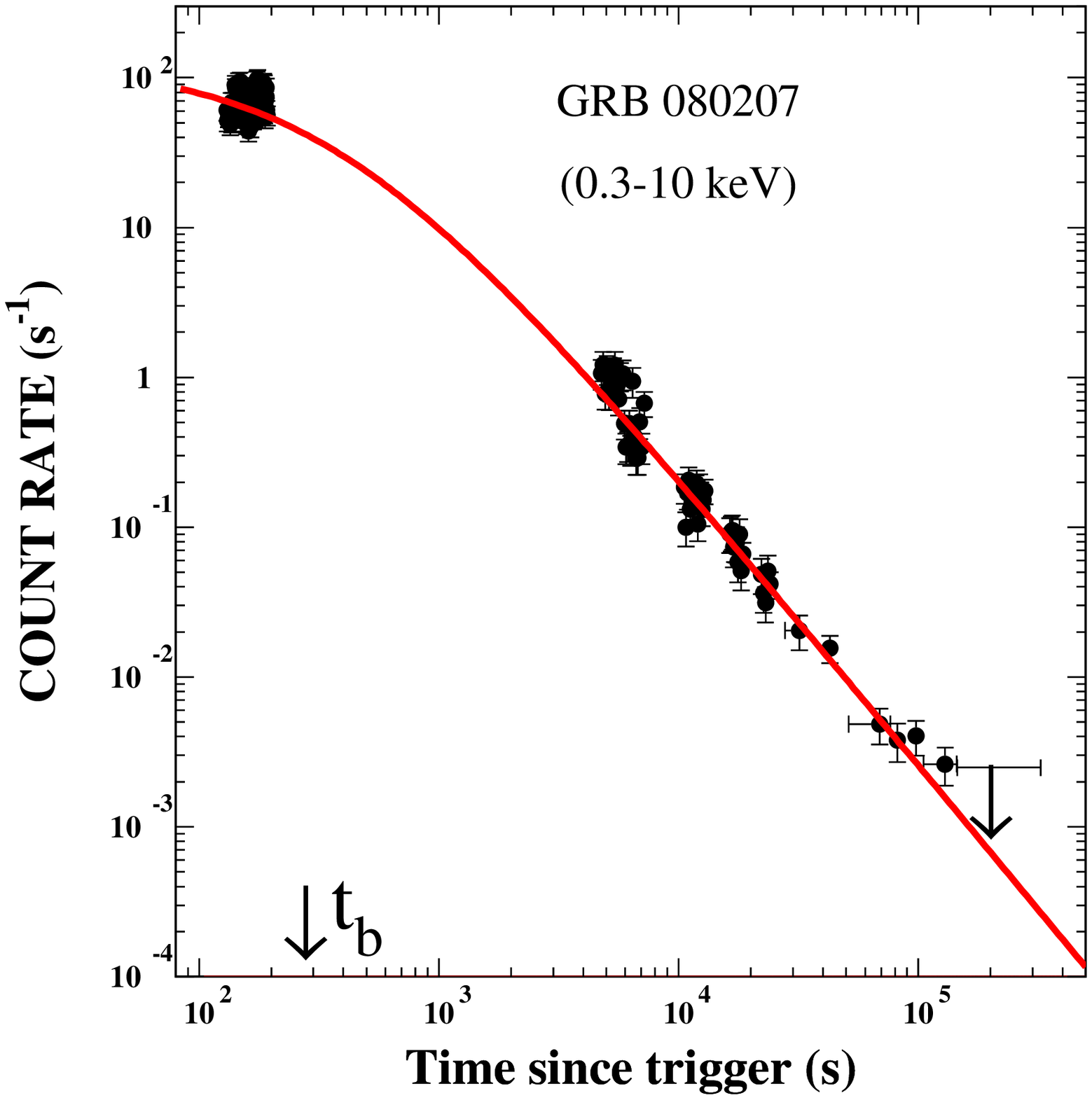,width=8.cm,height=6.cm}
\epsfig{file=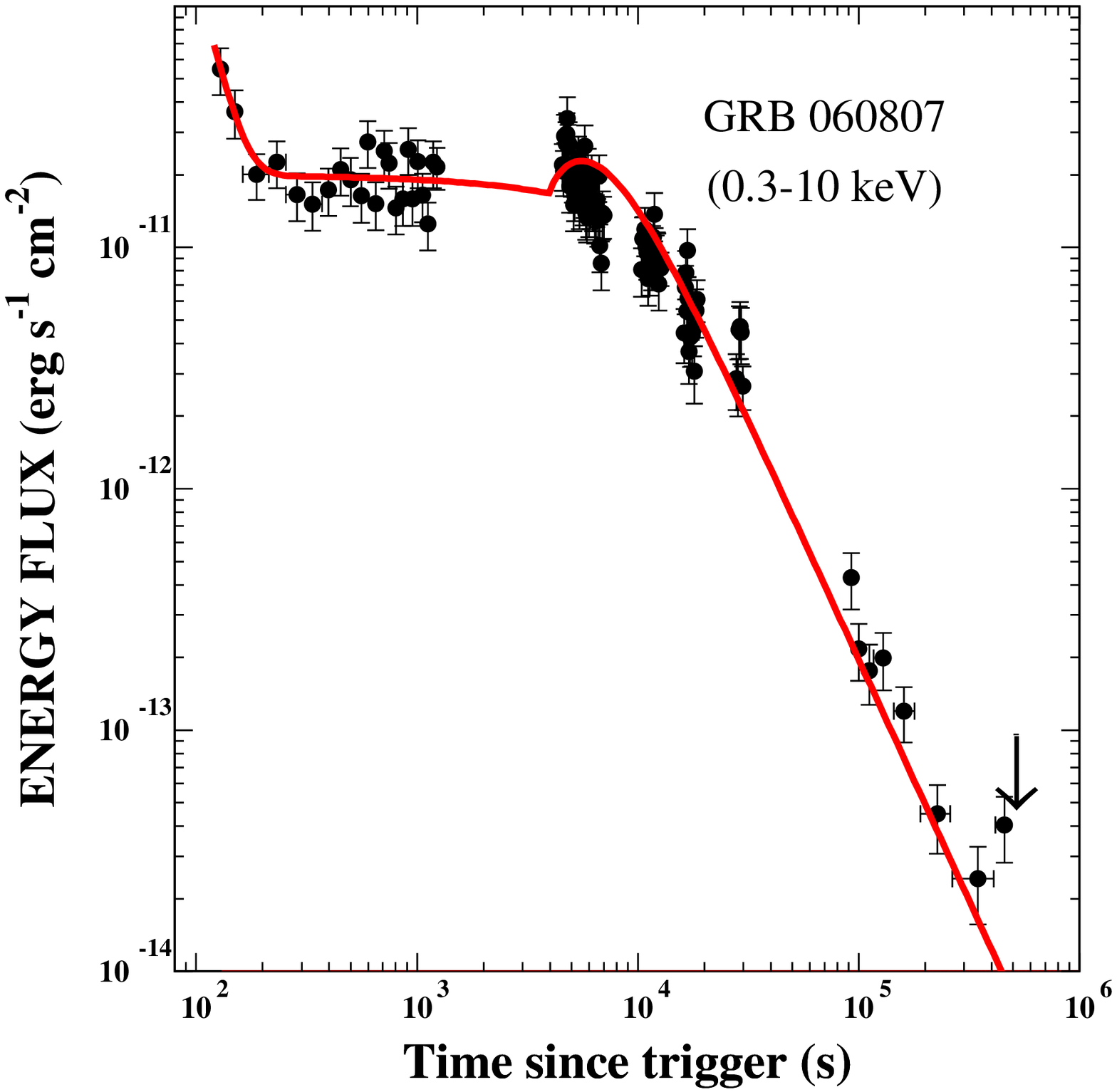,width=8.cm,height=6.cm}
}}
\vbox{
\hbox{
\epsfig{file=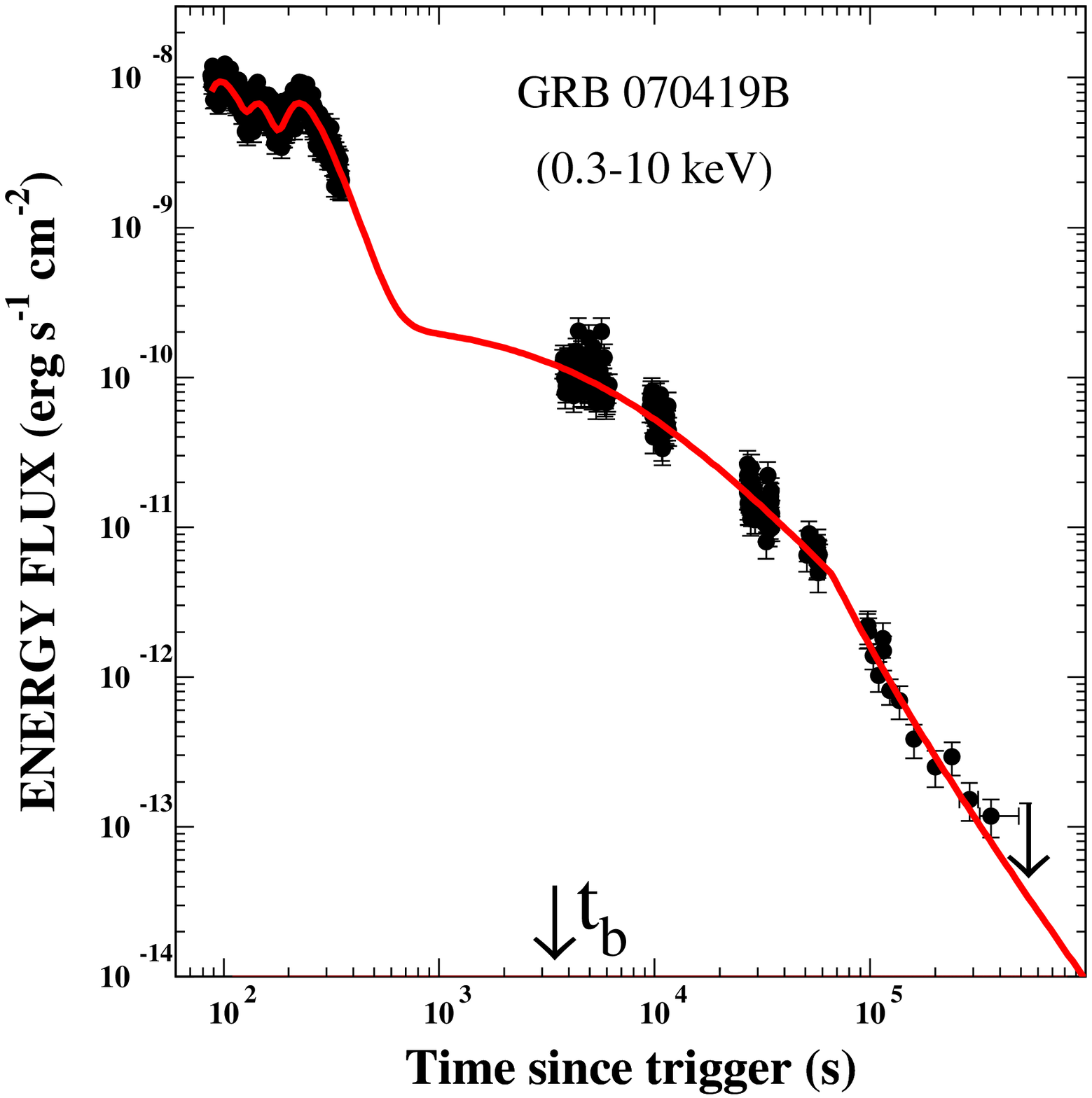,width=8.cm,height=6.cm}
\epsfig{file=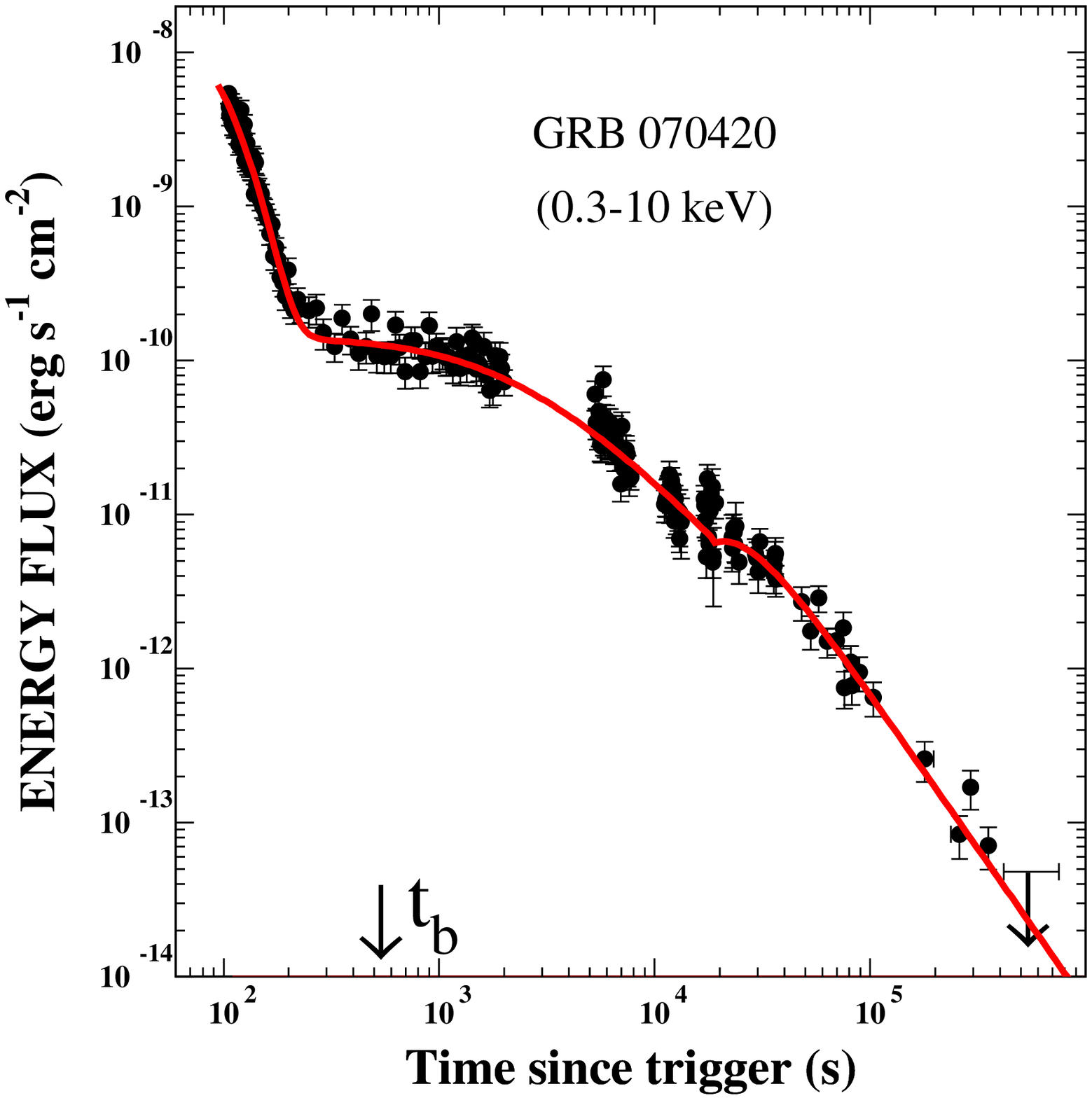,width=8.cm,height=6.cm}
}}
\caption{Comparison between  the Swift XRT-light curves 
with late-time decay index ($\alpha>2$) or a late-time flare and their CB-model 
description for:
{\bf Top left (a):} GRB 060813, with a steep decay.
{\bf Top right (b):} GRB 070521, with a steep  decay.
{\bf Middle left (c):} GRB 080207, with a steep  decay.
{\bf Middle right (d):} GRB 060807, with a  flare.
{\bf Bottom left (e):} GRB 070419B, with a  flare.
{\bf Bottom right (f):} GRB 070420,  with a  flare.}
\label{f7}
\end{figure}

\newpage
\begin{figure}[]
\centering
\vspace{-2cm}
\vbox{
\hbox{
\epsfig{file=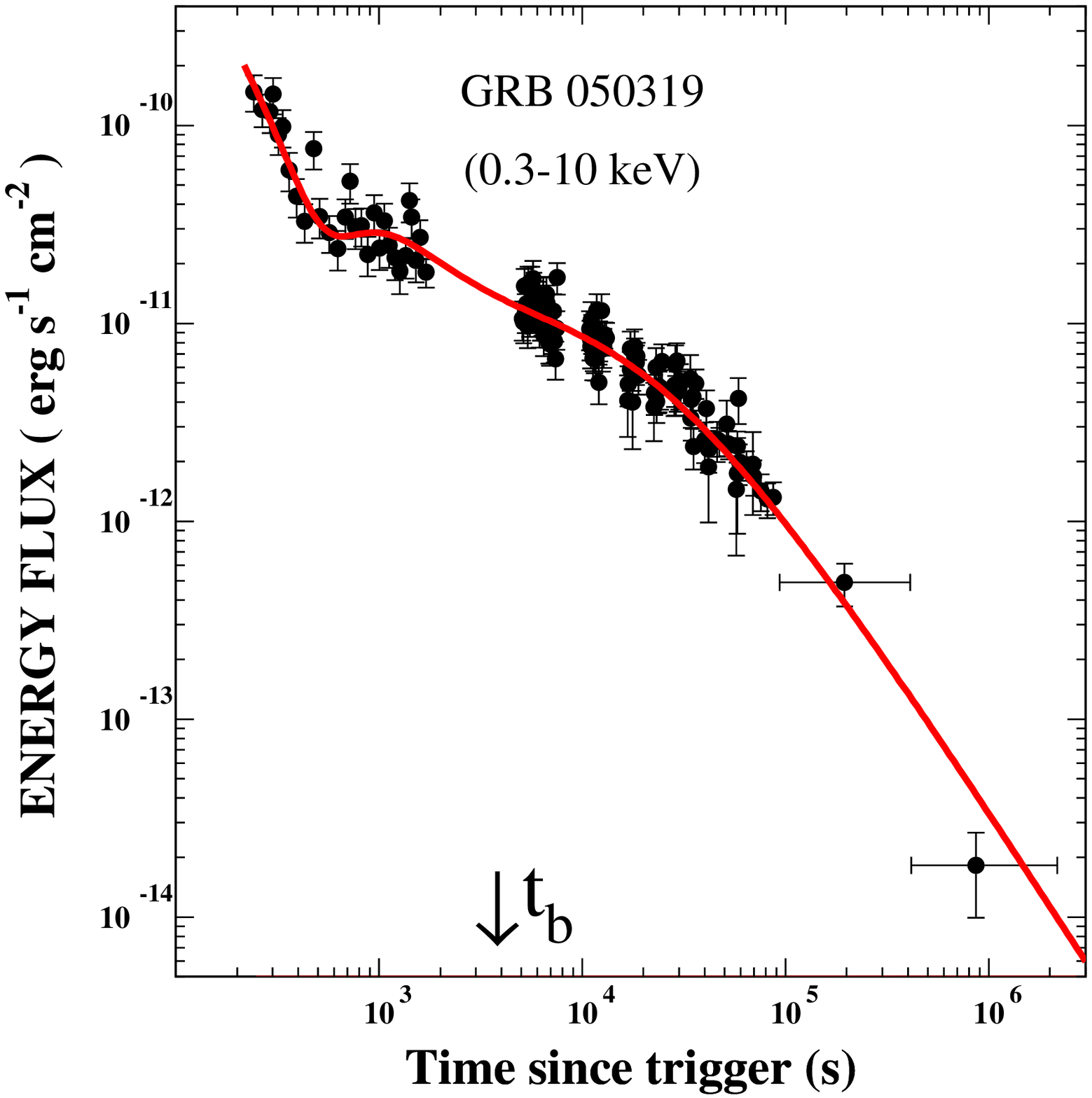,width=8.cm,height=6.cm}
\epsfig{file=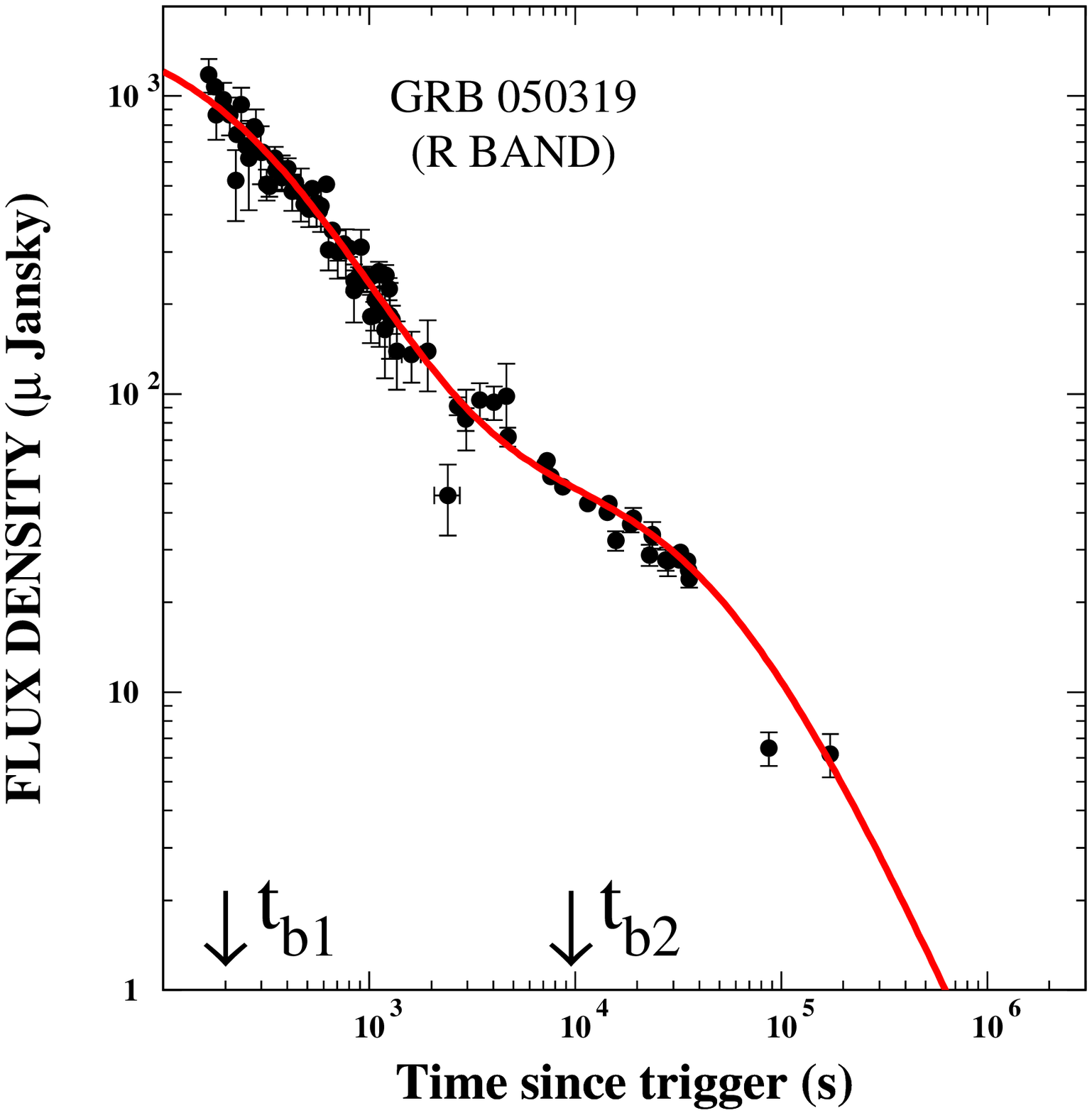,width=8.cm,height=6.cm}
}}
\vbox{
\hbox{
\epsfig{file=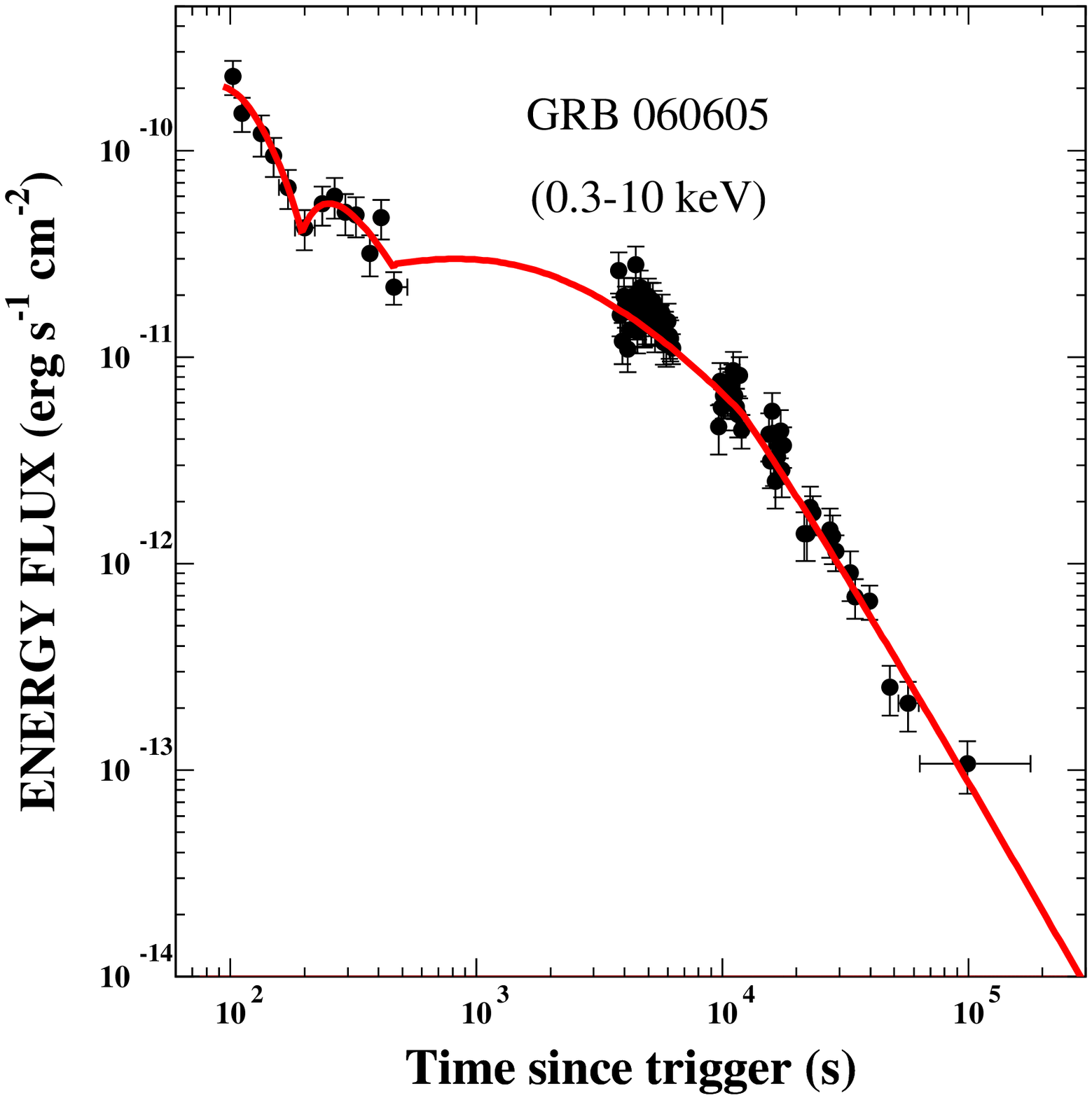,width=8.cm,height=6.cm}
\epsfig{file=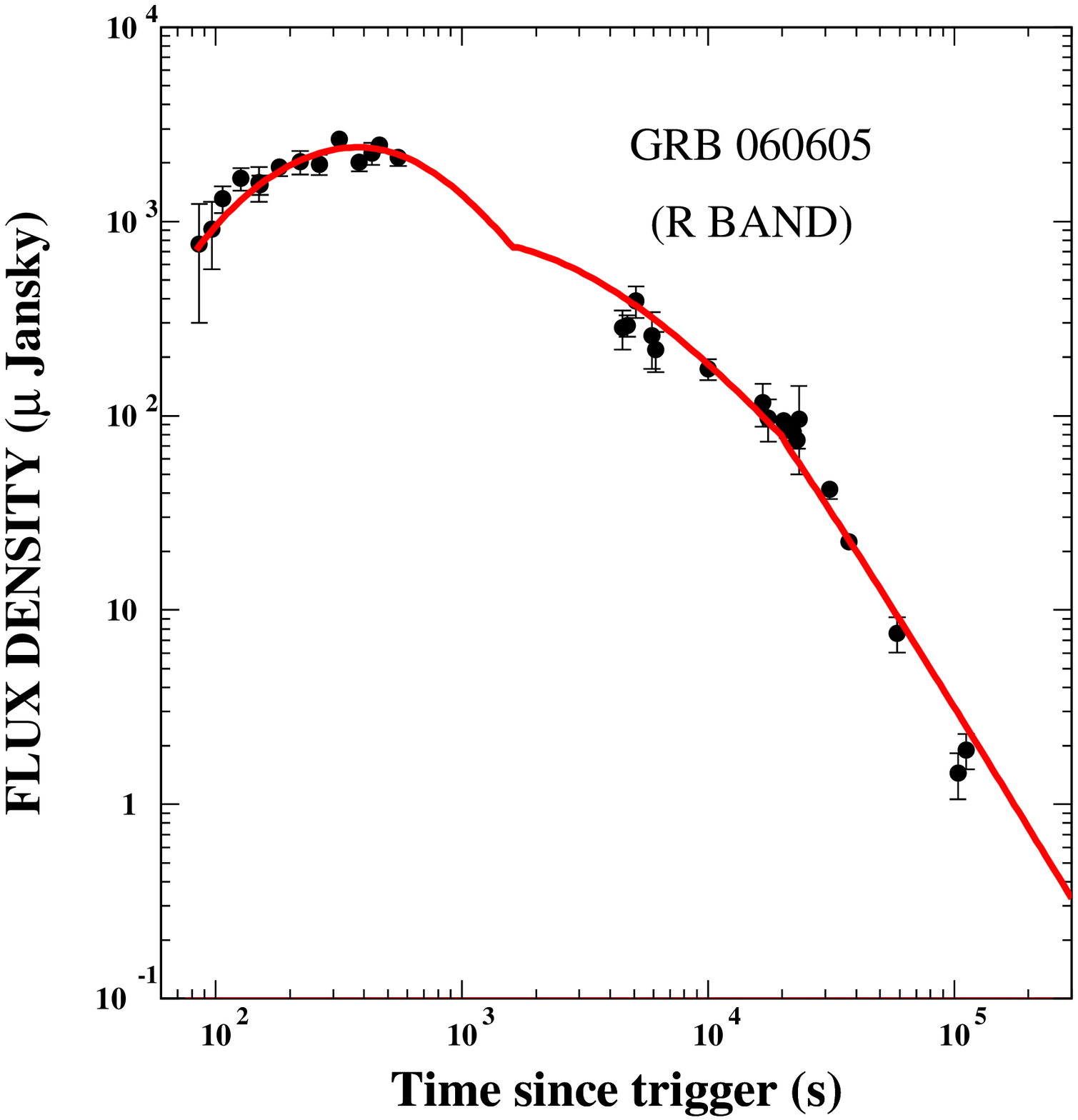,width=8.cm,height=6.cm}
}}
\vbox{
\hbox{
\epsfig{file=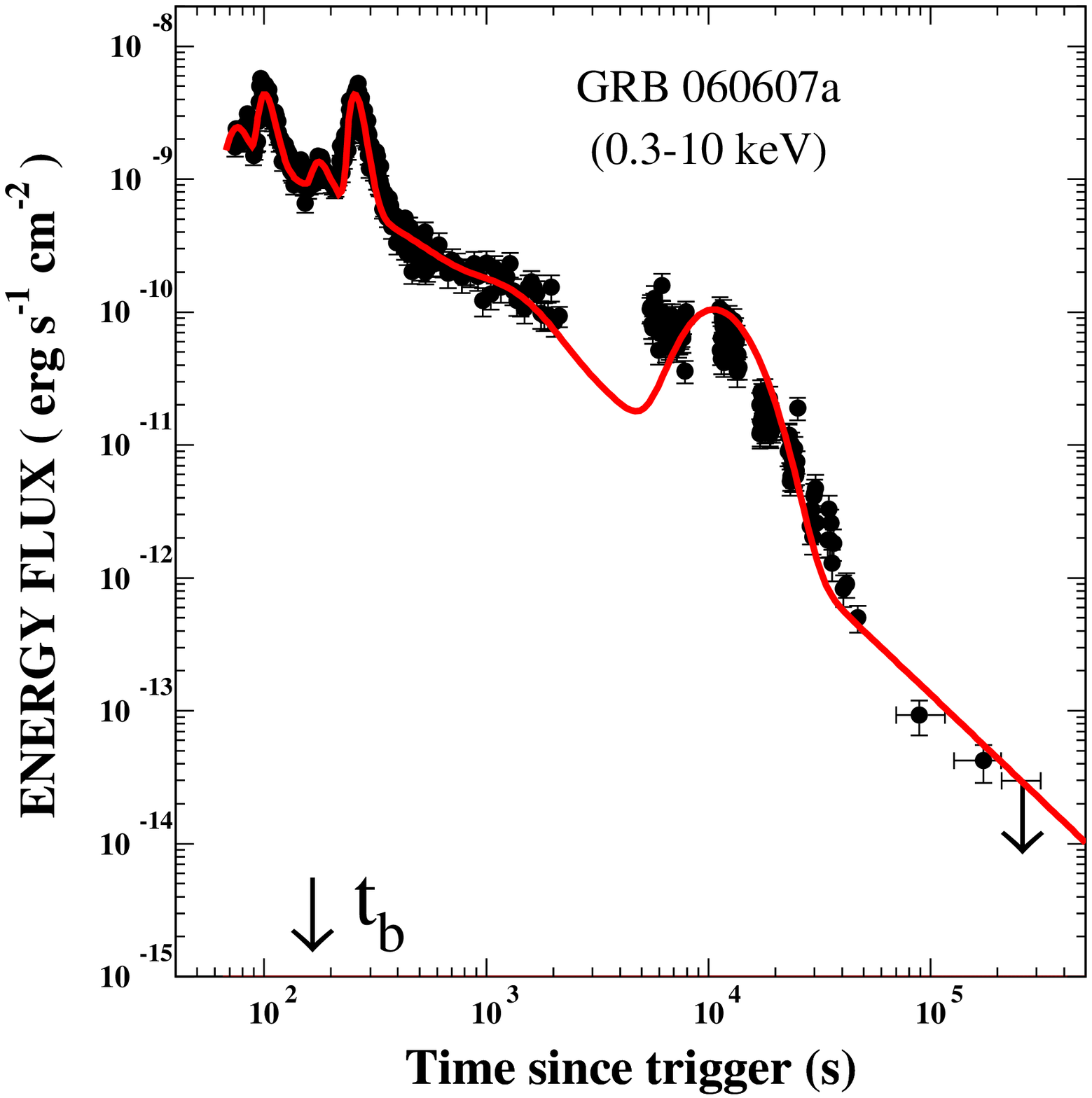,width=8.cm,height=6.cm}
\epsfig{file=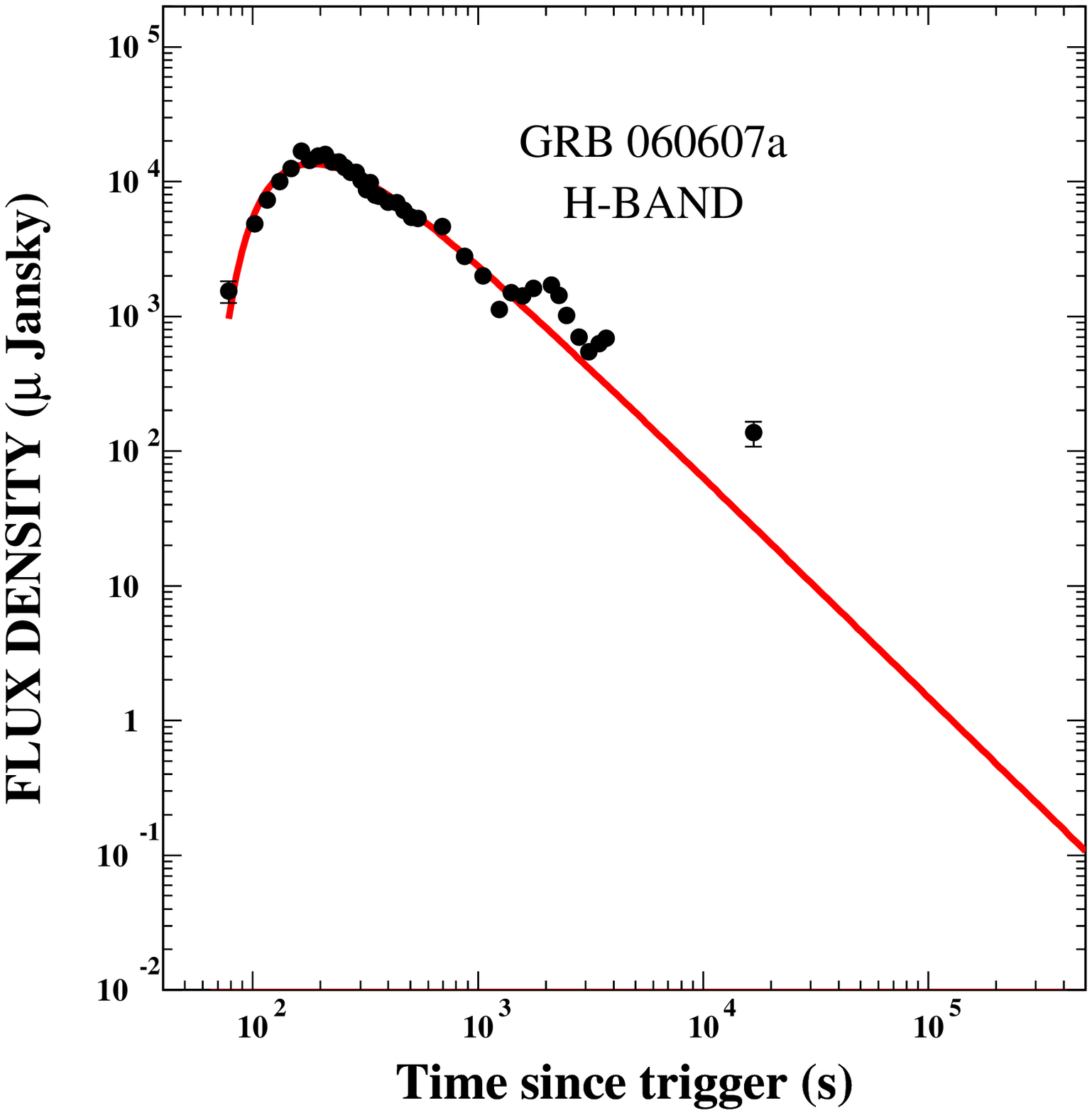,width=8.cm,height=6.cm}
}}
\caption{Comparison between complex chromatic light 
curves of GRBs  and their CB-model description for:
{\bf Top left (a):} The X-ray light curve  of GRB 050319.
{\bf Top right (b):} The $R$-band light curve 
of GRB 050319.
{\bf Middle left (c):} The X-ray light curve of GRB 060605.
{\bf Middle right (d):} The  $R_c$-band light curve of GRB 060605.
{\bf Bottom left (e):} The X-ray light curve of GRB 060607A.
{\bf Bottom right (f):}  The $H$-band light curve
of GRB 060607.
}
\label{f8}
\end{figure}

\newpage 
\begin{figure}[]
\vspace{-2.cm}
\hskip -1.cm 
\vbox{
\hbox{
\epsfig{file=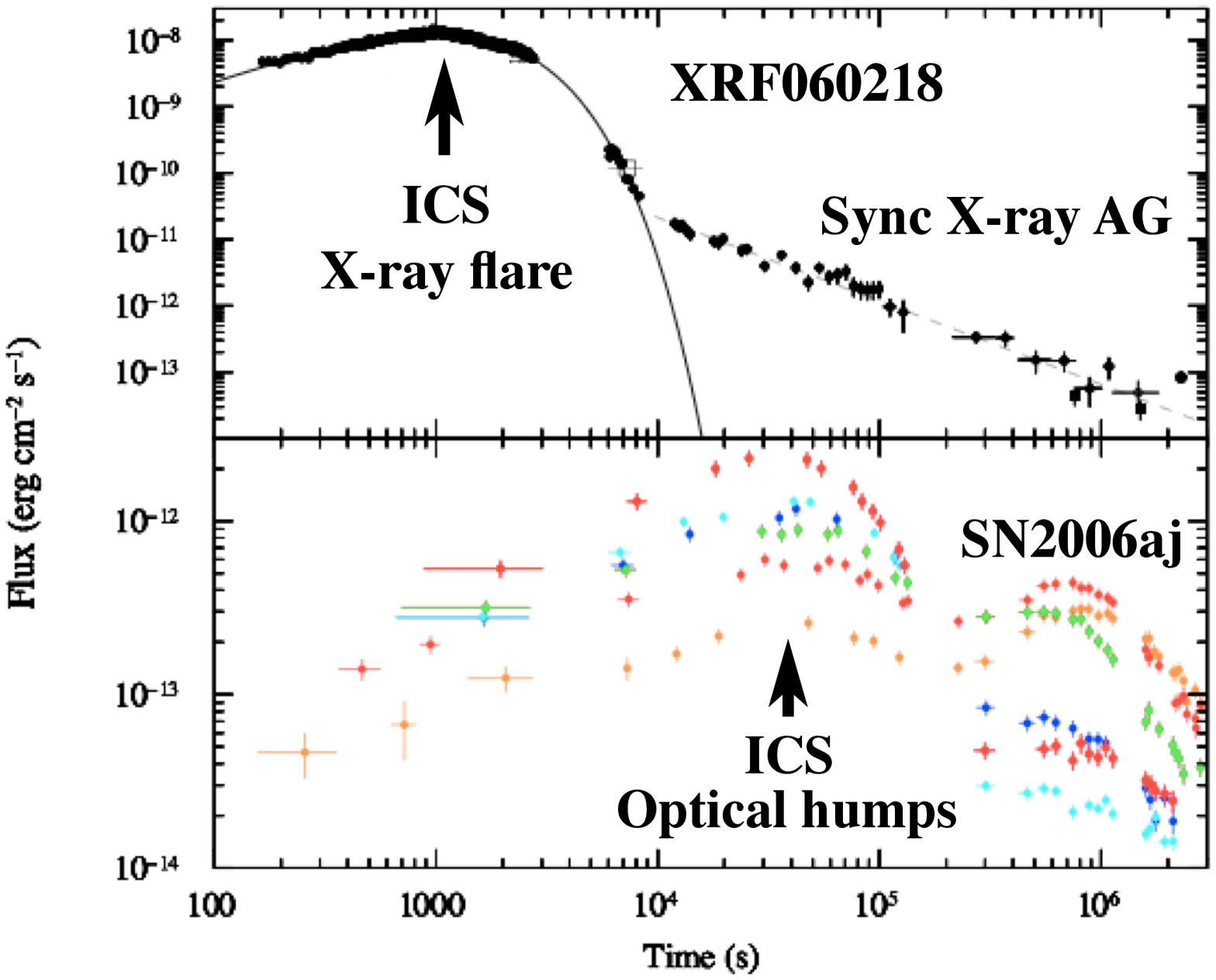,width=18.cm,height=12.cm } 
}
\vspace{-1.5 cm}
}
\vbox{
\hbox{
\epsfig{file=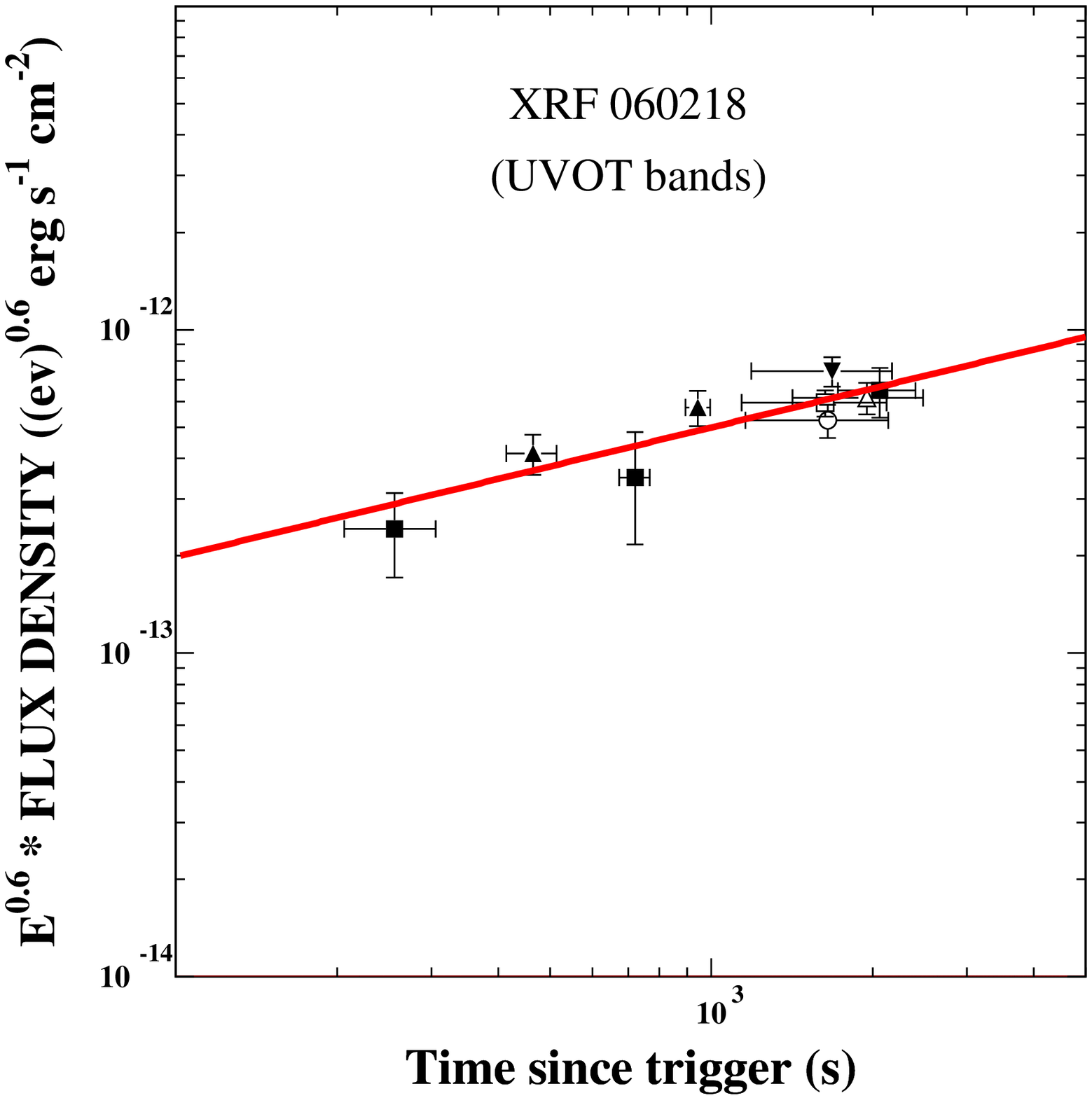,width=8.cm,height=6.cm}
\epsfig{file=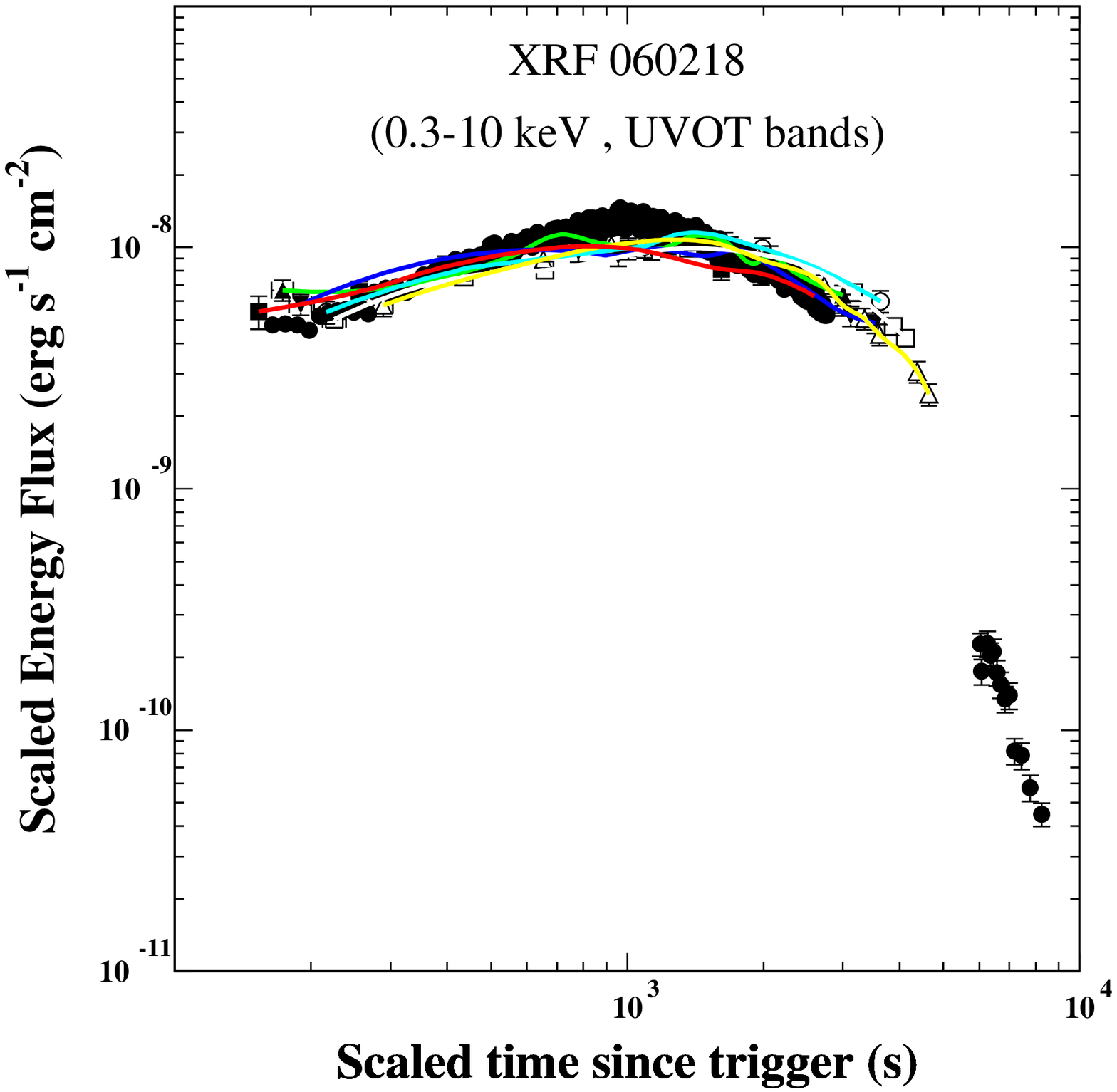,width=8.cm,height=6.cm}
}}
\caption{X-ray and {\it UVO} light curves of XRF 060218/SN2006aj.
{\bf (a) and (b): The top figure} (Campana et al.~2006b). Upper half {\bf (a)}:
The unabsorbed 0.3 -10 keV Swift-XRT light curve. The line 
is a sum of a cut-off power-law and a black body with 
fitted time-dependent radius and temperature. 
The dashed line is their best-fit power law for $t\!>\!10$ ks. The  
arrows indicate rough peak-flux times. 
Lower half {\bf (b)}: Energy fluxes 
corrected for reddening: red: $V$; 
green: $B$; blue: $U$, light 
blue: UVW1; magenta: UVM1 and yellow: UVW2. 
{\bf Bottom left (c)}: De-reddened energy flux densities 
multiplied by $\nu^{0.6}$, predicted to have the slope of the plotted line  ($\propto\!t^{0.4}$).
{\bf Bottom right (d)}: The unabsorbed and the de-redenned energy fluxes, 
divided by the band-width ratios, plotted as functions 
of $(E/E_X)^{1/2}\, t$.
}
\label{f9}
\end{figure}

\newpage 
\begin{figure}[] 
\centering 
\vspace{-1cm} 
\vbox{ 
\hbox{ 
\epsfig{file=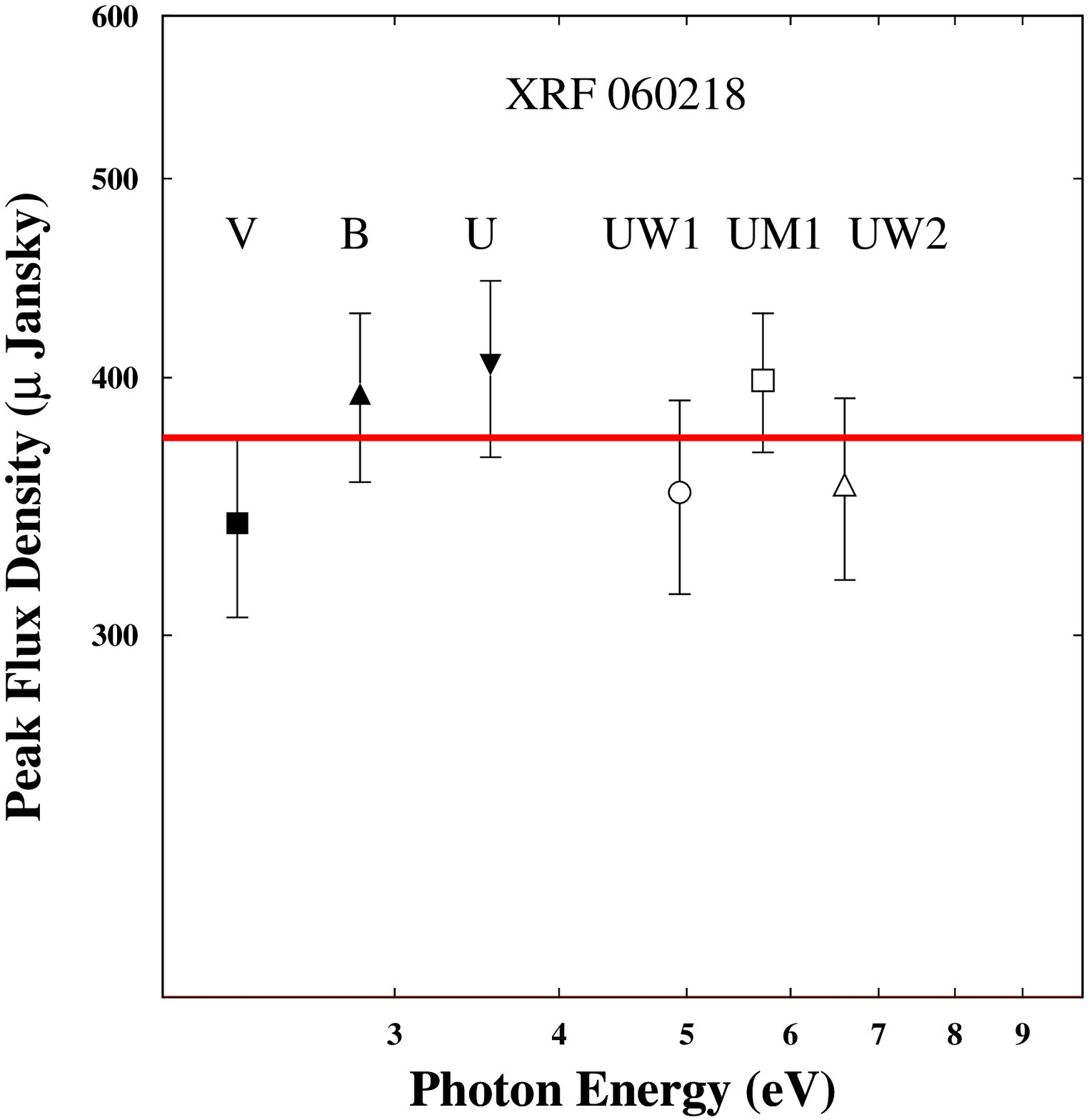,width=8.cm,height=6.cm} 
\epsfig{file=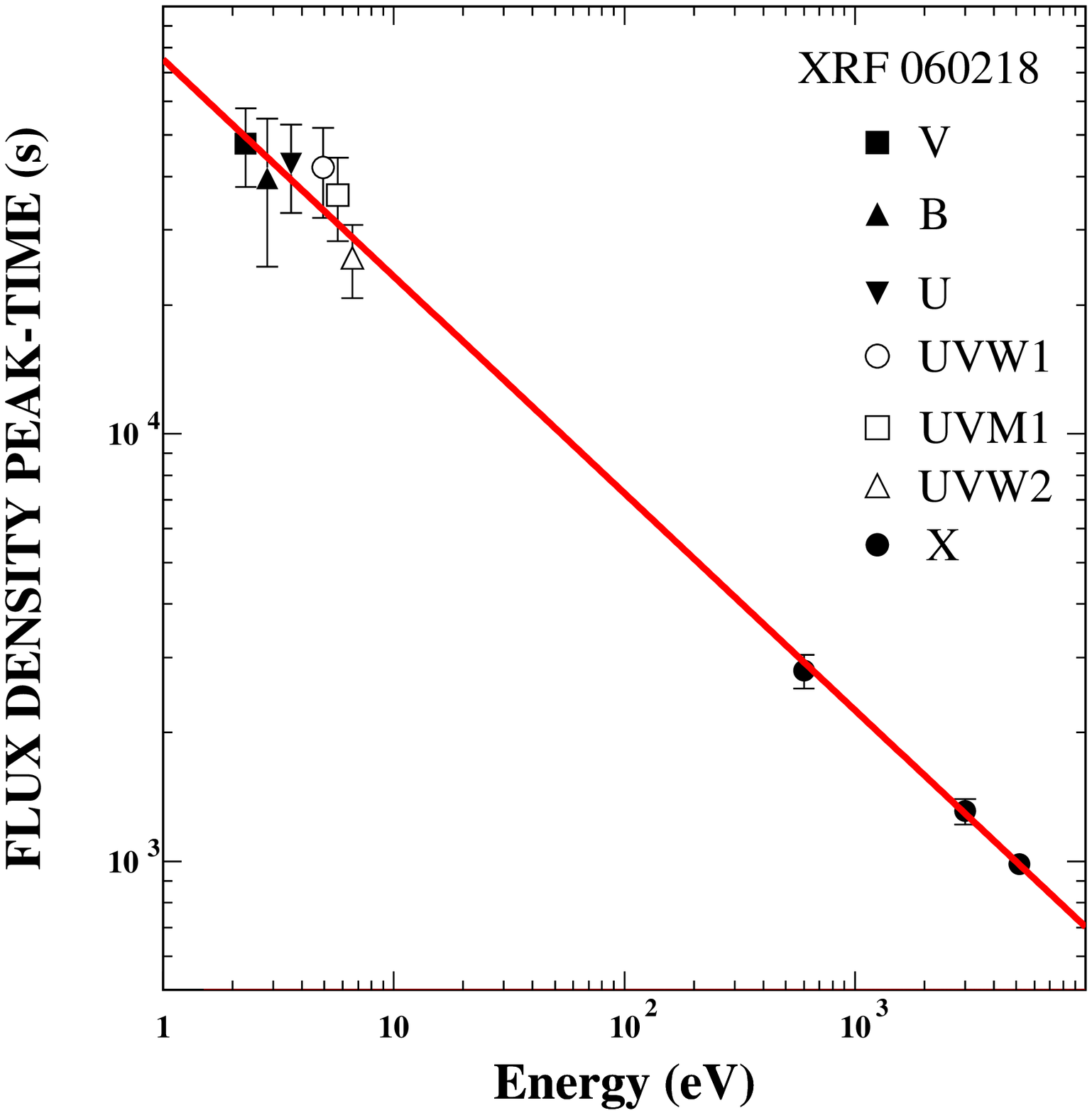,width=8.cm,height=6.cm} 
}} 
\vbox{ 
\hbox{ 
\epsfig{file=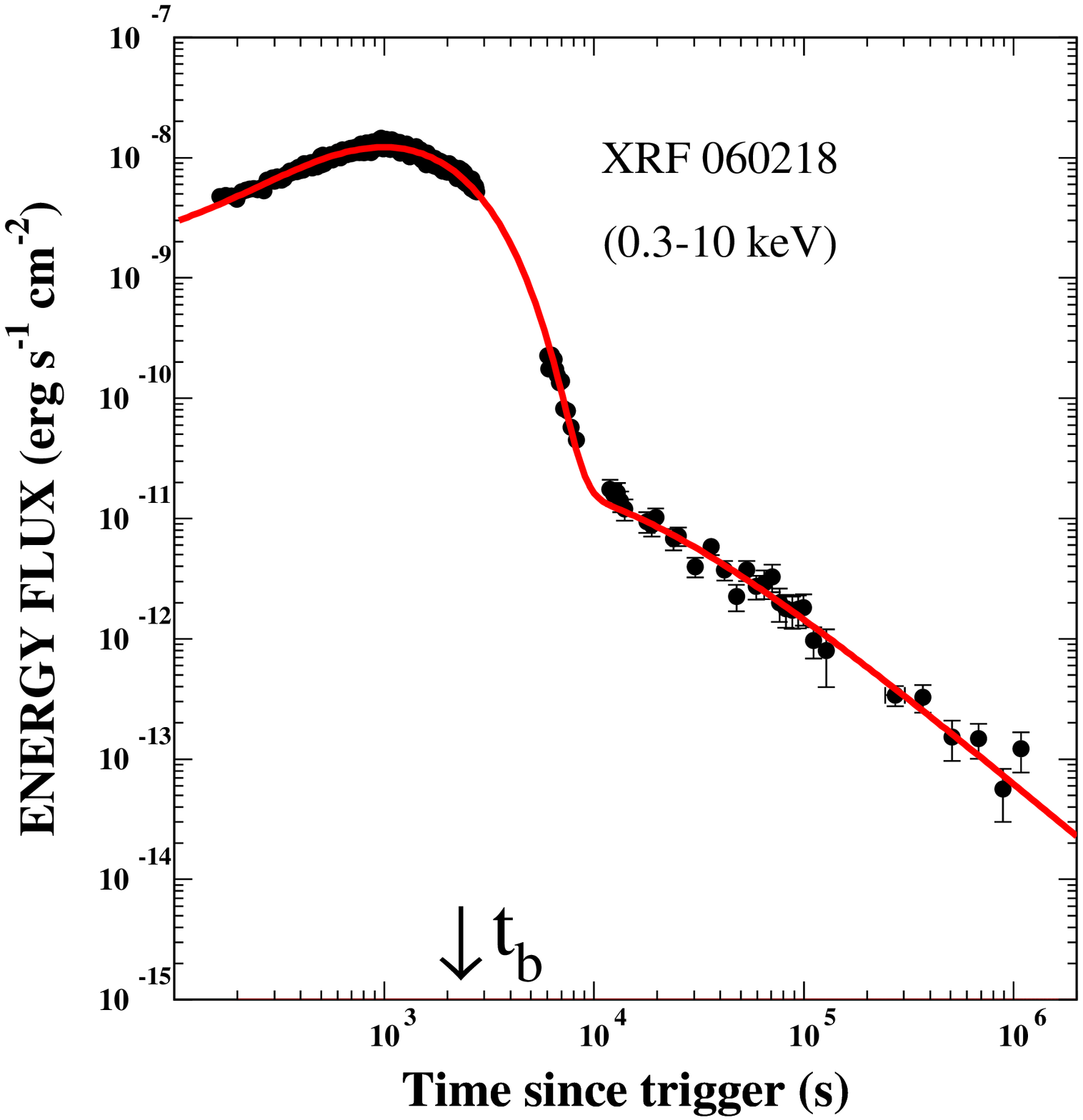,width=8.cm,height=6.cm} 
\epsfig{file=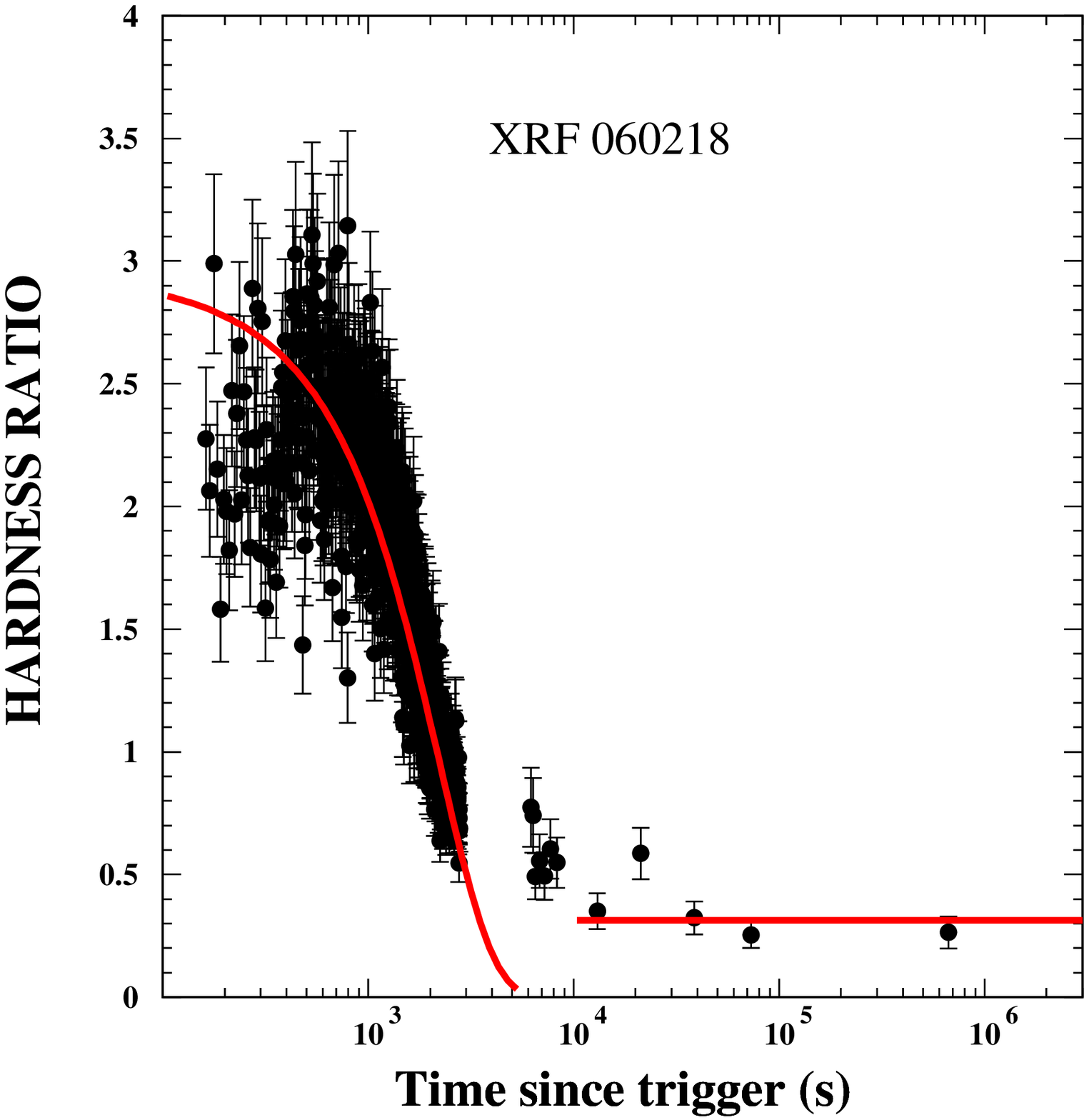,width=8.cm,height=6.cm} }} 
\vbox{ 
\hbox{ 
\epsfig{file=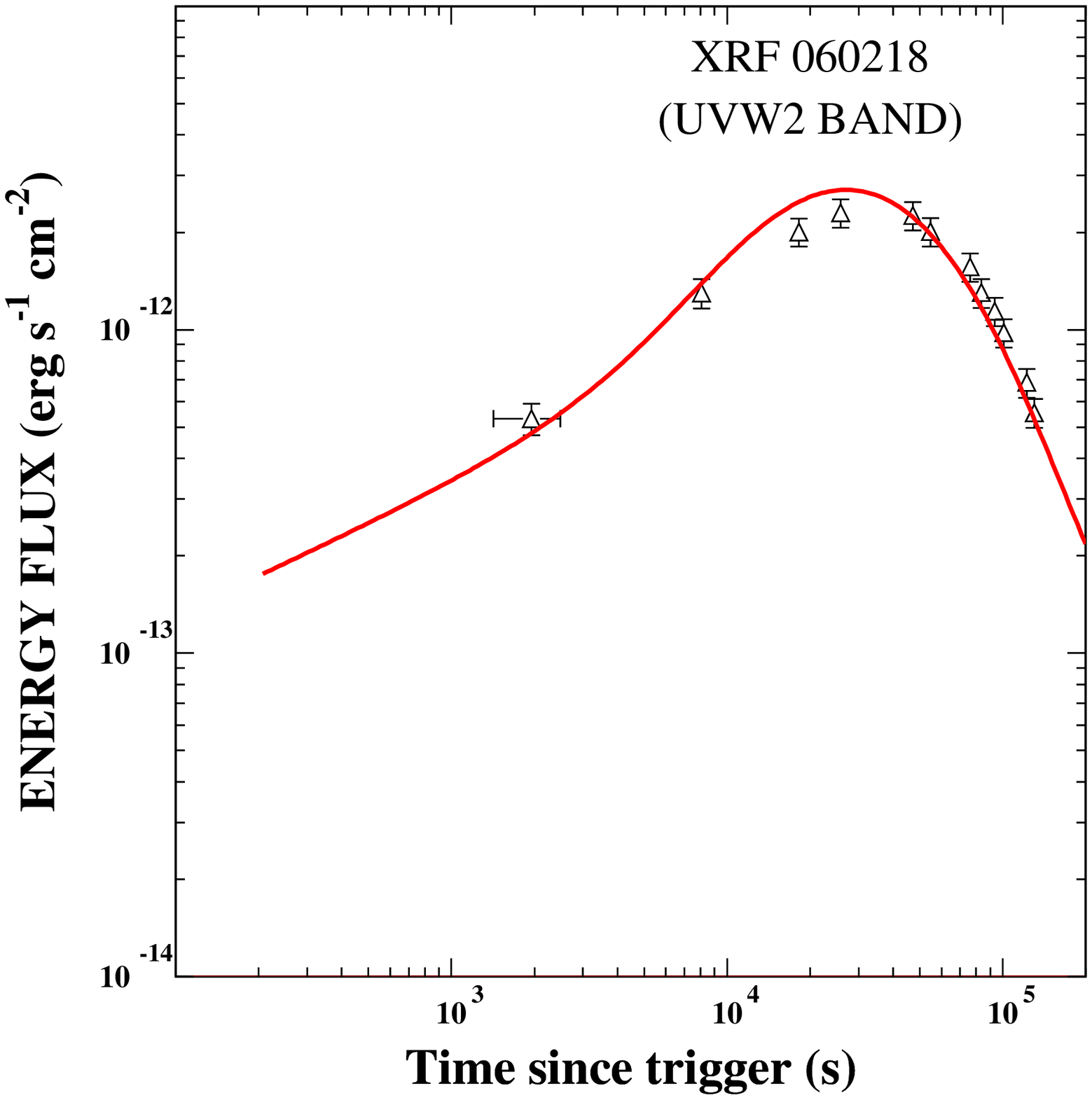,width=8.cm,height=6.cm} 
\epsfig{file=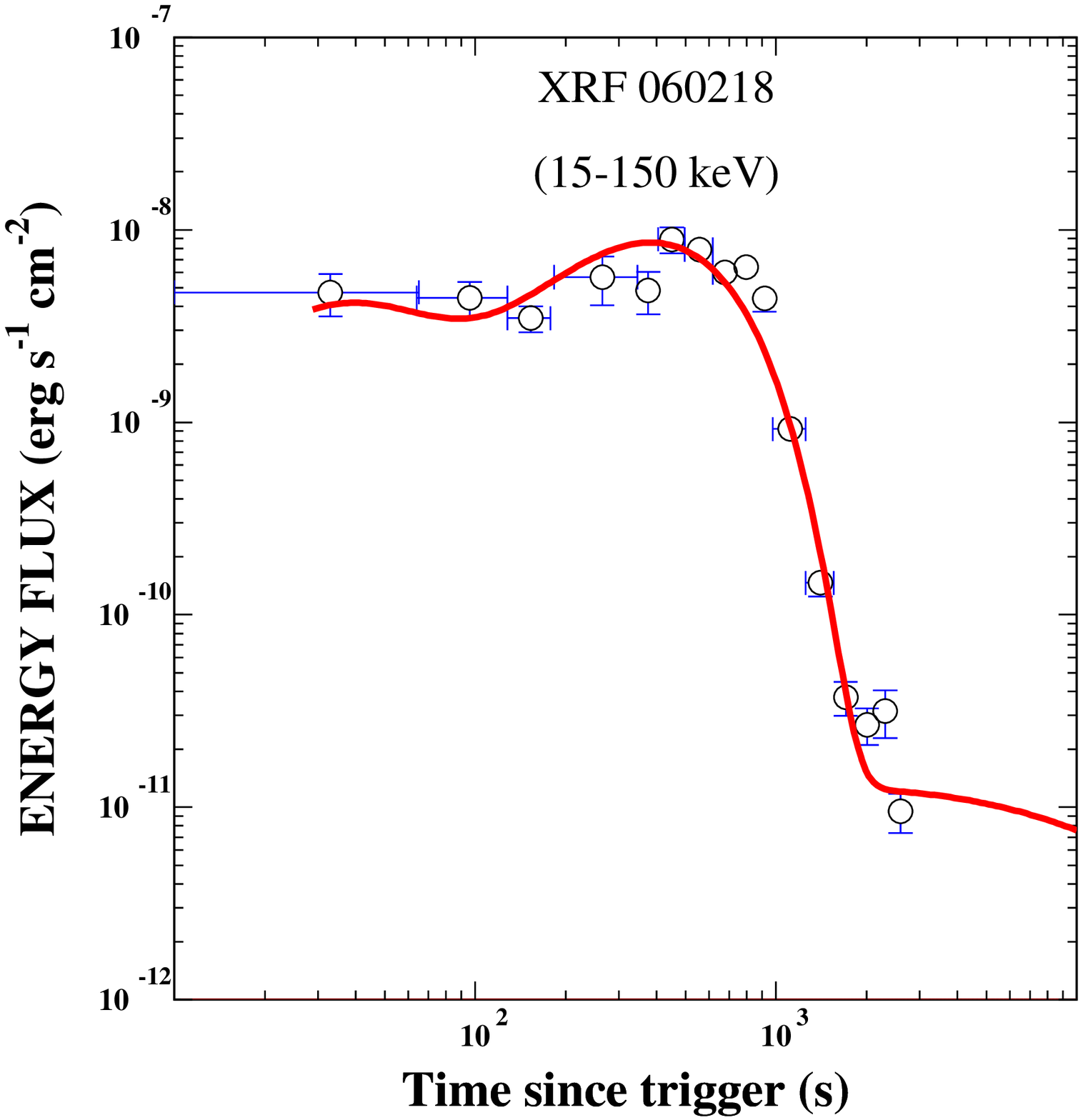,width=8.cm,height=6.cm} }} 
\caption{XRF 060218: data and CB-model predictions.
{\bf Top left (a)}: De-reddened UVOT PEF
divided by $E\, \Delta \lambda/\lambda$, plotted at the central 
energy of each band. The line is the  
prediction of the $E\, t^2$ law.  
{\bf Top right (b)}: PEF times
in XRT and UVOT filters, and the $E\,t^2$ law (red line's slope). 
{\bf Middle left (c)}: Swift
unabsorbed XRT light curve.
The  ICS $\to$ SR transition is at $\sim 9$ ks. 
{\bf Middle right (d)}: XRT hardness ratio.
{\bf Bottom left (e)}: UVW2 light curve.
{\bf Bottom Right (f):} BAT 30-150 keV  $\gamma$-ray light 
curve and its expected shape from 
the $E\,t^2$ law. 
An early peak, hinted by the hardness ratio,
was added.}

\label{f10}
\end{figure}

\end{document}